\documentclass[a4paper,11pt]{article}
\pdfoutput=1
\usepackage{jheppub}
\usepackage{amssymb,amsmath}
\usepackage{mathtools}
\usepackage{dsfont}
\usepackage{graphicx}
\usepackage{caption}
\usepackage{subcaption}
\usepackage{ytableau}
\usepackage{slashed}
\usepackage{enumerate}

\newcommand\br[1]{\{\mathbf{#1}\}}

\newcommand\nn{\nonumber}

\newcommand\Tstrut{\rule{0pt}{2.6ex}} 
\newcommand\Bstrut{\rule[-0.9ex]{0pt}{0pt}}
\usepackage{xcolor}

\begin{document}

\title{Supersymmetric Massive Gravity}

\author[a]{Laura Engelbrecht}
\emailAdd{ljohnson@phys.ethz.ch}
\author[b]{Callum R.~T.~Jones}
\emailAdd{cjones@physics.ucla.edu}
\author[c]{Shruti Paranjape}
\emailAdd{sparanjape@ucdavis.edu}

\affiliation[a]{
Institute for Theoretical Physics, ETH Zurich, Wolfgang-Pauli-Strasse 27, 8093, Zurich, Switzerland}
\affiliation[b]{
Mani L. Bhaumik Institute for Theoretical Physics, Department of Physics and Astronomy, University of California Los Angeles, Los Angeles, CA 90095, USA
}
\affiliation[c]{Center for Quantum Mathematics and Physics (QMAP)
Department of Physics \& Astronomy, University of California, Davis, CA 95616 USA
}

\abstract{We initiate a systematic study of the self-interactions of a massive spin-2 ``graviton" consistent with up to $\mathcal{N}=4$ supersymmetry. Using a recently developed massive on-shell superspace formalism, we construct the most general set of cubic massive graviton amplitudes in a form with all supersymmetry and Lorentz invariance manifest. We find that for $\mathcal{N}\geq 3$ supersymmetry, the family of consistent interactions coincide with those of the ghost-free dRGT model. For $\mathcal{N}=4$ (maximal) supersymmetry there is a single consistent cubic interaction which coincides with the unique structure required for the absence of asymptotic superluminality. Additionally, we discuss the structure of interactions in the high-energy limit, connections to supersymmetric Galileons and the possibility of a supersymmetric massive double copy.}

\maketitle
\flushbottom


\section{Introduction}
\label{sec:intro}

Supersymmetry is a powerful organizing principle for understanding the landscape of consistent models of quantum field theory and gravity. Phenomena of interest that can usually only be understood qualitatively, can -- for supersymmetric models -- often be brought under quantitative control. Even if the ultimate goal is to analyze the physics of a non-supersymmetric model, it is therefore often useful to first construct and study a closely related supersymmetric version.  

This strategy has been extremely profitable for understanding the dynamics of gauge theories \cite{Brink:1976bc,DHoker:1999yni} and gravity \cite{Freedman:1976xh,Deser:1976eh,Freedman:2012zz}. A general lesson is that models with \textit{maximal} supersymmetry\footnote{In this paper, all models are in $d=4$ (unless otherwise specified) with the associated counting of supersymmetries, \textit{e.g.} 16 supercharges corresponds to $\mathcal{N}=4$. }, such as $\mathcal{N}=4$ super Yang-Mills and $\mathcal{N}=8$ supergravity, are the most solvable, but also the most rigid. Even though the field content of these models is often vast ($4d$ $\mathcal{N}=8$ supergravity has $256$ propagating degrees of freedom compared to $2$ for non-supersymmetric Einstein gravity), the symmetry is so constraining that the particle spectrum is completely determined on general grounds and the number of free parameters in the Lagrangian remains small. Given the many remarkable properties of these models, some of which are apparent only in on-shell observables like the S-matrix, a strong case has been made that these are in some sense the \textit{simplest} quantum field theories \cite{Arkani-Hamed:2008owk}.

Inspired by this success, it is natural to suppose that supersymmetry may be an equally powerful tool for analyzing models of massive higher-spin states. As a concrete example, in this paper we initiate a systematic study of the possible supersymmetrizations of models of an isolated,  self-interacting massive spin-2 particle in $4d$ Minkowski spacetime, a \textit{massive graviton}. Previous studies of supersymmetric massive gravity include \cite{Gregoire:2004ic, Malaeb:2013lia,Malaeb:2013nra,Ondo:2016cdv,DelMonte:2016czb,Ferrara:2018iko, Ferrara:2018wlb, Ferrara:2020zef}. The study of massive spin-2 field theories has a very long history, beginning with the discovery of physical wave equations and an associated action principle by Fierz and Pauli \cite{Fierz:1939ix}. The proposed interpretation of these models as a long-distance modification of general relativity is briefly reviewed in Section \ref{dRGTreview}. 

In spite of the impressive scope of the literature on massive gravitons, many things remain unknown. Models of isolated massive higher-spin states are usually constructed as low-energy effective field theories (EFT) valid below some finite strong-coupling scale. The general problem of constructing UV completions of these EFTs satisfying the usual properties of locality, causality and unitarity remains unsolved. For the converse problem of demarcating the boundaries of the swampland (the space of otherwise healthy EFTs with no UV completion), various model-independent, bottom-up approaches have been pursued including: constraints from S-matrix positivity \cite{Alberte:2019xfh}, the absence of asymptotic superluminality \cite{Hinterbichler:2017qyt} and bounds on raising the strong coupling scale by tuning self-interactions \cite{Bonifacio:2018vzv,Bonifacio:2019mgk}. From the top-down, the open problem of constructing explicit, asymptotically Minkowski string vacua with a gap in the spin-2 spectrum (or proving a no-go theorem that such a construction is impossible\footnote{For example, see \cite{Klaewer:2018yxi} for recently proposed swampland criteria for massive spin-2 states.}) may be more tractable in the context of maximal supersymmetry. By contrast, for asymptotically AdS spacetimes, in which a weak-coupling spin-2 Higgs mechanism is possible \cite{Karch:2000ct,Karch:2001cw,Porrati:2001gx,Porrati:2003sa,Aharony:2003qf,Duff:2004wh,Aharony:2006hz,deRham:2006pe,Kiritsis:2008at,Apolo:2012gg,Gabadadze:2014rwa,Gabadadze:2015goa,Domokos:2015xka,Gabadadze:2017jom,Bachas:2018zmb}, explicit string vacua with a spin-2 gap are known \cite{Karch:2000ct,Karch:2001cw,Bachas:2018zmb}.

We are also motivated by the prospect of discovering hidden structures or symmetries in non-supersymmetric models of massive gravity. Even though a supersymmetric model necessarily requires an equal number of bosonic and fermionic degrees of freedom, when constructing solutions to the classical equations of motion (or calculating tree-level scattering amplitudes) it is always a consistent truncation to set the fermions to zero. There may be further possible truncations depending on the global symmetries of the model. As a consequence, supersymmetric arguments can play an important role in the classical dynamics of non-supersymmetric models\footnote{We remind the reader of two classic examples. $(i)$ The extremal Reissner-Nordstr\"om solution of the Einstein-Maxwell model admits Killing spinors, a remnant of its allowed embedding as a BPS solution of pure $\mathcal{N}=2$ supergravity. $(ii)$ The \textit{tree-level} (color-ordered) gluon amplitudes of pure Yang-Mills theory satisfy $A_n[+,+,...,+,+] = A_n[+,+,...,+,-]=0$ since these helicity sectors necessarily vanish in supersymmetric Yang-Mills as a consequence of the on-shell supersymmetry Ward identities. }. These hidden structures may only remain in bosonic models with special particle content and tunings of interactions. By constructing the most general \textit{supersymmetrizations} of the self-interactions of a massive graviton we may hope to discover such special tunings. 

The problem of constraining the dynamics of any model with arbitrarily extended supersymmetry is highly non-trivial; the additional difficulty of constructing models of interacting massive higher-spin states makes this problem essentially intractable with presently available tools. There is no known formalism that allows for the construction of generic models with the following properties simultaneously: $(i)$ manifest $\mathcal{N}\geq 4$ supersymmetry, $(ii)$ manifest Lorentz invariance and $(iii)$ a realization of the supersymmetry algebra that closes off-shell. We note that for $\mathcal{N}=1$ supersymmetry, the off-shell superfields corresponding to a massive graviton multiplet have been constructed \cite{Buchbinder:2002gh, Zinoviev:2018eok, Zinoviev:2002xn,Gates:2013tka}. One could imagine constructing the models presented in this paper using a version of harmonic superspace \cite{Galperin:2001seg}, with some of the supersymmetry non-manifest; we will not pursue this direction. Instead, we take inspiration from the success of the on-shell \textit{scattering amplitudes} program \cite{Elvang:2009wd}, and by-pass the construction of an off-shell effective action in favor of the on-shell (tree-level) S-matrix elements. Here we have many powerful tools at our disposal, especially a massive version of the spinor-helicity formalism \cite{Arkani-Hamed:2017jhn}, and a recently developed, manifestly Lorentz-covariant \textit{on-shell superspace} \cite{Herderschee:2019ofc}. Unlike its off-shell cousin, not only can on-shell superspace be (trivially) extended to arbitrary numbers of supercharges, but the resulting \textit{superamplitudes} are much simpler than the non-supersymmetric amplitudes! At the end of the calculation we can match the supersymmetric amplitudes to off-shell local operators and recover the constraints of supersymmetry. 

We would like emphasize that the techniques used in this paper are very general. We have focused on models of an isolated massive spin-2 particle as the simplest non-trivial example of a higher-spin state. We expect that on-shell superspace methods are a powerful approach to understanding the dynamics of supersymmetric massive higher-spin interactions in a wider variety of contexts. 

\subsection{Ghost-Free Massive Gravity}
\label{dRGTreview}
Much of the contemporary interest in models of massive gravity has followed from the discovery of the \textit{ghost-free} model of de Rham, Gabadadze and Tolley (dRGT) \cite{deRham:2010kj}. For a more comprehensive review of these developments see \cite{Hinterbichler:2011tt, deRham:2014zqa}. On a Minkowski background, dRGT massive gravity is described by an effective action obtained as a deformation of the Einstein-Hilbert action by a specially tuned potential of zero-derivative interactions,
\begin{equation} 
\label{nonlinlag}
S_{\text{dRGT}}[h]=\frac{M_{\text{P}}^2}2 \int\text{d}^4 x\sqrt{-g}\left( R[g] +{m^2}\left[S_2({\cal K})+\alpha_3 S_3({\cal K})+\alpha_4 S_4({\cal K})\right]\right),
\end{equation}
where  
\begin{align}
    g_{\mu\nu}&=\eta_{\mu\nu}+\frac{2}{M_{\text{P}}}h_{\mu\nu}\,, \nonumber\\
    S_n^{\rm }(\mathcal{K}) &= n!\, {\mathcal{K}^{[\mu_1}}_{\mu_1}{\mathcal{K}^{\mu_2}}_{\mu_2}\cdots {\mathcal{K}^{\mu_n]}}_{\mu_n}\,, \nonumber\\
    {\mathcal K^\mu}_{\nu} &= {\delta^\mu}_\nu-\sqrt{{\delta^\mu}_\nu-{h^\mu}_\nu}=-\sum_{n=1}^{\infty}\frac{(2n)!}{(1-2n)(n!)^2 4^n}{\left(h^n\right)^\mu}_\nu.
\end{align}
This model describes a self-interacting massive spin-2 field $h_{\mu\nu}$ in terms of 3 independent dimensionless parameters $(\alpha_3,\alpha_4,m/M_{\text{P}})$\footnote{Another commonly used set of parameters for the potential are $(c_3,d_5)$ \cite{Cheung:2016yqr}, which are related by $\alpha_3=-2c_3$ and $\alpha_4=-4d_5$.}. As shown in \cite{deRham:2010kj}, any deviation from this particular structure reintroduces an additional ghostly degree of freedom in the non-linear regime, known as the Boulware-Deser ghost \cite{Boulware:1972yco}. 

In this paper we will not assume \textit{a priori} that the massive graviton self-interactions are those of the dRGT model. In Section \ref{sec:susy3point}, this generality will allow us to observe that for cubic interactions, $\mathcal{N}\geq 3$ supersymmetry \textit{requires} the ghost-free tuning. Nonetheless, the problem of constructing supersymmetrizations of the dRGT model is an important motivation for this work and so we will briefly review some of the relevant features. 

The action (\ref{nonlinlag}) is suggestive of the interpretation of this model as a long-distance modification of Einstein gravity by an explicit breaking of diffeomorphism invariance. When the massive graviton is coupled to a heavy compact source of mass $M$, the effect of this breaking modifies the effective gravitational force on a test body. At distances large compared to the Compton wavelength of the massive graviton
\begin{equation}
    r\gtrsim r_C, \hspace{10mm} r_C = \frac{1}{m},
\end{equation}
the inclusion of a mass term leads to an expected exponential Yukawa suppression of the gravitational force. At intermediate distances 
\begin{equation}
    r_C \gtrsim r \gtrsim r_V, \hspace{10mm}r_V=\left(\frac{M}{M_\text{P}^2 m^2}\right)^{1/3},
\end{equation}
where $r_V$ is the so-called \textit{Vainshtein} radius \cite{VAINSHTEIN1972393}, linear classical effects dominate. In this regime we find the famous \textit{vDVZ discontinuity} \cite{vanDam:1970vg,Zakharov:1970cc}, the helicity-0 longitudinal mode of the massive graviton does not decouple and the dynamics of a test body do not agree with Einstein gravity, even for vanishingly small mass. At shorter distances
\begin{equation}
    r_V \gtrsim r \gtrsim r_Q, \hspace{10mm} r_Q=\left(\frac{1}{M_\text{P} m^2}\right)^{1/3},
\end{equation}
non-linear classical effects dominate. In this regime the helicity-0 mode is effectively screened due to strong self-interactions, and the dynamics of the test body \textit{agrees} with Einstein gravity \cite{VAINSHTEIN1972393,Babichev:2013usa}. Finally, at distances 
\begin{equation}
    r_Q \gtrsim r,
\end{equation}
quantum effects become important. In \cite{deRham:2013qqa} it was shown that while massive graviton loop effects will in general de-tune the ghost-free potential (\ref{nonlinlag}), the associated ghostly mode has a mass $m_{\text{ghost}}\sim M_\text{P}$ and the phenomenology at longer distances is not substantially modified.

The existence of a healthy EFT of a massive spin-2 particle that reproduces the phenomenology of general relativity in an appropriate regime is surprising, and there is a clear motivation to try and construct a UV completion. In contrast to the model-independent constraints described above, there have been additional bottom-up studies on the constraints imposed on the dRGT model by requiring standard causality and unitarity properties in the UV. Time advances in pp-wave solutions and associated closed time-like curves were analyzed in \cite{Camanho:2016opx}; the existence of an asymptotic Shapiro time-advance or time-delay, related to the sign of the eikonal phase, was studied in \cite{Hinterbichler:2017qyt,Bonifacio:2017nnt}. In both cases, avoiding apparently pathological behavior requires a specific tuning of the cubic dRGT parameter
\begin{equation}
\label{eq:alpha3}
    \alpha_3=-\frac{1}{2}.
\end{equation}
As first studied in \cite{Cheung:2016yqr}, with proposed refinements and clarifications \cite{Bellazzini:2017fep,deRham:2018qqo,Alberte:2019xfh,deRham:2017xox}, S-matrix positivity bounds related to analyticity, causality and unitarity in the UV, constrains the parameters $(\alpha_3,\alpha_4)$ to a small island\footnote{As discussed in \cite{Camanho:2016opx} there is a non-zero region of overlap between the causality and positivity bounds.}. It remains an open question how the notion of causality used in the derivation of the S-matrix positivity bounds is related to that used in arriving at \eqref{eq:alpha3}.

\subsection{Outline of this Paper}

In Section \ref{sec:multiplets}, we introduce the particle content of graviton supermultiplets with $\mathcal{N}=1,2,3,4$ supersymmetry. In Section \ref{sec:onshellWI}, we discuss the action of the supercharges on these multiplets and their Ward identities. Using the massive on-shell superspace formalism first introduced in \cite{Herderschee:2019ofc}, we construct graviton superfields in Section \ref{sec:onshellsuperfields} and find the most generic formula for a superamplitude that satisfies the supersymmetry Ward identities in Section \ref{sec:solvingWI}. 

In Section \ref{sec:cubicamps}, we introduce a basis of operators (and their corresponding amplitudes). In Section \ref{sec:susy3point}, we place $R$-symmetry and exchange symmetry constraints on this basis to determine which interactions are compatible with varying amounts of supersymmetry. 

In Section \ref{sec:HElim}, we study the high energy limit of supersymmetric massive gravity- first via massless limits of the multiplets in Section \ref{sec:masslesslimitmultiplets}, followed by a discussion of the high energy limit of massive spin-2 interaction terms and their (in)compatibility with massless supersymmetry Ward identites in Section \ref{sec:masslesscubic} .

In Section \ref{sec:doublecopy}, we first construct supersymmetric massive gluon multiplets and cubic interactions in Section \ref{sec:sYM}. We show how these double copy to massive graviton superfields and 3-point amplitudes in Section \ref{sec:DC}. We end with a Discussion section, which pays special attention to supersymmetric Galileon no-go theorems.

\section{Supersymmetry Ward Identities and On-Shell Superspace}
\label{sec:SWIsuperspace}
In this section, we first introduce the graviton supermultiplets studied in this paper in Section \ref{sec:multiplets}. Sections \ref{sec:onshellWI}, \ref{sec:onshellsuperfields} and \ref{sec:solvingWI} introduce on-shell superspace, construct explicit on-shell superfields and derive a simple expression for massive amplitudes in a generic supersymmetric model respectively.

\subsection{Massive Graviton Supermultiplets}
\label{sec:multiplets}
Particle states in a supersymmetric model may be organized into supermultiplets, irreducible representations of the supersymmetry algebra (\ref{susyalgebra}). The way in which supersymmetry constrains interactions among states depends on the multiplets present in the model. In this section we classify the multiplets which may contain a massive graviton. 

We need to impose some physical conditions to distinguish a \textit{massive graviton} multiplet from a generic multiplet containing massive spin-2 states. In this paper we make the following assumptions:
\begin{enumerate}[(i)]
    \item A massive graviton should be a singlet under all internal symmetries, i.e. it must have the same quantum numbers as a massless graviton. 
    \item A massive graviton multiplet contains no states with spin $>2$. Higher-spin states may be necessary to UV complete the model but they are assumed to be heavier than the massive graviton. 
    \item The lightest spin-2 state of a model of massive gravity should be non-degenerate, i.e. there should be a unique lightest spin-2 state identified as the massive graviton.
\end{enumerate}
Note that assumption (iii) already implies that the graviton cannot carry any additive charges, since this would require a distinct CPT conjugate state with opposite charge. Assumption (i) is not completely redundant however, since it also forbids discrete charges, for example it rules out the possibility that the massive graviton is a parity-odd pseudo-tensor. The requirement that the massive graviton is a singlet is necessary for the existence of cubic self-interactions. Assumption (ii) has an interesting status; the analogue for massless supergravity is a theorem rather than an assumption, since massless spin $>2$ states cannot consistently interact with a massless graviton \cite{Weinberg:1964ew}. For a massive graviton we make this assumption by analogy without a formal justification. Assumption (iii) almost implies (ii), in most cases the only way to construct a multiplet with a non-degenerate spin-2 is for the graviton to be the highest spin state.  

Note that assumptions (i) and (iii) exclude the possibility of obtaining a massive graviton (in the sense defined above) as a Kaluza-Klein (KK) mode from an $S^1$ compactification of a massless graviton in $5d$. Such models contain massless spin-2 KK zero-modes corresponding to $4d$ massless gravitons. Ignoring this, even the non-zero KK modes violate assumption (iii) since they always appear in pairs of left-moving and right-moving states. Furthermore, these massive KK modes are charged under the emergent $U(1)$ gauge field, the \textit{graviphoton}, violating assumption (i) and making cubic self-interactions impossible. 

On-shell supermultiplets are constructed by choosing a Clifford vacuum, defined as a spin-$s$ (sometimes called \textit{superspin}) multiplet of one-particle states satisfying
\begin{align}
    Q_{a\dot{\alpha}}^\dagger |\Omega^{I_1...I_{2s}}\rangle = 0\,,
\end{align}
where $a=1,2,...,\mathcal{N}$ is an R-symmetry index and $I_i=1,2$ is an index of the massive little group $SU(2)_{\text{LG}}$. Following \cite{Arkani-Hamed:2017jhn} we denote a spin-$s$ multiplet as a totally symmetric rank-$2s$ tensor of $SU(2)_{\text{LG}}$. Acting on $|\Omega^{I_1...I_{2s}}\rangle$ with the other supercharges $Q^a_\alpha$ in all possible ways generates the remaining states in the supermultiplet. Note that unlike in the case of massless supersymmetry, the Clifford vacuum is not necessarily the state in the multiplet with the lowest spin. 

The massive supermultiplets satisfying conditions (i)-(iii) were constructed some time ago \cite{Ferrara:1980ra,Ferrara:1981ep}. The results are summarized in Table \ref{tab:reptable} together with a decomposition of the component states into irreducible representations of the maximal R-symmetry group $SU(\mathcal{N})_R \times U(1)_R$. We will not review the details of this construction but will instead only make a few physically relevant comments.

For $\mathcal{N}>1$ supersymmetry, massive supermultiplets can be classified as either \textit{short} or \textit{long}. Short multiplets can arise if the supersymmetry algebra admits a central extension,
\begin{align}
    \{Q^a_\alpha,Q_{\beta}^{b}\} &= \epsilon_{\alpha\beta}Z^{ab} \nonumber\\
    \{Q^{\dagger}_{a\dot{\alpha}},Q_{b\dot{\beta}}^{\dagger}\} &= \epsilon_{\dot{\alpha}\dot{\beta}}Z_{ab}.
\end{align}
Since $Z^{ab}$ is a central charge, it commutes with all other operators in the super-Poincar\'e algebra (\ref{susyalgebra}), including the supercharges. This implies the existence of additional, additive quantum numbers $Z_i$, $i=1,...,\mathcal{N}/2$ carried by every component state of a supermultiplet. Such an algebra has a representation on physical states only if the mass $m$ and central charges $Z_i$ satisfy the BPS bounds,
\begin{align}
\label{BPS}
    m\geq |Z_i|\,.
\end{align}
At the special values $m=|Z_i|$, for some subset of $i$, there are fewer states in the multiplet than the generic (long multiplet) case, since some of the supercharges annihilate all of the states. The existence of an additive quantum number violates assumptions (i) and (iii); for a BPS multiplet with central charge $Z$ we would require a distinct multiplet with central charge $-Z$ to be consistent with the CPT theorem. We therefore conclude:
\begin{center}
    \noindent\fbox{
\begin{minipage}{14cm}
A massive graviton in a model with $\mathcal{N}>1$ supersymmetry, must be a component of a \textit{long} supermultiplet. 
\end{minipage}
}
\end{center}
For $\mathcal{N}$-extended supersymmetry, a long supermultiplet with superspin $s$ contains a \textit{unique} highest-spin state with $s_{\text{max}} = s+\mathcal{N}/2$. Since $s\geq 0$, assumption (ii) therefore leads to an upper bound on the number of supersymmetries:
\begin{center}
    \noindent\fbox{
\begin{minipage}{14cm}
Assuming all of the states in the multiplet have spin $\leq 2$, the maximal number of supersymmetries consistent with a massive graviton is $\mathcal{N}= 4$.
\end{minipage}
}
\end{center}
Finally, it is worth noting that in the context of these models $\mathcal{N}=3$ supersymmetry \textit{does not} automatically imply $\mathcal{N}=4$ supersymmetry. The multiplets given in Table \ref{tab:reptable} are already consistent with CPT and the $\mathcal{N}=3$ and $\mathcal{N}=4$ multiplets are clearly distinct. Furthermore, in Section \ref{sec:cubicstuff} we will see that the constraints on cubic massive graviton interactions are different in the two cases.

\begin{table}
    \begin{subtable}[h]{0.48\textwidth}
        \centering
\begin{tabular}{|c|c|c|c|}
		\hline
		Field & Spin & $U(1)_R$ & Dim.\\
		\hline
		\hline
		$\psi^{IJK}$ & $\frac32$ & 1 & \textbf{1}\\
		\hline
		$\gamma^{IJ}$ & 1 & 0 & \textbf{1}\\
		\hline
		$h^{IJKL}$ & 2 & 0 & \textbf{1}\\
		\hline
		$\tilde{\psi}^{IJK}$ & $\frac32$ & -1 & \textbf{1}\\
		\hline
	\end{tabular}
       \caption{$\mathcal{N}=1$ massive graviton multiplet}
       \label{tab:N1multiplet}
    \end{subtable}
    \hfill
    \begin{subtable}[h]{0.48\textwidth}
        \centering
	\begin{tabular}{|c|c|c|c|c|}
		\hline
		Field & Spin & $U(1)_R$ & $SU(2)_R$ & Dim.\\
		\hline
		\hline
		$\gamma^{IJ}$ & 1 & 2 & $\bullet$ & $\mathbf{1}$ \\
		\hline
		$\lambda^{aI}$ & $\frac12$ & 1 &\ytableausetup{boxsize=2mm,aligntableaux=top}\ydiagram{1} & $\mathbf{2}$\\
		\hline
		$\psi^{aIJK}$ & $\frac32$ & 1 &\ydiagram{1} & $\mathbf{2}$\\
		\hline
		$h^{IJKL}$ & 2 & 0 & $\bullet$ & $\mathbf{1}$\\
		\hline
		$\gamma^{abIJ}$ & 1 & 0 &\ydiagram{2} & $\mathbf{3}$\\
		\hline
		$V^{IJ}$ & 1 & 0 &$\bullet$ & $\mathbf{1}$\\
		\hline
		$\phi$ & 0 & 0 &$\bullet$ & $\mathbf{1}$\\
		\hline
		$\tilde{\lambda}^{aI}$ & $\frac12$ & -1 &\ydiagram{1} & $\mathbf{2}$\\
		\hline
		$\tilde{\psi}^{aIJK}$ & $\frac32$ & -1 &\ydiagram{1} & $\mathbf{2}$\\
		\hline
		$\tilde{\gamma}^{IJ}$ & 1 & -2 &$\bullet$ & $\mathbf{1}$ \\
		\hline
	\end{tabular}
        \caption{$\mathcal{N}=2$ massive graviton multiplet\\
        \hspace{5mm}}
        \label{tab:N2multiplet}
     \end{subtable}
    \begin{subtable}[h]{0.48\textwidth}
        \centering
	\begin{tabular}{|c|c|c|c|c|}
		\hline
		Field & Spin & $U(1)_R$ & $SU(3)_R$& Dim.\\
		\hline
		\hline
		$\lambda^{I}$ & $\frac12$ & 3 & \ytableausetup{boxsize=2mm,aligntableaux=top}
		$\bullet$&$\mathbf{1}$\\
		\hline
		$\phi^{a}$ & 0 & 2 & \ydiagram{1}&$\mathbf{3}$\\
		\hline
		$\gamma^{aIJ}$ & 1 & 2 & \ydiagram{1}&$\mathbf{3}$\\
		\hline
		$\lambda^{abI}$ & $\frac12$ & 1 &\ydiagram{2}&$\overline{\mathbf{6}}$\\
		\hline
		$\lambda^{I}_a$ & $\frac12$ & 1 &\ydiagram{1,1}&$\overline{\mathbf{3}}$\\[5pt]
		\hline
		$\psi^{IJK}_{a}$ & $\frac32$ & 1 &\ydiagram{1,1}&$\overline{\mathbf{3}}$\\[5pt]
		\hline
		$\phi^{abc}$ & 0 & 0 &\ydiagram{2,1}&$\mathbf{8}$\\[5pt]
		\hline
		$\gamma^{IJ}$ & 1 & 0 &$\bullet$&$\mathbf{1}$\\
		\hline
		$\gamma^{abcIJ}$ & 1 & 0 &\ydiagram{2,1}&$\mathbf{8}$\\[5pt]
		\hline
		$h^{IJKL}$ & 2 & 0 & $\bullet$&$\mathbf{1}$\\
		\hline
		$\lambda^{I}_{ab}$ & $\frac12$ & -1 &\ydiagram{2,2}&$\mathbf{6}$\\[5pt]
		\hline
		$\lambda^{aI}$ & $\frac12$ & -1 &\ydiagram{1}&$\mathbf{3}$\\[5pt]
		\hline
		$\psi^{aIJK}$ & $\frac32$ & -1 &\ydiagram{1}&$\mathbf{3}$\\
		\hline
		$\gamma^{IJ}_a$ & 1 & -2 &\ydiagram{1,1}&$\overline{\mathbf{3}}$\\[5pt]
		\hline
		$\phi_{a}$ & 0 & -2 &\ydiagram{1,1}&$\overline{\mathbf{3}}$\\[5pt]
		\hline
		$\tilde{\lambda}^{I}$ & $\frac12$ & -3 &$\bullet$&$\mathbf{1}$\\
		\hline
	\end{tabular}
       \caption{$\mathcal{N}=3$ massive graviton multiplet}
       \label{tab:N3multiplet}
    \end{subtable}
    \hfill
    \begin{subtable}[h]{0.48\textwidth}
        \centering
        \begin{tabular}{|c|c|c|c|c|}
		\hline
		Field & Spin & $U(1)_R$ & $SU(4)_R$ & Dim.\\
		\hline
		\hline
		$\phi$ & 0 & 4 & $\bullet$ & $\mathbf{1}$\\  
		\hline
		$\lambda^{aI}$ & $\frac12$ & 3 & \ytableausetup{boxsize=2mm,aligntableaux=top}
		\ydiagram{1} & $\mathbf{4}$\\
		\hline
		$\phi^{ab}$ & 0 & 2 &
		\ydiagram{2} & $\mathbf{10}$\\
		\hline
		$\gamma^{IJ}_{ab}$ & 1 & 2 & 
		\ydiagram{1,1} & $\mathbf{6}$\\[5pt]
		\hline
		$\lambda^{abcI}$ & $\frac12$ & 1 &\ydiagram{2,1}&$\overline{\mathbf{20}}$\\[5pt]
		\hline
		$\psi^{IJK}_{a}$ & $\frac32$ & 1 &\ydiagram{1,1,1}&$\overline{\mathbf{4}}$\\[10pt]
		\hline
		$\phi^{abcd}$ & 0 & 0 &\ydiagram{2,2}&$\mathbf{20}'$\\[5pt]
		\hline
		$\gamma^{abcdIJ}$ & 1 & 0 &\ydiagram{2,1,1}&$\mathbf{15}$\\[10pt]
		\hline
		$h^{IJKL}$ & 2 & 0 & $\bullet$&$\mathbf{1}$\\
		\hline
		$\psi^{aIJK}$ & $\frac32$ & -1 &\ydiagram{1}&$\mathbf{4}$\\
		\hline
		$\lambda^{abcdeI}$ & $\frac12$ & -1 &\ydiagram{2,2,1}&$\mathbf{20}$\\[10pt]
		\hline
		$\phi_{ab}$ & 0 & -2 &\ydiagram{2,2,2}&$\overline{\mathbf{10}}$\\[10pt]
		\hline
		$\gamma^{abIJ}$ & 1 & -2 &\ydiagram{1,1}&$\mathbf{6}$\\[5pt]
		\hline
		$\lambda^{I}_a$ & $\frac12$ & -3 &\ydiagram{1,1,1}&$\overline{\mathbf{4}}$\\[10pt]
		\hline
		$\tilde{\phi}$ & 0 & -4 &$\bullet$&$\mathbf{1}$\\
		\hline
	\end{tabular}
        \caption{$\mathcal{N}=4$ massive graviton multiplet}
        \label{tab:N4multiplet}
     \end{subtable}
     \caption{
     \label{tab:multiplets}
     On-shell content of massive graviton multiplets with $\mathcal{N}\leq 4$ supersymmetry. States are labelled with capital Latin indices $I,J,...$ corresponding to $SU(2)_{\text{LG}}$ and lowercase Latin indices $a,b,...$ corresponding to $SU(\mathcal{N})_{R}$. The last and second-to-last columns give the dimension and Young tableaux respectively of the $SU(\mathcal{N})_{R}$ representations in the conventions of \cite{Yamatsu:2015npn,Georgi:1982jb}. }
     \label{tab:reptable}
\end{table}

\subsection{On-Shell Supersymmetry Ward Identities}
\label{sec:onshellWI}
To derive the constraints of supersymmetry on the S-matrix, we first need an explicit description of the action of the supercharges on the asymptotic one-particle states. Our goal in this section is to motivate the introduction of on-shell superspace, and so we will present the details only of the simplest non-trivial case: the $\mathcal{N}=1$ massive graviton summarized in Table \ref{tab:N1multiplet}. 
\begin{align}
\label{N=1oneparticlesusyrep}
    &Q_\alpha |\psi^{IJK}(\vec{p})\rangle = \frac{1}{2\sqrt{6}}|p^{(I}]_\alpha |\gamma^{JK)}(\vec{p})\rangle + \sqrt{2}|p_L]_\alpha |h^{IJKL}(\vec{p})\rangle \nonumber\\
    &Q_\alpha |\gamma^{IJ}(\vec{p})\rangle = \sqrt{\frac{3}{2}}|p_K]_\alpha |\tilde{\psi}^{IJK}(\vec{p})\rangle\nonumber\\
    &Q_\alpha |h^{IJKL}(\vec{p})\rangle = \frac{1}{12\sqrt{2}}|p^{(I}]_\alpha |\tilde{\psi}^{JKL)}(\vec{p})\rangle\nonumber\\
    &Q_\alpha |\tilde{\psi}^{IJK}(\vec{p})\rangle = 0\nonumber\\
    &Q^\dagger_{\dot{\alpha}} |\psi^{IJK}(\vec{p})\rangle = 0\nonumber\\
    &Q^\dagger_{\dot{\alpha}} |\gamma^{IJ}(\vec{p})\rangle =\sqrt{\frac{3}{2}}\langle p_K|_{\dot{\alpha}} |\psi^{IJK}(\vec{p})\rangle \nonumber\\
    &Q^\dagger_{\dot{\alpha}} |h^{IJKL}(\vec{p})\rangle = -\frac{1}{12\sqrt{2}}\langle p^{(I}|_{\dot{\alpha}} |\psi^{JKL)}(\vec{p})\rangle\nonumber\\
    &Q^\dagger_{\dot{\alpha}} |\tilde{\psi}^{IJK}(\vec{p})\rangle = -\frac{1}{2\sqrt{6}}\langle p^{(I}|_{\dot{\alpha}} |\gamma^{JK)}(\vec{p})\rangle - \sqrt{2}\langle p_L|_{\dot{\alpha}} |h^{IJKL}(\vec{p})\rangle.
\end{align}
This explicit representation of the $\mathcal{N}=1$ supersymmetry algebra can be deduced from its off-shell realization on a linearized (free) theory with the field content of an $\mathcal{N}=1$ massive graviton multiplet \cite{Ondo:2016cdv, Buchbinder:2002gh, Zinoviev:2002xn}. In this form we can see the power and utility of the massive spinor formalism \cite{Arkani-Hamed:2017jhn}. The massive spinors give a uniform (and manifestly covariant) language in which we can relate bosons and fermions in different $SU(2)_{\text{LG}}$ representations. Similar one-particle representations exist for the $\mathcal{N}>1$ supersymmetric graviton multiplets, these can be most efficiently expressed in the language of on-shell superfields and will be described in detail in the following subsection.

When interactions are introduced, the off-shell supersymmetry transformations of the field operators will be modified (after integrating out possible auxiliary fields), but the action of the supercharges on on-shell asymptotic states (\ref{N=1oneparticlesusyrep}) will not. In a given supersymmetric model, the S-matrix operator commutes with the supercharges ($[Q_\alpha,S]=[Q^\dagger_{\dot{\alpha}},S]=0$), and if furthermore the supersymmetry is not spontaneously broken ($Q_\alpha|0\rangle = Q_{\dot{\alpha}}|0\rangle = 0$), then the associated S-matrix elements must satisfy a set of linear relations called \textit{on-shell} supersymmetry Ward identities (SWI) \cite{Grisaru:1976vm, Grisaru:1977px}. 

As an explicit example, using (\ref{N=1oneparticlesusyrep}) we can deduce the SWI satisfied by three-particle amplitudes in $\mathcal{N}=1$ supersymmetric massive gravity. Beginning with the vanishing matrix element
\begin{equation}
    \label{WIexampleuneval}
    \langle 0| [Q_\alpha, S]|h_1^{I_1 J_1 K_1 L_1}(\vec{p}_1), h_2^{I_2 J_2 K_2 L_2}(\vec{p}_2), \psi_3^{I_1 I_2 I_3}(\vec{p}_3)\rangle = 0,
\end{equation}
we extend the representation (\ref{N=1oneparticlesusyrep}) to multi-particle states by distributing over (possibly symmetric or anti-symmetric) tensor products in the standard way. In the all-ingoing momentum convention, (\ref{WIexampleuneval}) is equivalent to the following linear relation among three-particle scattering amplitudes
\begin{align}
    \label{WIexampleeval1}
    &\frac{1}{12\sqrt{2}}|1^{(I_1}]_\alpha A_3\left(\tilde{\psi}_1^{J_1 K_1 L_1)},h_2^{I_2 J_2 K_2 L_2}, \psi_3^{I_3 J_3 K_3}\right) \nonumber\\
    &+ \frac{1}{12\sqrt{2}}|2^{(I_2}]_\alpha A_3\left(h_1^{I_1 J_1 K_1 L_1},\tilde{\psi}_2^{J_2 K_2 L_2)}, \psi_3^{I_3 J_3 K_3}\right) \nonumber\\
    &+ \frac{1}{2\sqrt{6}}|3^{(I_3}]_\alpha A_3\left(h_1^{I_1 J_1 K_1 L_1}, h_2^{I_2 J_2 K_2 L_2}, \gamma_3^{J_3 K_3)}\right) \nonumber\\
    &+ \sqrt{2}|3_{L_3}]_\alpha A_3\left(h_1^{I_1 J_1 K_1 L_1},h_2^{I_2 J_2 K_2 L_2},h_3^{I_3 J_3 K_3 L_3}\right) = 0.
\end{align}
Other choices of incoming three-particle states lead to different relations, for example
\begin{equation}
    \langle 0| [Q_\alpha, S]|h_1^{I_1 J_1 K_1 L_1}(\vec{p}_1), h_2^{I_2 J_2 K_2 L_2}(\vec{p}_2), \tilde{\psi}_3^{I_1 I_2 I_3}(\vec{p}_3)\rangle = 0,
\end{equation}
leads to the SWI
\begin{align}
    \label{WIexampleeval2}
    &\frac{1}{12\sqrt{2}}|1^{(I_1}]_\alpha A_3\left(\tilde{\psi}_1^{J_1 K_1 L_1)},h_2^{I_2 J_2 K_2 L_2}, \tilde{\psi}_3^{I_3 J_3 K_3}\right) \nonumber\\
    &+ \frac{1}{12\sqrt{2}}|2^{(I_2}]_\alpha A_3\left(h_1^{I_1 J_1 K_1 L_1},\tilde{\psi}_2^{J_2 K_2 L_2)}, \tilde{\psi}_3^{I_3 J_3 K_3}\right) = 0.
\end{align}
Continuing in this way, from relations of the form (\ref{WIexampleuneval}) with all possible choices of incoming $n$-particle states, we can generate the complete set of supersymmetry constraints on $n$-particle scattering amplitudes in the form of a large system of linear equations. 

To solve these equations directly, meaning to identify a maximal set of linearly independent amplitudes in terms of which all others are uniquely fixed, is a difficult task. For massless supersymmetric models, with up to $\mathcal{N}=8$ supersymmetry, the resulting system of constraints can be solved \cite{Elvang:2009wd, Elvang:2010xn}.

One simplification that is common to both massless and massive supersymmetric models is that the SWI are block diagonal in irreducible representations of both the maximal R-symmetry group $SU(\mathcal{N})_R\times U(1)_R$ as well as any possible global symmetry realizable on the assumed particle spectrum. This remains true even if these would-be symmetries are explicitly broken by interactions. Intuitively, the SWI only impose the constraints of supersymmetry, and so should be compatible with any choice of additional optional symmetry. This means that in the SWI, amplitudes with external states in distinct symmetry sectors do not talk to one another and the identities can be solved independently. 

As a well known example, for massless multiplets we can always define $U(1)_R$ such that the the R-charge coincides with the helicity. In both supersymmetric Yang-Mills and supergravity, helicity is not a dynamically conserved quantity; nonetheless the SWI in these models are diagonal in $\text{N}^K$-MHV sectors, corresponding to eigenstates of the broken $U(1)_R$ symmetry. For massive supersymmetry the notion of an $\text{N}^K$-MHV sector no longer makes sense since helicity is not a Lorentz invariant quantum number for massive states, but the would-be $U(1)_R$ symmetry is still well-defined and plays the same role. This can be seen in the above examples, the component states of the $\mathcal{N}=1$ massive graviton multiplet can be assigned the $U(1)_R$ charges given in Table \ref{tab:N1multiplet}. The SWI (\ref{WIexampleeval1}) relates amplitudes which conserve $U(1)_R$-charge, while (\ref{WIexampleeval2}) relates amplitudes which each violate conservation of $U(1)_R$-charge by $-1$ units. 

For general $\mathcal{N}$, the unique highest spin state in the multiplet is the massive graviton and is therefore necessarily an R-singlet. The supersymmetry constraints on the cubic graviton interactions then arise only from R-singlet SWI. 
Without loss of generality, the problem of understanding the constraints of supersymmetry on scattering amplitudes at lowest multiplicity (cubic interactions) is therefore reduced to solving the SWIs in the R-singlet sector. As we will explain in the following subsections this problem can be essentially trivialized, even for arbitrarily extended supersymmetry, by rephrasing it in the language of on-shell superspace.

\subsection{Massive Graviton On-Shell Superfields}
\label{sec:onshellsuperfields}
The idea of an \textit{on-shell} superspace as a framework for making manifest arbitrarily extended supersymmetry for scattering amplitudes was introduced in \cite{Nair:1988bq}, see also \cite{Witten:2003nn}. The extension of these methods to massive supermultiplets was made in a non-covariant form in \cite{Boels:2011zz} and most recently in a fully Lorentz-covariant form in $4d$ \cite{Herderschee:2019ofc,Herderschee:2019dmc} based on the massive spinor formalism \cite{Arkani-Hamed:2017jhn}. Below we review the essential elements of massive on-shell superspace based on \cite{Herderschee:2019ofc,Herderschee:2019dmc}.

We define massive on-shell superspace by extending the field over which we form linear combinations in the asymptotic one-particle Hilbert space by an $SU(2)_{\text{LG}}$ doublet of anti-commuting c-number (Grassmann) valued parameters $\eta^I$. In this extended Hilbert space, we can form special linear combinations of the one-particle states comprising complete supersymmetry multiplets called \textit{on-shell superfields}. For example, for the $\mathcal{N}=1$ massive gravity multiplet we construct the massive graviton superfield
\begin{equation}
    \label{N=1mgravitonsuperfield}
    |\Psi^{IJK}(\vec{p},\eta)\rangle \equiv |\psi^{IJK}(\vec{p})\rangle+\frac{1}{4\sqrt{3}}\eta^{(I}|\gamma^{JK)}(\vec{p})\rangle+\eta_L|h^{IJKL}(\vec{p})\rangle+\frac{1}{2}\eta_L\eta^L|\tilde{\psi}^{IJK}(\vec{p})\rangle.
\end{equation}
The one-particle states or \textit{component fields} can be extracted from the superfield by acting with appropriate Grassmann differential projection operators, for example to project out the graviton 
\begin{align}
\label{projection}
    |h^{IJKL}(\vec{p})\rangle  \equiv \frac{1}{4}\frac{\partial}{\partial \eta_{(L}}|\Psi^{IJK)}(\vec{p},\eta)\rangle \biggr\vert_{\eta=0}.
\end{align}
To avoid clutter, it is conventional to use the simplified notation for states $|X^{I_1...I_s}(\vec{p},\eta)\rangle\sim X^{I_1...I_s}$, which we will follow for the rest of the paper. The key idea of on-shell superspace is that the action of supersymmetry can be represented in two different but equivalent ways. The first is to define the supercharges in the usual way as q-number operators, which act on the component states as in (\ref{N=1oneparticlesusyrep}) but commute with the Grassmann-valued coefficients in (\ref{N=1mgravitonsuperfield}). The second is to define the supercharges as anti-commuting c-number operators which act on the Grassmann-valued coefficients but commute with the component states. Explicitly, when acting on an $n$-particle state, the c-number supercharges are given by
\begin{equation}
    \label{cnumbersupercharges}
    Q^a_{\alpha} \equiv \sqrt{2}\sum_{i=1}^n|i_{I_i}]_\alpha \frac{\partial}{\partial \eta_{a i I_i}}, \hspace{10mm} Q^\dagger_{a\dot{\alpha}} \equiv -\sqrt{2}\sum_{i=1}^n\langle i^{I_i}|_{\dot{\alpha}} \eta_{a i I_i}.
\end{equation}
where $a=1,...,\mathcal{N}$ for $\mathcal{N}$-extended supersymmetry. It is a straightforward exercise to verify that (\ref{cnumbersupercharges}) reproduces the explicit representation (\ref{N=1oneparticlesusyrep}).

For the $\mathcal{N}\geq 2$ massive graviton multiplets, the strategy will \textit{not} be to begin with a representation derived from an off-shell Lagrangian describing the linearized model as we did for $\mathcal{N}=1$. Rather, we can construct the on-shell superfield directly by decomposing a generic Grassmann polynomial into the available $SU(2)_{\text{LG}}$ and $SU(\mathcal{N})_R$-covariant objects. From Table \ref{tab:multiplets}, we know which states must appear in each multiplet and, from the associated Young tableaux, their symmetry properties as $SU(\mathcal{N})_R$ tensors. The Grassmann monomial prefactors on each term must contract the R-indices to form a singlet, and also contract the $SU(2)_{\text{LG}}$ indices to reproduce the superspin of the multiplet. We find that there is a unique way to construct each term. The results are given below.
\begingroup
\allowdisplaybreaks
\begin{align}
\label{eq:N1sfield}
    &\hspace{-10mm}\mathbf{\mathcal{N}=1} \textbf{ massive graviton:}\nonumber\\
    &\nonumber\\
    \Psi^{IJK} &= \psi^{IJK}+\frac{1}{4\sqrt{3}}\eta^{(I}\gamma^{JK)}+\eta_Lh^{IJKL}+\frac{1}{2}\eta_L\eta^L\tilde{\psi}^{IJK}.\\
    &\nonumber\\
\label{eq:N2sfield}
    &\hspace{-10mm}\mathbf{\mathcal{N}=2} \textbf{ massive graviton:}\nonumber\\
    &\nonumber\\
    \Gamma^{IJ} &= \gamma^{IJ}+\frac1{\sqrt{6}}\eta_a^{(I}\lambda^{aJ)}+\eta_{aK}\psi^{aIJK}+\frac12 \epsilon^{ab}\eta_{aK}\eta_{bL}h^{IJKL}+\frac12 \eta_{aK}\eta_b^K\gamma^{abIJ}\nonumber\\
    &\hspace{5mm}+\frac18 \epsilon^{ab}\eta_a^{(I}\eta_{bK}V^{J)K}+\frac1{2\sqrt{3}}\epsilon^{ab}\eta_{a}^I\eta_{b}^J\phi+\frac14 \eta_{bK}\eta^{bK}\eta_{a}^{(I}\tilde{\lambda}^{aJ)}+\frac1{\sqrt{2}}\eta_{bK}\eta^{bK}\eta_{aL}\tilde{\psi}^{aIJL}\nonumber\\
    &\hspace{5mm}+\frac18\eta_{bK}\eta^{bK}\eta_{cL}\eta^{cL}\tilde{\gamma}^{IJ}.\\
    &\nonumber\\
\label{eq:N3sfield}
    &\hspace{-10mm}\mathbf{\mathcal{N}=3} \textbf{ massive graviton:}\nonumber\\
    &\nonumber\\
    \Lambda^{I} &= \lambda^I +\eta_a^I\phi^{a}+\eta_{aJ}\gamma^{aIJ}+\frac12 \eta_{aJ}\eta_b^{J}\lambda^{abI}+\frac12 \epsilon^{abc}\eta_a^I\eta_{bJ}\lambda_c^{J}+\frac12 \epsilon^{abc}\eta_{aK}\eta_{bJ}\psi_c^{IJK}\nonumber\\
    &\hspace{5mm}+\frac16 \eta_{aJ}\eta_b^{(J}\eta_c^{I)}\phi^{abc}+\frac{1}{2\sqrt{6}}\epsilon^{abc}\eta_a^I\eta_{bJ}\eta_{cK}\gamma^{JK}+\frac{1}{6\sqrt{2}}\eta_{aJ}\eta_b^{(J}\eta_{cK}\gamma^{abcK)I}\nonumber\\
    &\hspace{5mm}+\frac16\epsilon^{abc}\eta_{aJ}\eta_{bK}\eta_{cL}h^{IJKL}+\frac1{2\sqrt{2}}\epsilon^{abc}\eta_{a}^I\eta_{b}^J\eta_{c}^K\eta_{dJ}\lambda^{d}_K+\frac1{12}\epsilon^{abc}\epsilon^{def}\eta_{aJ}\eta_{bK}\eta_d^J\eta_e^K\lambda_{cf}^I\nonumber\\
    &\hspace{5mm}+\frac18 \epsilon^{abc}\eta_{aJ}\eta_{bK}\eta_{cL}\eta_{d}^L\psi^{dIJK}+\frac1{16}\epsilon^{abc}\eta_{aJ}\eta_{bK}\eta_{cL}\epsilon^{def}\eta_{d}^I\eta_{e}^J\gamma^{KL}_f\nonumber\\
    &\hspace{5mm}+\frac1{24\sqrt{2}}\epsilon^{abc}\eta_{a}^I\eta^J_{b}\eta^K_{c}\epsilon^{def}\eta_{dJ}\eta_{eK}\phi_f+\frac1{144}\epsilon^{abc}\eta_{a}^J\eta^K_{b}\eta^L_{c}\epsilon^{def}\eta_{dJ}\eta_{eK}\eta_{fL}
    \tilde{\lambda}^I.\\
    &\nonumber\\
\label{eq:N4sfield}
    &\hspace{-10mm}\mathbf{\mathcal{N}=4} \textbf{ massive graviton:}\nonumber\\
    &\nonumber\\
    \Phi &= \phi+\eta_{aI}\lambda^{aI}+\frac12 \eta_{aI}\eta_{b}^I\phi^{ab}+\frac1{2\sqrt{2}}\epsilon^{abcd}\eta_{aI}\eta_{bJ}\gamma^{IJ}_{cd}+\frac1{3\sqrt{2}}\eta_{aI}\eta_b^{(I}\eta_{cJ}\lambda^{abcJ)}\nonumber\\
    &\hspace{5mm}+\frac16\epsilon^{abcd}\eta_{aI}\eta_{bJ}\eta_{cK}\psi_d^{IJK}+\frac1{12}\eta_{aI}\eta_b^{(I}\eta_{cJ}\eta_d^{J)}\phi^{abcd}+\frac1{8\sqrt{6}}\eta_{aI}\eta_b^{(I}\eta_{cJ}\eta_{dK}\gamma^{abcdJK)}\nonumber\\
    &\hspace{5mm}+\frac1{24}\epsilon^{abcd}\eta_{aI}\eta_{bJ}\eta_{cK}\eta_{dL}h^{IJKL}+\frac1{30}\epsilon^{abcd}\eta_{aI}\eta_{bJ}\eta_{cK}\eta_{dL}\eta_{e}^L\psi^{eIJK}\nonumber\\
    &\hspace{5mm}+\frac{1}{24\sqrt{3}}\eta_{aI}\eta_b^{(I}\eta_{cJ}\eta_{d}^J\eta_{eK}\lambda^{K)abcde}+\frac1{144}\epsilon^{abcd}\epsilon^{efgh}\eta_{aI}\eta_{bJ}\eta_{cK}\eta_e^I\eta_f^J\eta_g^K\phi_{dh}\nonumber\\
    &\hspace{5mm}+\frac{1}{80}\epsilon^{abcd}\eta_{aI}\eta_{bJ}\eta_{cK}\eta_{dL}\eta_e^I\eta_f^J\gamma^{efKL}+\frac1{360}\epsilon^{abcd}\epsilon^{efgh}\eta_{aI}\eta_{bJ}\eta_{cK}\eta_{dL}\eta_e^I\eta_f^J\eta_g^K\lambda_h^{L}\nonumber\\
    &\hspace{5mm}+\frac1{2880}\epsilon^{abcd}\epsilon^{efgh}\eta_{aI}\eta_{bJ}\eta_{cK}\eta_{dL}\eta_e^I\eta_f^J\eta_g^K\eta_h^{L}\tilde{\phi}.
\end{align}
\endgroup

The numerical pre-factors on the component states are fixed by requiring that the superfield completeness relation agrees with the sum of the completeness relations for each component state. A more detailed discussion of this, as well as the relative phases between the terms is given in Appendix \ref{appendix:superfields}.

From these superfields, we can work backwards to deduce the explicit form of the representation, as in (\ref{N=1oneparticlesusyrep}). Clearly the superfield is a much more compact way of encoding this information. Likewise there is a more compact superspace representation of the SWI constraints given by \textit{superamplitudes}. These are defined the same way as ordinary scattering amplitudes, but with ordinary $n$-particle states replaced with $n$-particle on-shell superfields. For example a cubic $\mathcal{N}=1$ massive graviton superamplitude is extracted from the superspace S-matrix element
\begin{align}
    &\langle 0 | S |\Psi_1^{I_1 J_1 K_1}(\vec{p}_1,\eta_1),\Psi_2^{I_2 J_2 K_2}(\vec{p}_2,\eta_2),\Psi_3^{I_3 J_3 K_3}(\vec{p}_3,\eta_3) \rangle   \nonumber\\
    &\equiv i(2\pi)^4 \delta^{(4)}\left(p_1+p_2+p_3\right)\mathcal{A}_3\left(\Psi_1^{I_1 J_1 K_1}(\vec{p}_1,\eta_1),\Psi_2^{I_2 J_2 K_2}(\vec{p}_2,\eta_2),\Psi_3^{I_3 J_3 K_3}(\vec{p}_3,\eta_3) \right).
\end{align}
In this paper we will use $\mathcal{A}$ to denote superamplitudes and $A$ to denote component or ordinary amplitudes. As before, the component amplitudes are extracted from the superamplitudes by acting with projection operators (\ref{projection}).

\subsection{General Form of a Massive Superamplitude}
\label{sec:solvingWI}

A major advantage of using the on-shell superspace formalism is that it is very simple to directly construct a completely general (and non-redundant) expression for the superamplitude. Such an expression represents a solution to the complicated system of SWIs in the sense that by acting with projection operators (\ref{projection}) each of the component amplitudes can be expressed in terms a minimal set of independent functions. The approach in this section is similar in spirit to \cite{Elvang:2009wd, Elvang:2010xn} for massless supersymmetry.

A general superamplitude $\mathcal{A}_n\left(\{\eta\}\right)$ is an order $2n\mathcal{N}$ polynomial in the Grassmann parameters $\eta_{ai}^{I_i}$, where $i=1,...,n$ and $a=1,...,\mathcal{N}$. As discussed above, in this formalism the $Q$-supercharges are represented by differential operators and the $Q^\dagger$-supercharges are represented multiplicatively (\ref{cnumbersupercharges}), we will call this the $\eta$-representation. The SWIs are encoded in the statement that the superamplitude is annihilated by both kinds of supercharges, 
\begin{align}
\label{superconstraints}
    Q^a_\alpha \cdot \mathcal{A}_n\left(\{\eta\}\right) = 0, && Q^{\dagger}_{a\dot{\alpha}} \cdot \mathcal{A}_n\left(\{\eta\}\right) = 0.
\end{align}
The main result of this section is summarized below:

\vspace{5mm}
\noindent\fbox{
\begin{minipage}{14cm}
The most general form of a superamplitude with only massive external states is
\begin{equation}
\label{solSWI}
  \mathcal{A}_n\left(\{\eta_i\}\right) = \delta^{(2\mathcal{N})}\left(Q^\dagger\right)G\left(\{\eta_{in}\}\right),
\end{equation}
where $G\left(\{\eta_{in}\}\right)$ is an arbitrary polynomial in the $2(n-2)\mathcal{N}$ Grassmann variables 
\begin{equation}
    \label{invarianteta}
    \eta_{a,in}^{I_i} \equiv \eta_{a,i}^{I_i} - \frac{1}{m_n}[i^{I_i}n_{I_n}]\eta_{a,n}^{I_n}, \hspace{10mm} i=1,...,n-2,
\end{equation}
and the supersymmetric delta function is
\begin{equation}
\label{susydelta}
    \delta^{(2\mathcal{N})}\left(Q^\dagger\right) = \prod_{a=1}^\mathcal{N} \left(\sum_{i<j} \langle i_{I_i} j_{I_j}\rangle \eta_{a,i}^{I_i}\eta_{a,j}^{I_j}+\frac{1}{2}\sum_{i=1}^n m_i \eta_{a,i I_i}\eta_{a,i}^{I_i}\right).
\end{equation}
\end{minipage}
}\\
\vspace{3mm}\\
The statement that such an expression satisfies (\ref{superconstraints}) can be straightforwardly verified. The action of the $Q^\dagger$-supercharges is trivialized by the identity $Q^\dagger \delta^{(2\mathcal{N})}\left(Q^\dagger\right) = 0$. Less obviously, the $Q$-supercharge also separately annihilates both $\delta^{(2\mathcal{N})}\left(Q^\dagger\right)$ and  $G\left(\{\eta_{in}\}\right)$. The latter follows from repeatedly applying the product rule and
\begin{equation}
    Q^a_{\alpha} \cdot \eta_{b,in}^{I_i}  = 0.
\end{equation}
Expressions of the form (\ref{solSWI}) therefore give \textit{a} solution of the contraints (\ref{superconstraints}), the more non-trivial claim that we establish in this section is that \textit{all} solutions can be written in this form.

As a first step we will review the argument that annihilation of the superamplitude by $Q^\dagger$ requires the amplitude to be proportional to $\delta^{(2\mathcal{N})}\left(Q^\dagger\right)$ and explain how the explicit expression  (\ref{susydelta}) is obtained. For simplicity we will present the argument for $\mathcal{N}=1$ supersymmetry, the extension to $\mathcal{N}\geq 1$ is straightforwardly obtained by iterating in the R-index. We begin by defining a second, physically equivalent, representation of the superamplitude by taking a Grassmann Fourier transform
\begin{equation}
    \tilde{\mathcal{A}}_n\left(\{\eta^\dagger\}\right) \equiv \prod_{i=1}^n\int \text{d}^2 \eta_i e^{\eta_i \eta_i^\dagger } \mathcal{A}_n\left(\{\eta\}\right),
\end{equation}
where $\text{d}^2 \eta \equiv \frac{1}{2}\epsilon^{IJ} \text{d} \eta_I \text{d}\eta_J$. In this $\eta^\dagger$-representation, the supercharges act on the superamplitude as
\begin{equation}
    Q_\alpha \equiv -\sqrt{2}\sum_{i=1}^n|i^{I_i}]_{\alpha} \eta^\dagger_{i I_i}, \hspace{10mm} Q^\dagger_{\dot{\alpha}} \equiv \sqrt{2}\sum_{i=1}^n\langle i_{I_i}|_{\dot{\alpha}} \frac{\partial}{\partial \eta^\dagger_{i I_i}}. 
\end{equation}
In these variables the identity of the multiplicative and differential supercharges are inverted. Again, in this notation the SWIs are encoded in the conditions $Q_\alpha \cdot \tilde{\mathcal{A}}_n\left(\{\eta^\dagger\}\right) = Q^\dagger_{\dot{\alpha}} \cdot \tilde{\mathcal{A}}_n\left(\{\eta^\dagger\}\right) = 0$. That the superamplitude is annihilated by a linear differential operator is equivalent to the existence of a superspace translation invariance of the form
\begin{equation}
    \tilde{\mathcal{A}}_n\left(\eta_1^\dagger,...,\eta_n^\dagger\right) = \tilde{\mathcal{A}}_n\left(\eta_1^\dagger+\langle \epsilon 1\rangle,..., \eta_n^\dagger+\langle\epsilon n \rangle\right),
\end{equation}
for an arbitrary Grassmann valued spinor $|\epsilon\rangle^{\dot{\alpha}}$. We can use this freedom to eliminate $\eta_n^\dagger$ by choosing 
\begin{equation}
    \langle \epsilon|_{\dot{\alpha}} = \frac{1}{m_n} \eta_n^{\dagger I_n} \langle n_{I_n}|_{\dot{\alpha}}.
\end{equation}
To make use of this, we express the original $\eta$-representation superamplitude as the inverse Fourier transform 
\begin{equation}
    \mathcal{A}_n\left(\{\eta\}\right) \equiv \prod_{i=1}^n\int \text{d}^2 \eta_i^\dagger e^{-\eta_i \eta_i^\dagger } \tilde{\mathcal{A}}_n\left(\{\eta^\dagger\}\right).
\end{equation}
By making a change of variables in the Grassmann integral 
\begin{equation}
    \hat{\eta}_i^{\dagger I_i} \equiv \eta_i^{\dagger I_i} - \frac{1}{m_n} \eta_n^{\dagger I_n} \langle n_{I_n} i^{I_i}\rangle,
\end{equation}
it follows that the integrand factor $\tilde{\mathcal{A}}_n\left(\{\hat{\eta}^\dagger\}\right)$ is independent of $\hat{\eta}_n^\dagger$. The integration over $\hat{\eta}_n^\dagger$ reduces to a (universal) integral of an exponential, which can be straightforwardly evaluated 
\begin{align}
    \int \text{d}^2 \hat{\eta}^\dagger_n e^{-\left(\eta_{nI_n} - \sum_{i=1}^{n-1}\frac{1}{m_i}\langle i^{I_i} n_{I_n}\rangle \eta_{i I_i}\right)\hat{\eta}_n^{\dagger I_n}} \propto \delta^{(2)}\left(Q^\dagger\right).
\end{align}
The most general solution to the condition $Q^\dagger \cdot \mathcal{A}_n\left(\{\eta\}\right) = 0$ is then of the form
\begin{equation}
    \mathcal{A}_n\left(\{\eta\}\right) = \delta^{(2)}\left(Q^\dagger\right) G\left(\{\eta\}\right),
\end{equation}
for some polynomial in $\eta_i$, $i=1,...,n$. Due to the delta function, for a given $\mathcal{A}_n\left(\{\eta\}\right)$, the function $G\left(\{\eta\}\right)$ is non-unique, this allows us to make a slightly stronger statement. Since $Q^\dagger$ is a linear function of the $\eta_i$ in this representation, we can always choose to fix the ambiguity by removing one of the $\eta_i$ from $G\left(\{\eta\}\right)$ by adding terms proportional to $Q^\dagger$. 

In the second step we use a similar argument. By explicit calculation, $Q \cdot \delta^{(2)}\left(Q^\dagger\right) = 0$, and so the remaining supersymmetry constraint is $Q \cdot G\left(\{\eta\}\right)\propto Q^\dagger$. Since $Q^\dagger$ contains all of the $\eta_i$, the right-hand-side of this equation is actually zero if we define $G\left(\{\eta\}\right)$ to be independent of one of them as described above. Without loss of generality we will choose this to be $\eta_{n-1}$. In this case the remaining supersymmetry constraint becomes a homogeneous, linear differential equation
\begin{equation}
    Q \cdot G\left(\{\eta\}\right) = 0.
\end{equation}
As described above, this implies a superspace translation invariance 
\begin{equation}
    G\left(\eta_1, ... , \eta_{n-2}, \eta_{n}\right) = G\left(\eta_1 + [\epsilon 1], ... , \eta_{n-2}+ [\epsilon, n-2], \eta_{n} + [\epsilon n]\right),
\end{equation}
for arbitrary $[\epsilon|_{\dot{\alpha}}$. Using this freedom we can choose 
\begin{equation}
    [\epsilon|_{\dot{\alpha}} = - \frac{1}{m_n}[n_{I}|_{\dot{\alpha}}\eta_n^{I},
\end{equation}
which gives 
\begin{equation}
    G\left(\eta_1, ... , \eta_{n-2}, \eta_{n}\right) = G\left(\eta_{1,n} , ... , \eta_{n-2,n},0\right).
\end{equation}
As promised, the function $G\left(\{\eta\}\right)$ is a function of the $n-2$ supersymmetry invariant combinations (\ref{invarianteta}). 

\section{Cubic Interactions}
\label{sec:cubicstuff}
 In this section, we discuss cubic amplitudes in models of supersymmetric massive gravity. In Section \ref{sec:cubicamps}, we introduce a convenient basis of kinematic objects and use this to enumerate a basis of cubic interactions for a massive graviton. In Section \ref{sec:susy3point}, we present our main result, the use of on-shell superspace to derive a complete classification of supersymmetrizable cubic massive graviton interactions.

\subsection{Massive Graviton Self-Interactions}
\label{sec:cubicamps}
We construct the most general set of cubic interactions among massive particles. Rather than proceeding from an off-shell effective action and calculating 3-particle amplitudes with Feynman rules, we will instead construct the on-shell objects directly. By a corollary of the S-matrix equivalence theorem \cite{Arzt:1993gz}, kinematically independent scattering amplitude contact terms are in one-to-one correspondence with local operators modulo field redefinition and integration-by-parts. 

We find that a convenient way to construct a complete and non-redundant basis of such cubic on-shell objects, is to construct them as polynomials in the following building blocks
\begin{align}
\label{braces}
    \{i^I j^J\} \equiv [i^I|\slashed{p}_1 \slashed{p}_2|j^J],
\end{align}
which we will call \textit{spinor braces}. A justification of this choice is given in Appendix \ref{bracesequivalent}. For a  $SU(2)_{\text{LG}}$ tensor of given rank associated to each particle, there are finitely many polynomials that can be constructed from the objects (\ref{braces}). However, these are not all independent, there is a redundancy in this construction corresponding to the following identity\footnote{This identity can be derived straightforwardly by using $\epsilon^{\alpha \beta} \epsilon^{\gamma \delta} + \epsilon^{\alpha \gamma} \epsilon^{\delta \beta} + \epsilon^{\alpha \delta} \epsilon^{\beta \gamma} = 0$ and $\epsilon^{\dot{\alpha} \dot{\beta}} \epsilon^{\dot{\gamma} \dot{\delta}} + \epsilon^{\dot{\alpha} \dot{\gamma}} \epsilon^{\dot{\delta} \dot{\beta}} + \epsilon^{\dot{\alpha} \dot{\delta}} \epsilon^{\dot{\beta} \dot{\gamma}} = 0$ in all possible ways on a product of two braces (\ref{braces}). Interestingly, even though there are multiple ways of applying these, there is only one inequivalent brace relation. This redundancy is related the the usual ambiguity associated with Schouten identities.}
\begin{align}
\label{braceschoutenA}
    &\{i^I j^J\}\{k^K l^L\}+\{i^I k^K\}\{l^L j^J\}-\{i^I k^K\}\{j^J l^L\}\nonumber\\
    &\hspace{10mm}-\{k^K i^I\}\{l^L j^J\}+\{k^K i^I\}\{j^J l^L\}-\{i^I l^L\}\{k^K j^J\}=0.
\end{align}

In a physical scattering amplitude, a massive external state is an irreducible representation of $SU(2)_{\text{LG}}$, which in this notation corresponds to a totally symmetric tensor. Following \cite{Arkani-Hamed:2017jhn}, it is convenient in this case to suppress the little-group indices and write the associated particle label in bold. Finally, we sometimes also need to further impose that the particles we are describing are identical bosons/fermions. In a scattering amplitude, (Fermi) Bose symmetry is the same as relabelling (anti-)symmetry. For example, if particles 1 and 2 are identical bosons then
\begin{equation}
    A_3\left(1,2,3\right) = A_3\left(2,1,3\right) = A_3\left(1,2,3\right)\biggr\vert_{1\leftrightarrow 2}.
\end{equation}
To implement this on the expressions built out of spinor braces we need to know how they transform under an exchange of labels. These transformations are given in Appendix \ref{BFrelation}. 

While the spinor-braces are particularly useful for constructing 3-point superamplitudes, it is often useful to write bosonic amplitudes in terms of more familiar components; polarization tensors and momentum vectors. Here, we express them in the formalism outlined in \cite{Bonifacio:2017nnt}. Since the graviton polarization tensors are symmetric, transverse, and traceless, it is convenient to express them using auxiliary polarization vectors, $z_a^\mu$, which are null, $z_a^2=0$, and transverse, $p_a\cdot z_a=0$. We can then make the identification $\epsilon_a^{\mu\nu}= z^\mu_a z^\nu_a$.

Even-parity cubic spin-2 amplitudes can then be built out of a set of 6 building blocks $\{(z_1\cdot z_2),\,(z_2\cdot z_3),\,(z_1\cdot z_3),\,(p_2\cdot z_1),\,(p_3\cdot z_2),\,(p_1\cdot z_3)\}$ and odd-parity amplitudes can be constructed by including the additional building blocks $\{\epsilon (p_1p_2z_1z_2),\,\epsilon (p_1p_2z_1z_3),\,\epsilon (p_1p_2z_2z_3)\}$,
where contractions with an antisymmetric Levi-Civita tensor are denoted $\epsilon(v_1v_2v_3v_4)\equiv\epsilon_{\mu\nu\alpha\beta}v_1^\mu v_2^\nu v_3^\alpha v_4^\beta$. Other structures such as $(p_3\cdot z_1)$  or $\epsilon(p_1z_1z_2z_3)$ can be related to these through Schouten identities and momentum conservation. Amplitudes in this notation can be translated into massive spinor variables by making the replacement
\begin{equation}    
\label{ztoe}
    z_i^\mu \rightarrow\frac{1}{2\sqrt{2}m}\langle \mathbf{i}|\overline{\sigma}_\mu|\mathbf{i}] ,
\end{equation}
It will also be useful to translate the above building blocks directly into spinor brace notation. The explicit expressions are given in Appendix \ref{zptospinor}. 

Using the spinor braces, we can construct a basis of linearly independent cubic massive graviton amplitudes. Resolving the redundancy (\ref{braceschoutenA}) and imposing total Bose symmetry, we find that there are 6 independent interactions. This result is well-known \cite{Bonifacio:2017nnt}, and can be straightforwardly shown to be a reorganization of the following cubic amplitudes
\begingroup
\allowdisplaybreaks
\begin{align}
\label{cubicamps}
    \mathcal{B}_1&=\frac{m^2}{M_{\text{P}}}(z_1\cdot z_2)(z_1\cdot z_3)(z_2\cdot z_3)\,,\nn\\
    \mathcal{B}_2&=\frac{1}{M_{\text{P}}}\left[(z_2\cdot z_3)(p_2\cdot z_1)+(z_1\cdot z_3)(p_3\cdot z_2)+(z_1\cdot z_2)(p_1\cdot z_3)\right]^2\,,\nn\\\mathcal{B}_3&=\frac{1}{M_{\text{P}}}\left[(z_2\cdot z_3)^2(p_2\cdot z_1)^2+(z_1\cdot z_3)^2(p_3\cdot z_2)^2+(z_1\cdot z_2)^2(p_1\cdot z_3)^2\right]\,,\nn\\
    \mathcal{B}_4&=\frac{1}{m^4M_{\text{P}}}(p_2\cdot z_1)^2(p_3\cdot z_2)^2(p_1\cdot z_3)^2\,,\nn\\
    \mathcal{B}_5&=\frac{1}{M_{\text{P}}}\left[(z_1\cdot z_3)(z_2\cdot z_3)\epsilon(p_1p_2z_1z_2)-(z_1\cdot z_2)(z_2\cdot z_3)\epsilon(p_1p_2z_1z_3)\right.\nonumber\\
    &\hspace{20mm}\left.+(z_1\cdot z_2)(z_1\cdot z_3)\epsilon(p_1p_2z_2z_3)\right]\,,\nn\\
    \mathcal{B}_6&=\frac{1}{m^4M_{\text{P}}}(p_2\cdot z_1)(p_3\cdot z_2)(p_1\cdot z_3)\left[(p_1\cdot z_3)\epsilon(p_1p_2z_1z_2)-(p_3\cdot z_2)\epsilon(p_1p_2z_1z_3)\right.\nonumber\\
    &\hspace{20mm}\left.+(p_2\cdot z_1)\epsilon(p_1p_2z_2z_3)\right]\,,
\end{align}
\endgroup
where $\mathcal{B}_{1,2,3,4}$ are parity-even and $\mathcal{B}_{5,6}$ are parity-odd. A particular combination of these interactions give the massive spin-2 interactions in superstring theory and bosonic string theory \cite{Lust:2021jps}.

The advantage of the basis (\ref{cubicamps}) over the spinor brace expressions is that they can be more simply related to a basis of local operators:
\begingroup
\allowdisplaybreaks
\begin{align}
\label{cubicoperators}
    \mathcal{L}_1&=\frac{m^2}{3M_{\text{P}}} {h_\mu}^\nu {h_\nu}^\lambda {h_\lambda}^\mu&\rightarrow& \,\hspace{10mm}2\mathcal{B}_1\,,\nn\\
    \mathcal{L}_2&=\frac{M_{\text{P}}^2}{2} \sqrt{-g}R|_{(3)}&\rightarrow&\, \hspace{10mm}-6\mathcal{B}_1+2\mathcal{B}_2\,,\nn\\
    \mathcal{L}_3&=\frac{1}{M_{\text{P}}}\epsilon^{\mu\nu\lambda\rho}\epsilon^{\alpha\beta\gamma\delta}\partial_\mu\partial_\alpha h_{\nu\beta}h_{\lambda\gamma}h_{\rho\delta}&\rightarrow&\,\hspace{10mm}12\mathcal{B}_1-2\mathcal{B}_2-2\mathcal{B}_3\,,\nn\\
    \mathcal{L}_4&=\frac{M_{\text{P}}^2}{m^4} \sqrt{-g}{R_{\mu\nu}}^{\alpha \beta}{R_{\alpha\beta}}^{\lambda \rho}{R_{\lambda\rho}}^{\mu \nu}|_{(3)}&\rightarrow& \,\hspace{10mm}24\mathcal{B}_1-12\mathcal{B}_3+48\mathcal{B}_4\,,\nn\\
    \mathcal{L}_5&= \frac{1}{M_{\text{P}}}\epsilon^{\mu\nu\lambda\rho}\partial_\mu h_{\nu\alpha}\partial_\lambda h_{\rho \beta}h^{\alpha \beta}&\rightarrow&\, \hspace{10mm}2\mathcal{B}_5\,,\nn\\
    \mathcal{L}_6&= \frac{M_{\text{P}}^2}{m^4}\sqrt{-g}{\widetilde{R}_{\mu\nu}}\,^{\alpha \beta}{R_{\alpha\beta}}^{\lambda \rho}{R_{\lambda\rho}}^{\mu \nu}|_{(3)}&\rightarrow&\,\hspace{10mm}-8\mathcal{B}_5-32\mathcal{B}_6\,,
\end{align}
\endgroup
 where the amplitudes are computed by expanding around a flat background metric, $g_{\mu\nu}=\eta_{\mu\nu}+\frac{2}{M_{\text{P}}}h_{\mu\nu}$ and $|_{(3)}$ denotes taking the $\mathcal{O}\left(h^3\right)$ part of the operator. 

\subsection{Cubic Superamplitudes}
\label{sec:susy3point}
In this section, we bring together everything we have discussed so far to build cubic superamplitudes of graviton supermultiplets and to subsequently determine which linear combinations of the operators (\ref{cubicoperators}) are consistent with varying amounts of supersymmetry. 

We begin by noting that as per the discussion in Section \ref{sec:solvingWI}, at cubic order the non-universal part of the superamplitude is a function of a single, $Q$-invariant Grassmann variable that we choose to be 
\begin{equation}
\label{eta12}
    \eta^I_{12,a} = 
\eta_{1,a}^{I} - \frac{1}{m}[1^{I}2_{J}]\eta_{2,a}^{J}.
\end{equation}
Our goal is to construct the most general superamplitude that contributes to the cubic massive graviton component amplitude. Without loss of generality we can restrict our analysis to the $SU(\mathcal{N})_R\times U(1)_R$ singlet sector. As discussed at the end of Section \ref{sec:onshellWI}, this does not mean we are assuming that the interactions of the model preserve R-symmetry, rather we use the fact that at cubic order the supersymmetry constraints do not mix amplitudes in different R-sectors.\footnote{At multiplicity $n>3$ this is no longer true. Even if an operator contributes local on-shell matrix elements in the R-singlet sector, such as $h^4$, these may mix with non-local tree diagrams containing vertices from non-singlet sectors. The special property of $n=3$ is that there are no such non-local contributions.} From the quantum numbers of the Clifford vacuum states in Table \ref{tab:multiplets}, we find that the massive graviton superfields carry $\mathcal{N}$ units of $U(1)_R$ charge. Since the supersymmetric delta function (\ref{susydelta}) carries $2\mathcal{N}$ units of $U(1)_R$ charge, the R-singlet sector contribution to the function $G(\{\eta\})$ appearing in (\ref{solSWI}) for cubic superamplitudes must be a polynomial of degree $\mathcal{N}$ in $\eta_{12}$. Furthermore the Clifford vacuum state is an $SU(\mathcal{N})_R$ singlet, and so in the contribution to $G(\{\eta\})$ from the R-singlet sector, the R-indices on $\eta_{12,a}$ must be contracted with an invariant tensor. This uniquely fixes the form of the R-singlet contribution to the cubic superamplitude to have the form
\begin{equation}
    \mathcal{A}_3\left(\Phi^{\{I_1\}}\Phi^{\{I_2\}}\Phi^{\{I_3\}}\right)\biggr\vert_{\text{R-singlet}}=\delta^{(2\mathcal{N})}\left(Q^\dagger\right) {F^{\{I_1\}\{I_2\}\{I_3\}}}_{J_1...J_{\mathcal{N}}} \epsilon^{a_1...a_{\mathcal{N}}}\eta_{12,a_1}^{J_1}...\eta_{12,a_{\mathcal{N}}}^{J_{\mathcal{N}}}\,.
\end{equation}
Note that this means that $F$ is fully symmetric in $J_1, \cdots J_{\mathcal{N}}$. Here $\{I_i\}$ collectively denotes the external indices of the on-shell superfield $\Phi$. Because of our choice of supersymmetry invariant (\ref{eta12}), the internal indices always correspond to particle 1.
The problem of constructing the general form of the R-singlet sector superamplitude is reduced to constructing a single, commuting c-number \textit{F-function}. The procedure for constructing this function is as follows:
\begin{enumerate}
    \item Write down the most general expression for $F$ using spinor-braces (\ref{braces}). External indices $\{I_i\}$ are supressed (bolded), while internal indices $J_1,...,J_\mathcal{N}$ are explicit. 
    \item Resolve the redundancy associated with the identities (\ref{braceschoutenA}). The result is the most general $F$-function for \textit{distinguishable} superfields.
    \item Impose \textit{super-statistics} constraints to enforce Bose/Fermi symmetry for indistinguishable superfields.
\end{enumerate}
What we call super-statistics constraints are the requirement that the superamplitude is totally symmetric for a bosonic Clifford vacuum ($\mathcal{N}=2\text{ and }4$) and totally anti-symmetric for a fermionic Clifford vacuum ($\mathcal{N}=1\text{ and }3$). This condition is equivalent to requiring Bose/Fermi symmetry for each component amplitude containing pairs of identical states. Explicitly for the $F$-functions, the super-statistics constraints for $1\leftrightarrow 2$ exchange are
\begin{align}
\label{superstatistics12}
F_{\mathcal{N}=1}^{\{I_1\}\{I_2\}\{I_3\} L_1}&= -\left. \left(F_{\mathcal{N}=1}\right|_{1\leftrightarrow 2}\right)^{\{I_2\} \{I_1\} \{I_3\} L_2}  \frac1m [2_{L_2}1^{L_1}]\,,\nn\\
F_{\mathcal{N}=2}^{\{I_1\} \{I_2\} \{I_3\} K_1 L_1}&= \left. \left(F_{\mathcal{N}=2}\right|_{1\leftrightarrow 2}\right)^{\{I_2\} \{I_1\} \{I_3\} K_2 L_2}  \frac1{m^2} [2_{K_2}1^{K_1}][2_{L_2}1^{L_1}]\,,\nn\\
F_{\mathcal{N}=3}^{I_1 I_2 I_3 J_1 K_1 L_1}&= -\left. \left(F_{\mathcal{N}=3}\right|_{1\leftrightarrow 2}\right)^{I_2 I_1 I_3 J_2 K_2 L_2}  \frac1{m^3} [2_{J_2}1^{J_1}][2_{K_2}1^{K_1}][2_{L_2}1^{L_1}]\,,\nn\\
F_{\mathcal{N}=4}^{I_1 J_1 K_1 L_1}&= \left. \left(F_{\mathcal{N}=4}\right|_{1\leftrightarrow 2}\right)^{I_2 J_2 K_2 L_2}  \frac1{m^4}[2_{I_2}1^{I_1}] [2_{J_2}1^{J_1}][2_{K_2}1^{K_1}][2_{L_2}1^{L_1}]\,.
\end{align}
Similarly, the super-statistics constraints for $1\leftrightarrow 3$ exchange
\begin{align}
    F^{\{I_1\}\{I_2\}\{I_3\} L_1}_{\mathcal{N}=1}&=\left.\left(F_{\mathcal{N}=1}\right|_{1\leftrightarrow 3}\right)^{\{I_3\}\{I_2\}\{I_1\} L_3}\frac{1}{m} \langle3_{L_3} 1^{L_1}\rangle \,,\nonumber\\
    F^{\{I_1\}\{I_2\}\{I_3\}K_1 L_1}_{\mathcal{N}=2}&=\left.\left(F_{\mathcal{N}=2}\right|_{1\leftrightarrow 3}\right)^{\{I_3\}\{I_2\}\{I_1\}K_3 L_3}\frac{1}{m^2} \langle3_{K_3} 1^{K_1}\rangle \langle3_{L_3} 1^{L_1}\rangle\,,\nonumber\\
    F^{I_1 I_2 I_3 J_1 K_1 L_1}_{\mathcal{N}=3}&=\left.\left(F_{\mathcal{N}=3}\right|_{1\leftrightarrow 3}\right)^{I_3 I_2 I_1 J_3 K_3 L_3}\frac{1}{m^3} \langle3_{J_3} 1^{J_1}\rangle \langle3_{K_3} 1^{K_1}\rangle \langle3_{L_3} 1^{L_1}\rangle\,,\nonumber\\
     F^{I_1 J_1 K_1 L_1}_{\mathcal{N}=4}&=\left.\left(F_{\mathcal{N}=4}\right|_{1\leftrightarrow 3}\right)^{I_3 J_3 K_3 L_3}\frac{1}{m^4} \langle3_{I_3} 1^{I_1}\rangle \langle3_{J_3} 1^{J_1}\rangle \langle3_{K_3} 1^{K_1}\rangle \langle3_{L_3} 1^{L_1}\rangle\,.
\end{align}
By composing these pairs of conditions, the resulting superamplitude is automatically (anti-)symmetric under $2\leftrightarrow 3$.

After constructing the most general $F$-function, we can project out the physical component amplitudes. We are specifically interested in the constraints on the cubic massive graviton amplitudes which take the general form
\begin{align}
\label{eq:generic3h}
A_3\left(h,h,h\right) = b_1 \mathcal{B}_1+b_2 \mathcal{B}_2+b_3 \mathcal{B}_3+b_4 \mathcal{B}_4+b_5 \mathcal{B}_5+b_6 \mathcal{B}_6\,,
\end{align}
where $\mathcal{B}_i$ are defined in \eqref{cubicamps}. To extract this from the superamplitude we use the massive graviton projectors 
\begin{align}
    h_{\mathcal{N}=1}^{IJKL} &\propto  \frac{\partial}{\partial \eta_{(I}}\Psi^{JKL)} \biggr \vert_{\eta=0}\nonumber\\
    h_{\mathcal{N}=2}^{IJKL} &\propto  \epsilon_{ab}\frac{\partial^2}{\partial \eta_{(I,a}\partial \eta_{J,b}}\Gamma^{KL)} \biggr \vert_{\eta=0}\nonumber\\
    h_{\mathcal{N}=3}^{IJKL} &\propto  \epsilon_{abc}\frac{\partial^3}{\partial \eta_{(I,a}\partial \eta_{J,b}\partial \eta_{K,c}}\Lambda^{L)} \biggr \vert_{\eta=0}\nonumber\\
    h_{\mathcal{N}=4}^{IJKL} &\propto  \epsilon_{abcd}\frac{\partial^4}{\partial \eta_{(I,a}\partial \eta_{J,b}\partial \eta_{K,c}\partial \eta_{L),d}}\Phi\biggr \vert_{\eta=0},
\end{align}
where the proportionality factors are not important for this calculation. Finally, by matching the projection to the general form (\ref{eq:generic3h}) and using (\ref{cubicoperators}) we derive the constraints of supersymmetry on cubic massive graviton interactions. The results of this calculation are summarized in Table \ref{tab:susycubicgraviton}. Below we give the explicit form of the $F$-function in each case and comment on the physical implications of the results.

\begin{table}
    \centering
     \begin{tabular}{|c|c|c|c|c|}
    \hline
    & $\mathcal{N}=1$ & $\mathcal{N}=2$
    & $\mathcal{N}=3$
    & $\mathcal{N}=4$\\
    \hline
    Parameters in ansatz & 24 & 19& 8 & 1\\
    \hline
    Parameters after super-statistics & 4 & 5& 2& 1\\
    \hline
    Parameters in $A_3(h,h,h)$ & 4& 4& 2&1\\
    \hline
    Supersymmetrizable amplitudes & $\mathcal{B}_1$,$\mathcal{B}_2$,$\mathcal{B}_3$,$\mathcal{B}_5$ &$\mathcal{B}_1$,$\mathcal{B}_2$,$\mathcal{B}_3$,$\mathcal{B}_5$& $\mathcal{B}_1$,$\mathcal{B}_2$&
    $\mathcal{B}_2$\\
    \hline
    Supersymmetrizable operators & $\mathcal{L}_1$,$\mathcal{L}_2$,$\mathcal{L}_3$,$\mathcal{L}_5$ & $\mathcal{L}_1$,$\mathcal{L}_2$,$\mathcal{L}_3$,$\mathcal{L}_5$ & $\mathcal{L}_1$,$\mathcal{L}_2$ & $3\mathcal{L}_1+\mathcal{L}_2$\\
    \hline
    \end{tabular}
    \caption{Summary of supersymmetry constraints on cubic massive spin-2 interactions. The explicit form of the amplitudes $\mathcal{B}_i$ are given in (\ref{cubicamps}), and the correspondence with local operators $\mathcal{L}_i$ in (\ref{cubicoperators}). In general, arbitrary linear combinations of the listed amplitudes (operators) are consistent with supersymmetry, except for $\mathcal{N}=4$ where only a specific linear combination of operators $\mathcal{L}_1$ and $\mathcal{L}_2$ is supersymmetrizable.  }
    \label{tab:susycubicgraviton}
\end{table}

\paragraph{$\mathbf{\mathcal{N}=1}$ massive gravity:}
\begingroup
\allowdisplaybreaks
\begin{align}
&F_{\mathcal{N}=1}^{ L}\nn\\
&=\beta_1 \bigg(\{\mathbf{33}\} (\{1^L\mathbf{3}\}-2 \{\mathbf{3}1^L\}) \{\mathbf{12}\}^3+3 \{\mathbf{13}\}^2 \{\mathbf{23}\} \{1^L\mathbf{2}\} \{\mathbf{12}\}+\{\mathbf{11}\}
(\{\mathbf{12}\} ((3 \{\mathbf{3}1^L\}\nonumber\\
&\hspace{1cm}-\{1^L\mathbf{3}\})\{\mathbf{23}\}^2-2 (\{\mathbf{2}1^L\} \{\mathbf{33}\}+\{\mathbf{32}\} \{\mathbf{3}1^L\})
\{\mathbf{23}\}+\{\mathbf{22}\} \{\mathbf{33}\} (2 \{\mathbf{3}1^L\}-\{1^L\mathbf{3}\})) \nn \\
&\hspace{1cm}+\{\mathbf{13}\} \{\mathbf{23}\}(\{\mathbf{23}\}(2
\{\mathbf{2}1^L\}-\{1^L\mathbf{2}\})-\{\mathbf{22}\} (\{\mathbf{3}1^L\}+\{1^L\mathbf{3}\})))\bigg)\nn\\
&+\beta_2 \bigg(\{\mathbf{33}\} (2 \{\mathbf{3}1^L\}-\{1^L\mathbf{3}\}) \{\mathbf{12}\}^3+3 \{\mathbf{13}\} \{\mathbf{23}\}(\{1^L\mathbf{3}\}-2 \{\mathbf{3}1^L\})
\{\mathbf{12}\}^2\nonumber\\
&\hspace{1cm}+(3 \{\mathbf{23}\} \{\mathbf{2}1^L\} \{\mathbf{13}\}^2+\{\mathbf{11}\} ((6 \{\mathbf{3}1^L\}-5 \{1^L\mathbf{3}\})\{\mathbf{23}\}^2-(\{\mathbf{2}1^L\}\{\mathbf{33}\}\nonumber\\
&\hspace{1cm}+\{\mathbf{32}\} \{\mathbf{3}1^L\}) \{\mathbf{23}\}+\{\mathbf{22}\} \{\mathbf{33}\} (\{1^L\mathbf{3}\}-2
\{\mathbf{3}1^L\}))) \{\mathbf{12}\}\nonumber\\
&\hspace{1cm}+\{\mathbf{11}\} \{\mathbf{13}\} \{\mathbf{23}\} (\{\mathbf{23}\} (\{1^L\mathbf{2}\}-2 \{\mathbf{2}1^L\})+\{\mathbf{22}\}
(\{\mathbf{3}1^L\}+\{1^L\mathbf{3}\}))\bigg) \nonumber\\
&+\beta_3 \bigg((\{\mathbf{13}\} (-\{\mathbf{2}1^L\} \{\mathbf{33}\}+2 \{\mathbf{32}\} \{\mathbf{3}1^L\}-2 \{\mathbf{23}\}(\{\mathbf{3}1^L\}-\{1^L\mathbf{3}\}))+\{\mathbf{33}\}
\{\mathbf{21}\} \{\mathbf{3}1^L\}\nonumber\\
&\hspace{1cm}-\{\mathbf{12}\} \{1^L\mathbf{3}\})) \{\mathbf{12}\}^2+\{\mathbf{11}\} \{\mathbf{23}\} (-\{\mathbf{1}1^L\} \{\mathbf{22}\}
\{\mathbf{33}\}+\{\mathbf{12}\} (\{\mathbf{2}1^L\} \{\mathbf{33}\}\nonumber\\
&\hspace{1cm}-2 \{\mathbf{32}\} \{\mathbf{3}1^L\}+2 \{\mathbf{23}\}
(\{\mathbf{3}1^L\}-\{1^L\mathbf{3}\}))+\{\mathbf{13}\} \{\mathbf{22}\} \{1^L\mathbf{3}\})\bigg)\nonumber\\
&+\beta_4 \bigg(\{\mathbf{11}\} \{\mathbf{22}\} (-2 \{\mathbf{1}1^L\} \{\mathbf{23}\} \{\mathbf{33}\}+\{\mathbf{21}\} \{\mathbf{3}1^L\}+\{\mathbf{12}\} (\{\mathbf{3}1^L\}-\{1^L\mathbf{3}\}))
\{\mathbf{33}\}\nonumber\\
&\hspace{1cm}+\{\mathbf{13}\} (\{\mathbf{32}\} \{\mathbf{3}1^L\}+\{\mathbf{23}\} (\{1^L\mathbf{3}\}-\{\mathbf{3}1^L\})))\bigg)\,.
\end{align}
\endgroup
Projecting onto the cubic graviton amplitude and matching to
\eqref{eq:generic3h},
\begingroup
\allowdisplaybreaks
\begin{align}
    b_1=&-192\, m^{14}M_{\text{P}}\left(\beta_1 +2\beta_2-\beta_3-2\beta_4\right)\,,\nn\\
    b_2=&-64\, m^{14}M_{\text{P}}\left(\beta_1 +2\beta_2+2\beta_3+2\beta_4\right)\,,\nn\\
    b_3=&64\, m^{14}M_{\text{P}}\left(\beta_1 +2\beta_2-2\beta_4\right)\,,\nn\\
    b_5=&96\, i\, m^{14}M_{\text{P}}\beta_1\,,\nn\\
    b_4=&b_6=0\,.
\end{align}
\endgroup
Translating this into a statement about the local operator basis (\ref{cubicoperators}):
\vspace{3mm}
\begin{center}
    \noindent\fbox{
\begin{minipage}{14cm}
For $\mathcal{N}=1$ supersymmetry, arbitrary linear combinations of the operators $\mathcal{L}_1$, $\mathcal{L}_2$, $\mathcal{L}_3$ and $\mathcal{L}_5$ are supersymmetrizable. 
\end{minipage}
}
\end{center}
\vspace{3mm}
Conversely, this result shows that for any amount of supersymmetry the operators $\mathcal{L}_4$ and $\mathcal{L}_6$ must vanish. These operators correspond to the two inequivalent contractions of $(\text{Riemann})^3$. An identical statement is true for massless supergravity where these operators lead to forbidden \textit{all-plus} helicity amplitudes. This suggests that the two results might be connected in an appropriate massless limit; we confirm this intuition in Section \ref{sec:masslesscubic}. 
\vspace{2mm}
\paragraph{$\mathbf{\mathcal{N}=2}$ massive gravity:}

\begingroup
\allowdisplaybreaks
\begin{align}
&F_{\mathcal{N}=2}^{ K L}\nn\\
&= \beta_1 \bigg(\{\textbf{11}\} (\{\textbf{23}\} (\{\textbf{2}1^{L}\} \{\textbf{3}1^{K}\} + \{\textbf{2}1^{K}\} \{\textbf{3}1^{L}\} - \{\textbf{32}\} \{1^{K}1^{L}\}) + \{\textbf{22}\} (-\{\textbf{3}1^{K}\} \{\textbf{3}1^{L}\}\nn\\
&\hspace{10mm}+ \{\textbf{33}\} \{1^{K}1^{L}\})) + \{\textbf{13}\}^2 \{\textbf{2}1^{K}\} (-\{\textbf{2}1^{L}\} + \{1^{L}\textbf{2}\}) + \frac12 \{\textbf{12}\} \{\textbf{23}\} (-\{\textbf{1}1^{L}\} \{\textbf{3}1^{K}\}\nn\\
&\hspace{10mm}+ \{\textbf{1}1^{K}\} (\{\textbf{3}1^{L}\} - 2 \{1^{L}\textbf{3}\})) + \{\textbf{12}\}^2 \{\textbf{3}1^{K}\} \{1^{L}\textbf{3}\} + \frac12 \{\textbf{12}\} \{\textbf{13}\} (-2 \{\textbf{2}1^{K}\} \{1^{L}\textbf{3}\}\nn\\
&\hspace{10mm}+ \{\textbf{23}\} (\{1^{K}1^{L}\} + \{1^{L}1^{K}\}))\bigg)\nonumber\\
&+ \beta_2 \bigg(\{\textbf{13}\}^2 \{\textbf{2}1^K\} (\{\textbf{2}1^L\} - \{1^L \textbf{2}\}) + \frac12 \{\textbf{12}\} \{\textbf{23}\}(\{\textbf{1}1^L\} \{\textbf{3}1^K\} - 
    \{\textbf{1}1^K\} \{\textbf{3}1^L\})\nonumber\\
&\hspace{10mm}- \{\textbf{12}\}^2 \{\textbf{3}1^K\} \{1^L\textbf{3}\} - \frac12 \{\textbf{12}\} \{\textbf{13}\}(-2 \{\textbf{2}1^K\} \{1^L\textbf{3}\} + 
    \{\textbf{23}\} (\{1^K1^L\} + \{1^L1^K\}))\nonumber\\
    &\hspace{10mm}+ \{\textbf{11}\} (-\{\textbf{23}\} (\{\textbf{2}1^L\} \{\textbf{3}1^K\}+ \{\textbf{2}1^K\} \{\textbf{3}1^L\} - \{\textbf{32}\} \{1^K1^L\})+ \{\textbf{22}\} (\{\textbf{3}1^K\} \{\textbf{3}1^L\}\nonumber\\
    &\hspace{10mm} + \{\textbf{33}\} \{1^L1^K\}))+ \{\textbf{12}\} \{\textbf{23}\} 
    \{\textbf{1}1^K\} \{1^L\textbf{3}\}\bigg)\nn\\
&+\beta_3 \bigg(-\{\textbf{12}\}^2 \{\textbf{3}1^K\} (\{\textbf{3}1^L\} - 2 \{1^L\textbf{3}\}) + \{\textbf{11}\} \{\textbf{23}\} (\{\textbf{2}1^L\} \{\textbf{3}1^K\} + \{\textbf{23}\} \{1^K1^L\}\nonumber\\
&\hspace{10mm}+ \{\textbf{2}1^K\} \{1^L\textbf{3}\}) + \{\textbf{12}\} (-2 \{\textbf{1}1^L\} \{\textbf{23}\} \{\textbf{3}1^K\} + \{\textbf{13}\} \{\textbf{2}1^K\} \{\textbf{3}1^L\} \nonumber\\
&\hspace{10mm}- 2 \{\textbf{13}\} \{\textbf{23}\} \{1^K1^L\}+ 2 \{\textbf{21}\} \{\textbf{33}\} \{1^K1^L\} + \{\textbf{1}1^K\} \{\textbf{23}\} (\{\textbf{3}1^L\}- \{1^L\textbf{3}\})\nonumber\\
&\hspace{10mm}- 2 \{\textbf{13}\} \{\textbf{2}1^K\} \{1^L\textbf{3}\} + 2 \{\textbf{13}\} \{\textbf{23}\} \{1^L1^K\}+ \{\textbf{13}\} \{\textbf{32}\} \{1^K1^L\})\bigg) \nonumber\\
&+\beta_4 \bigg(\{\textbf{11}\} \{\textbf{23}\}(\{\textbf{23}\} \{1^L1^K\}-\{\textbf{2}1^L\} \{\textbf{3}1^K\} - \{\textbf{2}1^K\} \{1^L\textbf{3}\}) + \frac12  \{\textbf{12}\} (\{\textbf{13}\} \{\textbf{23}\} \{1^K1^L\}\nn\\
&\hspace{10mm}+3 \{\textbf{1}1^L\} \{\textbf{23}\} \{\textbf{3}1^K\}- 2 \{\textbf{13}\} \{\textbf{2}1^K\} \{\textbf{3}1^L\} - 2 \{\textbf{21}\} \{\textbf{33}\} \{1^K1^L\} \nn\\
&\hspace{10mm} + 4 \{\textbf{13}\} \{\textbf{2}1^K\} \{1^L\textbf{3}\} - 5 \{\textbf{13}\} \{\textbf{23}\} \{1^L1^K\}- \{\textbf{1}1^K\} \{\textbf{23}\} (\{\textbf{3}1^L\}- 2 \{1^L\textbf{3}\})) \nn\\
&\hspace{10mm}+ \{\textbf{12}\}^2 (\{\textbf{3}1^K\} (\{\textbf{3}1^L\}- 2 \{1^L\textbf{3}\}) + \{\textbf{33}\} \{1^L1^K\})\bigg)\nonumber\\
&+\beta_5 \bigg( \frac{1}{2} \{\textbf{12}\} (\{\textbf{1}1^L\} \{\textbf{23}\} \{\textbf{3}1^K\} + 2 \{\textbf{12}\} \{\textbf{33}\} \{1^K1^L\}- 2 \{\textbf{13}\} \{\textbf{32}\} \{1^K1^L\} \nonumber\\
&\hspace{10mm}+ \{\textbf{13}\} \{\textbf{23}\} \{1^K1^L\} -
   \{\textbf{1}1^K\} \{\textbf{23}\} \{\textbf{3}1^L\} - 2 \{\textbf{21}\} \{\textbf{33}\} \{1^K1^L\}\nn\\
   &\hspace{10mm}- \{\textbf{13}\} \{\textbf{23}\} \{1^L1^K\})\bigg).
\end{align}
\endgroup
Projecting onto the cubic graviton amplitude and matching to \eqref{eq:generic3h},
\begin{align}
    b_1=&-96\, m^{12}\, M_{\text{P}} (\beta_1 + 3 \beta_2 + 
   2 \beta_3 + \beta_4 - \beta_5)\,,\nonumber\\
    b_2=&-32\, m^{12}\, M_{\text{P}} (\beta_1 - 5 \beta_2 - 
    5 \beta_4 - \beta_5)\,,\nonumber\\
    b_3=& 32\, m^{12}\, M_{\text{P}} (\beta_1 + 3 \beta_2 + 
    4 \beta_3 - \beta_4 - \beta_5)\,,\nonumber\\
    b_5=& 16\, i\, m^{12}\, M_{\text{P}} (\beta_1 - \beta_2 + 
    4 \beta_3 - \beta_4 - \beta_5)\,, \nonumber\\
    b_4=&b_6=0\,.
\end{align}
Translating this into a statement about the local operator basis (\ref{cubicoperators}):
\vspace{3mm}
\begin{center}
    \noindent\fbox{
\begin{minipage}{14cm}
For $\mathcal{N}=2$ supersymmetry, arbitrary linear combinations of the operators $\mathcal{L}_1$, $\mathcal{L}_2$, $\mathcal{L}_3$ and $\mathcal{L}_5$ are supersymmetrizable. 
\end{minipage}
}
\end{center}
\vspace{3mm}
The same operators consistent with $\mathcal{N}=1$ supersymmetry are also consistent with $\mathcal{N}=2$. Note that there are 5 parameters in the $F$-function, but only 4 parameters in the cubic graviton amplitude, unlike $\mathcal{N}=1$ where there are 4 in each. This is not a contradiction, the extra parameter contributes to other component amplitudes such as $A_3\left(h, \gamma, \tilde{\gamma} \right)$.

\paragraph{$\mathbf{\mathcal{N}=3}$ massive gravity:}
\begingroup
\allowdisplaybreaks
\begin{align}
    &F_{\mathcal{N}=3}^{JKL}\nn\\
    &= \beta_1\bigg(-2 \{\textbf{1}1^L\} \{\textbf{23}\} \{1^J1^K\} + \{\textbf{13}\} \{\textbf{2}1^L\} \{1^J1^K\} +  \{\textbf{1}1^K\} \{\textbf{23}\} \{1^J1^L\}\nn\\
    &\hspace{1cm} - 3 \{\textbf{13}\} \{\textbf{2}1^K\} \{1^J1^L\} +  2 \{\textbf{12}\} \{\textbf{3}1^K\} \{1^J1^L\} + \{\textbf{13}\} \{\textbf{2}1^J\} \{1^K1^L\}\nn\\
    &\hspace{1cm} +  2 \{\textbf{21}\} \{\textbf{3}1^J\} \{1^K1^L\} +  2 \{\textbf{13}\} \{1^J\textbf{2}\} \{1^K1^L\} -  2 \{\textbf{12}\} \{1^J\textbf{3}\} \{1^K1^L\}\nn\\
    &\hspace{1cm} -  \{\textbf{1}1^K\} \{\textbf{23}\} \{1^L1^J\} + \{\textbf{13}\} \{\textbf{2}1^K\} \{1^L1^J\} -  \{\textbf{12}\} \{\textbf{3}1^K\} \{1^L1^J\}\nn\\
    &\hspace{1cm} +  \{\textbf{12}\} \{\textbf{3}1^J\} \{1^L1^K\} +  \{\textbf{1}1^J\} (3 \{\textbf{2}1^K\} \{\textbf{3}1^L\} - \{\textbf{23}\} (4 \{1^K1^L\} + \{1^L1^K\}))\bigg)\nn\\
    & +\beta_2 \bigg(-\{\textbf{1}1^L\} \{\textbf{23}\} \{1^J1^K\} + (-\{\textbf{1}1^J\} \{\textbf{23}\} + \{\textbf{21}\} \{\textbf{3}1^J\} + \{\textbf{13}\} \{1^J\textbf{2}\}) \{1^K1^L\}\nn\\
    &\hspace{1cm} + \{\textbf{12}\} (\{\textbf{3}1^L\} \{1^J1^K\} - \{1^J\textbf{3}\} \{1^K1^L\})\bigg)\,.
\end{align}
\endgroup
Projecting onto the cubic graviton amplitude and matching to \eqref{eq:generic3h},
\begin{align}
    b_1=&-192\, m^{10}\, M_{\text{P}} \beta_1\,,\nonumber\\
    b_2=& 128\, m^{10}\, M_{\text{P}} (7 \beta_1 + 2 \beta_2) \,,\nonumber\\
    b_3=&b_4=b_5=b_6=0\,.
\end{align}
Translating this into a statement about the local operator basis (\ref{cubicoperators}):
\vspace{3mm}
\begin{center}
    \noindent\fbox{
\begin{minipage}{14cm}
For $\mathcal{N}=3$ supersymmetry, arbitrary linear combinations of the operators $\mathcal{L}_1$ and $\mathcal{L}_2$ are supersymmetrizable. 
\end{minipage}
}
\end{center}
\vspace{3mm}
Interestingly this two parameter family of interactions coincides with the ghost-free interactions of the dRGT model. This result may indicate that the decoupling of the Boulware-Deser ghost (at cubic order) is a consequence of $\mathcal{N}\geq 3$ supersymmetry. 

\paragraph{$\mathbf{\mathcal{N}=4}$ massive gravity:}
\begin{align}
    &F_{\mathcal{N}=4}^{I_1 J_1 K_1 L_1} = \beta_1\bigg( \{1^{I_1}1^{K_1}\} \{1^{L_1}1^{J_1}\}+\{1^{J_1}1^{I_1}\} \{1^{K_1}1^{L_1}\}\bigg)\,.
\end{align}
Projecting onto the cubic graviton amplitude and matching to \eqref{eq:generic3h},
\begin{align}
    b_2=& 256\, m^{8}\, M_{\text{P}} \beta_1 \,,\nonumber\\
    b_1=&b_3=b_4=b_5=b_6=0\,.
\end{align}
Translating this into a statement about the local operator basis (\ref{cubicoperators}):
\vspace{3mm}
\begin{center}
    \noindent\fbox{
\begin{minipage}{14cm}
For $\mathcal{N}=4$ (maximal) supersymmetry, only the linear combination $3\mathcal{L}_1 +\mathcal{L}_2$ is supersymmetrizable.
\end{minipage}
}
\end{center}
\vspace{3mm}
The remarkable simplicity of the above expression for the $F$-function clearly illustrates the constraining power of maximal supersymmetry. In this case the supersymmetrizable cubic graviton interaction is a one-parameter sub-family of the ghost-free interactions. In the parametrization of the dRGT potential (\ref{nonlinlag}) this interaction corresponds to the value
\begin{equation}
\label{eq:splvalue}
   \alpha_3 = -\frac{1}{2} \hspace{5mm}\text{or}\hspace{5mm} c_3 =\frac{1}{4}.
\end{equation}
This particular cubic interaction is very special, it was previously identified as the unique choice consistent with the absence of closed-timelike-curves  \cite{Camanho:2016opx} and also the absence of asymptotic superluminality  \cite{Bonifacio:2017nnt}. It also corresponds to the special dRGT parameter value found in the partially massless decoupling limit of dRGT in de Sitter, where the strong coupling scale is raised \cite{DeRham:2018axr}. 

\section{High-Energy Limit}
\label{sec:HElim}

The kinematic structure of scattering amplitudes for massive states often simplifies in the limit of large center-of-mass energy, $E \gg m$. In this limit the irreducible massive graviton supermultiplets become reducible, break-up into various massless supermultiplets and we regain some of the simplifications of working with massless helicity states. 

By a generalization of the Goldstone boson equivalence theorem \cite{Cornwall:1974km,Vayonakis:1976vz,Lee:1977eg}, we expect the high-energy behavior to be described by a massless effective field theory which non-linearly realizes an (extended) shift symmetry for the longitudinal modes. As is well-known \cite{deRham:2010ik}, in non-supersymmetric dRGT massive gravity this high-energy or \textit{decoupling limit} theory describes the interactions of a massless spin-2 \textit{tensor} mode $h_{\mu\nu}$, a spin-1 \textit{vector} mode $A_\mu$ and a massless spin-0 \textit{scalar} mode $\phi$. The scalar mode is often referred to as the \textit{Galileon} since the decoupling limit theory provides a non-linear realization of the Galileon algebra $\mathfrak{Gal}(4,1)$, an \.{I}n\"{o}n\"{u}-Wigner contraction of the (4+1)d Poincar\'{e} algebra \cite{Goon:2012dy}. Explicitly this symmetry acts on the Galileon as
 \begin{equation}
     \phi \rightarrow \phi + a + b_\mu x^\mu.
 \end{equation}
An alternative perspective on the decoupling limit is given by introducing
\textit{Stückelberg fields} in the effective action \cite{Arkani-Hamed:2002bjr,Schwartz:2003vj}
\begin{align}
h_{\mu\nu}\rightarrow&\;h_{\mu\nu}+\frac{1}{m}(\partial_\mu A_\nu+\partial_\mu A_\nu)+\frac{1}{m^2}(2\partial_\mu\partial_\nu\phi-\partial_\mu A^\alpha\partial_\nu A_\alpha)\nn\\
&-\frac{1}{m^3}(\partial_\mu A^\alpha \partial_\nu\partial_\alpha \phi+\partial_\mu\partial^\alpha\phi\partial_\nu A_\alpha)-\frac{1}{m^4}\partial_\mu\partial^\alpha\phi\partial_\nu\partial_\alpha\phi+...
\end{align}
where $h_{\mu\nu}$, $A_\mu$ and $\phi$ on the right-hand side are precisely the tensor, vector and scalar modes introduced above.
In the dRGT model (\ref{nonlinlag}), the decoupling limit is defined as a double-scaling limit
\begin{equation}
    m\rightarrow 0,\quad M_\text{P}\rightarrow\infty,\quad \text{with} \quad 
    \Lambda_3\equiv \left(m^2 M_{\text{P}}\right)^{1/3}\,\,\textrm{fixed}.
\end{equation}
It is often simpler to analyze the decoupling limit theory than the full massive model. For instance, the decoupling of the Boulware-Deser ghost was first proven in this limit \cite{deRham:2010ik} and only later extended to full massive gravity \cite{Hassan:2011hr}. We expect similar logic to apply to the problem of analyzing the constraints of supersymmetry. If the high-energy limit of an interaction term is inconsistent with $\mathcal{N}$-extended supersymmetry, then this operator cannot be consistent with $\mathcal{N}$-supersymmetry at any scale\footnote{This argument is purely based on a simplification due to special kinematics. It does not require that the decoupling limit model is a consistent limit of an underlying UV completion. Indeed there are various arguments indicating that new states and interactions must appear before the strictly massless high-energy limit is reached \cite{Tolley:2020gtv}. The converse of this argument is also obviously invalid, consistency with supersymmetry in the high-energy limit is not sufficient to guarantee supersymmetry at low-energies where the masses are important.}. 

The precise form of the constraints of supersymmetry on the high-energy interactions depends on which supermultiplets the massless states belong to. Since the resulting massless limits of the massive graviton supermultiplets contain many helicity $0$ and helicity $1$ states across massless multiplets of varying superspin, identifying which of these are the vector and Galileon modes is a non-trivial problem. As we explain in the following subsection, in some cases this can be determined by careful bookkeeping of R-symmetry representations and CPT in the massless limit. 

\subsection{Massless Limit of Massive Supermultiplets}
\label{sec:masslesslimitmultiplets}

In the high-energy limit, massive supermultiplets effectively break-up into massless multiplets, usually with superspin degeneracies. As a consequence, the high-energy limit may admit additional global symmetries not present at low-energies. All statements about the high-energy limit in this section should be understood as approximate statements, valid at leading order in a $1/m$ expansion. The presence of global symmetries makes the definition of R-symmetry in the massless limit ambiguous since we are always free to form arbitrary linear combinations of R- and global symmetry generators. These emergent global symmetries and their relation to the low-energy or \textit{massive} R-symmetry are a useful organizing principle for understanding the dynamics of the high-energy limit. 

Our main application of these global symmetries will be to identify the massless supermultiplets containing the Galileon and vector modes of the original massive graviton. To do this we make use of the fact that these modes have the same quantum numbers as the graviton, i.e. they are in the $\mathbf{1}_0$ representation of the massive R-symmetry group $SU(\mathcal{N})_R\times U(1)_R$ and are CPT even. In some cases we will see that this is enough to uniquely identify the states.

The general approach we take is to begin with the action of the \textit{naive} massless R-symmetry group $SU(\mathcal{N})_R^{m=0} \times U(1)_R^{m=0}$ appropriate for an isolated massless supermultiplet (and its CPT conjugate). The $U(1)_R^{m=0}$ charge is chosen to be $0$ for the highest helicity state of a positive helicity multiplet, $-1$ for the next-to-highest and so on. The quantum numbers of the negative helicity multiplets are fixed by CPT. For CPT self-conjugate multiplets, such as the $\mathcal{N}=4$ vector multiplet, the $U(1)_R^{m=0}$ charge of the highest helicity state is chosen to be $\mathcal{N}/2$ and for the lowest helicity state $-\mathcal{N}/2$. Our goal is to identify a global symmetry $SU(\mathcal{N})_{\text{global}}\times U(1)_{\text{global}}$ such that the massive R-symmetry group is the diagonal subgroup
\begin{equation}    
\label{diagonal}
    SU(\mathcal{N})_R^{m=0} \times U(1)_R^{m=0} \times SU(\mathcal{N})_{\text{global}}\times U(1)_{\text{global}} \xrightarrow[]{\text{diagonal}} SU(\mathcal{N})_R \times U(1)_R.
\end{equation}
There is an immediately obvious reason why such a mechanism is required, that is, why the naive R-symmetry group $SU(\mathcal{N})_R^{m=0} \times U(1)_R^{m=0}$ cannot be preserved when the mass is turned on. In general,  the naive massless R-symmetry group defined above is \textit{chiral}, meaning there are positive helicity states which transform in a complex representation, with the CPT conjugate negative helicity states transforming in the complex conjugate representation. Such symmetries are only consistent for strictly massless states for which helicity is a Lorentz-invariant quantum number. When the mass is turned on however, only non-chiral or \textit{vector-like} symmetries can be preserved. In such cases we require a non-trivial $SU(\mathcal{N})_{\text{global}}\times U(1)_{\text{global}}$ which also acts as a chiral symmetry, with the preserved massive R-symmetry given by the vector-like diagonal subgroup (\ref{diagonal}).

 There are finitely many possible ways $SU(\mathcal{N})_{\text{global}}$ can act on the degenerate multiplets (together with a general parametrization of the $U(1)_{\text{global}}$ charge assignments), and so by considering each possibility in turn we can discover the correct representation corresponding to the known, massive $SU(\mathcal{N})_R \times U(1)_R$ representations described in Table \ref{tab:multiplets}. The results of this section are summarized in Table \ref{tab:decoupling}. These results are consistent with those presented in \cite{Ferrara:2018wqd}.
 
\begin{table}
    \begin{center}
    \begin{tabular}{|c|c|c|c|}
    \hline
        & $h^\pm$ & $v^\pm$ & $\phi$  \\
    \hline
    \hline
       $\mathcal{N}=1$  & graviton & vector + gravitino & chiral\\
    \hline
       $\mathcal{N}=2$  & graviton & vector + gravitino & hyper + vector' \\
    \hline
       $\mathcal{N}=3$  & graviton & vector + gravitino & vector'\\
    \hline
       $\mathcal{N}=4$  & graviton & gravitino &vector\\
    \hline
    \end{tabular}
    \end{center}
    \caption{   
    \label{tab:decoupling}
    Massless supermultiplets containing the tensor $h$, vector $v$ and scalar (Galileon) $\phi$ modes of the massive graviton in the high-energy limit. In some cases $+$ is used to denote a possible admixture, when representation theory is insufficient to identify a unique multiplet. The notation $\text{vector}/\text{vector}$' is used to distinguish distinct massless vector multiplets.}
    \end{table}

\newpage
\noindent \textbf{$\mathbf{\mathcal{N}=1}$ massive graviton:}

\vspace{3mm}
\noindent In this case there is only a question of the $U(1)_{\text{global}}$ charge assignments, we will denote these below the corresponding  massless multiplet.
\begin{align}
    &\left(\mathcal{N}=1 \text{ massive graviton}\right) \nonumber\\ &\xrightarrow[]{m=0} \underbrace{\left(\mathcal{N}=1 \text{ massless graviton}\right)_+}_{0} \oplus \underbrace{\left(\mathcal{N}=1 \text{ massless graviton}\right)_-}_{0} \nonumber\\
    &\hspace{10mm} \oplus \underbrace{\left(\mathcal{N}=1 \text{ massless gravitino}\right)_+}_{1} \oplus \underbrace{\left(\mathcal{N}=1 \text{ massless gravitino}\right)_-}_{-1}\nonumber\\
    &\hspace{10mm} \oplus \underbrace{\left(\mathcal{N}=1 \text{ massless vector}\right)_+}_0 \oplus \underbrace{\left(\mathcal{N}=1 \text{ massless vector}\right)_-}_{0}\nonumber\\
    &\hspace{10mm} \oplus \underbrace{\left(\mathcal{N}=1 \text{ massless chiral}\right)_+}_{1} \oplus \underbrace{\left(\mathcal{N}=1 \text{ massless chiral}\right)_-}_{-1}.
\end{align}
Here the subscript $\pm$ denotes the positive or negative helicity massless multiplet. For example, the $+$ helicity $\mathcal{N}=1$ massless chiral multiplet consists of a single $+1/2$-helicity state and a single $0$-helicity state with $U(1)_R^{m=0}$ charges $0$ and $-1$ respectively. 

In total, this model contains 2 helicity-0 states. Given that the original massive multiplet contained a single massive spin-2 particle and a single massive spin-1 particle, these must correspond to the scalar longitudinal modes of each. Now we enumerate the $U(1)_R^{m=0}\times U(1)_{\text{global}}$ representations of the helicity-0 states, one from each of the positive and negative helicity chiral multiplets, together with the induced representation of the diagonal subgroup 
\begin{equation}
    (-1,1)\oplus (1,-1) \rightarrow (0)\oplus (0).
\end{equation}
As expected both of these states are $U(1)_R$ singlets. To identify the Galileon mode we use the additional fact that the Galileon is a CPT even state. Since CPT acts by interchanging the positive and negative chiral multiplets, we can always form CPT even and odd combinations. In this case, \textit{the even combination must be the Galileon}.

We can try to apply similar logic to the helicity-1 modes, the corresponding induced representations are 
\begin{equation}
    (-1,1)\oplus (0,0) \rightarrow (0)\oplus (0),
\end{equation}
where the first state on the left-hand-side corresponds to the $+$ helicity gravitino multiplet and the second to the $+$ helicity vector multiplet. Here CPT does not give any additional constraints and symmetry alone is not sufficient to identify the vector mode of the $\mathcal{N}=1$ massive graviton.

\vspace{2mm}
\begin{center}
    \noindent\fbox{
\begin{minipage}{14cm}
In the decoupling limit of the $\mathcal{N}=1$ massive graviton, the Galileon mode is uniquely identified with a CPT even linear combination of massless \textit{chiral} multiplets.
\end{minipage}
}
\end{center}

\newpage
\noindent \textbf{$\mathbf{\mathcal{N}=2}$ massive graviton:}

\vspace{3mm}
\noindent
The $SU(2)_{\text{global}}\times U(1)_{\text{global}}$ representations in the massless limit are found to be 
\begin{align}
    &\left(\mathcal{N}=2 \text{ massive graviton}\right) \nonumber\\ &\xrightarrow[]{m=0} \underbrace{\left(\mathcal{N}=2 \text{ massless graviton}\right)_+}_{\mathbf{1}_0} \oplus \underbrace{\left(\mathcal{N}=2 \text{ massless graviton}\right)_-}_{\mathbf{1}_0} \nonumber\\
    &\hspace{10mm}\oplus \underbrace{2\times\left(\mathcal{N}=2 \text{ massless gravitino}\right)_+}_{\mathbf{2}_1} \oplus \underbrace{2\times\left(\mathcal{N}=2 \text{ massless graviton}\right)_-}_{\mathbf{2}_{-1}} \nonumber\\
    &\hspace{10mm}\oplus \underbrace{2\times\left(\mathcal{N}=2 \text{ massless vector}\right)_+}_{\mathbf{1}_2\oplus \mathbf{1}_{0}} \oplus \underbrace{2\times\left(\mathcal{N}=2 \text{ massless vector}\right)_-}_{\mathbf{1}_{-2}\oplus \mathbf{1}_{0}} \nonumber\\
    &\hspace{10mm}\oplus \underbrace{2\times\left(\mathcal{N}=2 \text{ massless hyper}\right)}_{\mathbf{2}_0} .
\end{align}
The induced diagonal representations of the helicity-0 modes are:
\begin{align}
    &(\mathbf{1}_{-2},\mathbf{1}_2)\oplus(\mathbf{1}_{-2},\mathbf{1}_{0})\oplus (\mathbf{1}_{2},\mathbf{1}_{-2})\oplus (\mathbf{1}_{2},\mathbf{1}_{0})\oplus (\mathbf{2}_{0},\mathbf{2}_0) \nonumber\\
    &\hspace{10mm}\rightarrow \mathbf{1}_{0}\oplus \mathbf{1}_{-2}\oplus \mathbf{1}_{0}\oplus \mathbf{1}_{2}\oplus \mathbf{1}_{0}\oplus \mathbf{3}_0,
\end{align}
where the states on the left-hand-side are labelled by their quantum numbers under the group $ SU(2)_R^{m=0} \times U(1)_R^{m=0} \times SU(2)_{\text{global}} \times U(1)_{\text{global}} $. Here we find 3 different states with the quantum numbers $\mathbf{1}_0$ of the massive graviton. One of these arises from the pair of hyper multiplets and two more from the $\mathbf{1}_2 \in SU(2)_{\text{global}}\times U(1)_{\text{global}}$ part of the vector multiplets. In the hyper multiplet case the state is CPT even and in the vector multiplet case we can form a CPT even linear combination as we did for $\mathcal{N}=1$. So we have exhausted the analysis of symmetries and failed to uniquely identify the Galileon mode of the $\mathcal{N}=2$ massive graviton; in general it could be a linear admixture of states in vector and hyper multiplets\footnote{It is possible that this is a feature and not a bug. An $\mathcal{N}=2$ massive graviton could arise from a ``Higgsing" of diffeomorphism invariance in different ways in different UV models, analogous to the way massive vector bosons can arise from the Coulomb branch or the Higgs branch of the moduli space of vacua in $\mathcal{N}=2$ gauge theories. We may require an explicit UV completion to disambiguate which multiplet the Galileon belongs to.}. 

The induced diagonal representations of the helicity-1 modes are:
\begin{equation}
    (\mathbf{1}_{-2},\mathbf{1}_0)\oplus (\mathbf{2}_{-1},\mathbf{2}_1)\oplus (\mathbf{1}_{0},\mathbf{1}_2)\oplus (\mathbf{1}_{0},\mathbf{1}_{0})\rightarrow \mathbf{1}_{-2}\oplus \mathbf{1}_{0}\oplus \mathbf{3}_{0}\oplus \mathbf{1}_2\oplus \mathbf{1}_{0}.
\end{equation}
We can construct CPT even combinations of states from both gravitino and vector multiplets and so cannot uniquely identify the vector mode of the $\mathcal{N}=2$ massive graviton. 

\vspace{2mm}
\begin{center}
    \noindent\fbox{
\begin{minipage}{14cm}
In the decoupling limit of the $\mathcal{N}=2$ massive graviton, neither the Galileon nor vector multiplets can be uniquely identified without further dynamical input.
\end{minipage}
}
\end{center}

\newpage
\noindent \textbf{$\mathbf{\mathcal{N}=3}$ massive graviton:}

\vspace{3mm}
\noindent The $SU(3)_{\text{global}}\times U(1)_{\text{global}}$ representations in the massless limit are found to be 
\begin{align}
    &\left(\mathcal{N}=3 \text{ massive graviton}\right) \nonumber\\ &\xrightarrow[]{m=0} \underbrace{\left(\mathcal{N}=3 \text{ massless graviton}\right)_+}_{\mathbf{1}_0} \oplus \underbrace{\left(\mathcal{N}=3 \text{ massless graviton}\right)_-}_{\mathbf{1}_0}  \nonumber\\
    &\hspace{10mm} \oplus \underbrace{3\times\left(\mathcal{N}=3 \text{ massless gravitino}\right)_+}_{\overline{\mathbf{3}}_1} \oplus \underbrace{3\times\left(\mathcal{N}=3 \text{ massless gravitino}\right)_-}_{\mathbf{3}_{-1}} \nonumber\\
    &\hspace{10mm} \oplus \underbrace{4\times\left(\mathcal{N}=3 \text{ massless vector}\right)_+}_{\mathbf{1}_0\oplus \mathbf{3}_2} \oplus \underbrace{4\times\left(\mathcal{N}=3 \text{ massless vector}\right)_-}_{\mathbf{1}_0\oplus \overline{\mathbf{3}}_{-2}}.
\end{align}
The induced diagonal representations of the helicity-0 modes are:
\begin{align}
    &(\mathbf{1}_{-3},\overline{\mathbf{3}}_1)\oplus(\mathbf{1}_3,\mathbf{3}_{-1})\oplus(\overline{\mathbf{3}}_{-2},\mathbf{1}_0)\oplus(\overline{\mathbf{3}}_{-2},\mathbf{3}_2)\oplus(\mathbf{3}_{2},\mathbf{1}_0)\oplus(\mathbf{3}_2,\overline{\mathbf{3}}_{-2}) \nonumber\\
    &\hspace{10mm}\rightarrow \mathbf{1}_0\oplus \mathbf{1}_0\oplus\mathbf{3}_2\oplus \mathbf{3}_2\oplus\overline{\mathbf{3}}_{-2}\oplus\overline{\mathbf{3}}_{-2}\oplus\mathbf{8}_0\oplus\mathbf{8}_0\,,
\end{align}
where the states on the left-hand-side are labelled by their quantum numbers under the group $ SU(3)_R^{m=0} \times U(1)_R^{m=0} \times SU(3)_{\text{global}} \times U(1)_{\text{global}} $. There are two helicity-0 states with the quantum numbers of the massive graviton $\mathbf{1}_0$, which arise from the positive and negative helicity vector multiplets. These multiplets are exchanged by CPT, so we can uniquely identify the Galileon mode as the CPT even linear combination. 

The induced diagonal representations of the helicity-1 modes are:
\begin{align}
    &(\overline{\mathbf{3}}_{-2},\mathbf{1}_0)\oplus(\mathbf{3}_{-1},\overline{\mathbf{3}}_1)\oplus(\mathbf{1}_0,\mathbf{1}_0)\oplus(\mathbf{1}_0,\mathbf{3}_2)\rightarrow \mathbf{1}_0\oplus\mathbf{1}_0\oplus\mathbf{3}_2\oplus\overline{\mathbf{3}}_{-2}\oplus\mathbf{8}_0.
\end{align}
In this case, again, there are multiple states with helicity-1 and the quantum numbers of the massive graviton $\mathbf{1}_0$, so we are unable to uniquely identify the vector mode of the $\mathcal{N}=3$ massive graviton.

\vspace{2mm}
\begin{center}
    \noindent\fbox{
\begin{minipage}{14cm}
In the decoupling limit of the $\mathcal{N}=3$ massive graviton, the Galileon mode is uniquely identified with a CPT even linear combination of states in massless \textit{vector} multiplets.
\end{minipage}
}
\end{center}

\vspace{3mm}
\noindent \textbf{$\mathbf{\mathcal{N}=4}$ massive graviton:}

\vspace{3mm}
\noindent The $SU(4)_{\text{global}}\times U(1)_{\text{global}}$ representations in the massless limit are found to be 
\begin{align}
    &\left(\mathcal{N}=4 \text{ massive graviton}\right) \nonumber\\ &\xrightarrow[]{m=0} \underbrace{\left(\mathcal{N}=4 \text{ massless graviton}\right)_+}_{\mathbf{1}_0} \oplus  \underbrace{\left(\mathcal{N}=4 \text{ massless graviton}\right)_-}_{\mathbf{1}_0}  \nonumber\\
    &\hspace{10mm} \oplus  \underbrace{4\times\left(\mathcal{N}=4 \text{ massless gravitino}\right)_+}_{\overline{\mathbf{4}}_1} \oplus  \underbrace{4\times\left(\mathcal{N}=4 \text{ massless gravitino}\right)_-}_{\mathbf{4}_{-1}} \nonumber\\
    &\hspace{10mm} \oplus  \underbrace{6\times\left(\mathcal{N}=4 \text{ massless vector}\right)}_{\mathbf{6}_0}.
\end{align}
The induced diagonal representations of the helicity-0 modes are:
\begin{align}
    &(\mathbf{1}_{-4},\mathbf{1}_0)\oplus (\mathbf{1}_4,\mathbf{1}_0) \oplus (\overline{\mathbf{4}}_{-3},\overline{\mathbf{4}}_1) \oplus (\mathbf{4}_3,\mathbf{4}_{-1})\oplus (\mathbf{6}_0,\mathbf{6}_0) \nonumber\\
    &\rightarrow \mathbf{1}_{-4}\oplus \mathbf{1}_4\oplus\mathbf{1}_0 \oplus \mathbf{6}_{-2} \oplus \mathbf{6}_2 \oplus \mathbf{10}_2 \oplus \overline{\mathbf{10}}_{-2} \oplus \mathbf{15}_0 \oplus \mathbf{20}'_0.
\end{align}
There is a unique helicity-0 state with the quantum numbers of the massive graviton $\mathbf{1}_0$ which arises from the CPT self-conjugate vector multiplet, which therefore must be the Galileon mode. More specifically, we denote the scalar fields in the vector multiplet as $\phi_{ij,ab}$ where $i,j=1,...,4$ correspond to the anti-symmetric $\mathbf{6}$ representation of $SU(4)_{\text{global}}$ and $a,b=1,...,4$ correspond to the anti-symmetric $\mathbf{6}$ representation of $SU(4)_{R}^{m=0}$. The Galileon is uniquely identified as the singlet part of the diagonal representation. 

The induced diagonal representations of the helicity-1 modes are:
\begin{equation}
    (\mathbf{6}_{-2},\mathbf{1}_0)\oplus (\mathbf{4}_{-1},\overline{\mathbf{4}}_{1})\oplus (\mathbf{1}_{2},\mathbf{6}_0)\rightarrow \mathbf{1}_0\oplus \mathbf{6}_{-2} \oplus \mathbf{6}_2 \oplus \mathbf{15}_0.
\end{equation}
Here there is a unique helicity-1 mode with the quantum numbers of the massive graviton $\mathbf{1}_0$ which arises from the positive helicity gravitino multiplet. 

\vspace{2mm}
\begin{center}
    \noindent\fbox{
\begin{minipage}{14cm}
In the decoupling limit of the $\mathcal{N}=4$ massive graviton, the Galileon mode is uniquely identified with a linear combination of massless \textit{vector} multiplets given by the operator
\begin{equation}
    \phi^{(\mathcal{N}=4\text{ Galileon})} \propto \epsilon^{ijab} \phi_{ij,ab}.
\end{equation}
The vector mode is uniquely identified with a linear combination states in massless \textit{gravitino} multiplets.
\end{minipage}
}
\end{center}

\subsection{Cubic Interactions in the Massless Limit}
\label{sec:masslesscubic}
In this section, we will study the constraints of massless supersymmetry on the cubic interactions of the tensor, vector and scalar (Galileon) modes of the massive graviton in the massless limit. We use this to rule out possible supersymmetrizations of the massive graviton cubic interactions presented in Section \ref{sec:cubicamps}. Details of the procedure for taking the high-energy limit with massive spinors are reviewed in Appendix \ref{appendix:conventions}.

We now take the massless limit of all the massive spin-2 cubic self-interactions presented in Section \ref{sec:cubicamps}. These interactions only contribute non-trivially to certain massless amplitudes. We find the high energy limit for our various amplitudes to be (plus permutations of the particle labels):
\begingroup
\allowdisplaybreaks
\begin{align}
   \mathcal{B}_1 &\xrightarrow[]{\text{HE}}\left\{
\begin{array}{ll}
      (h^-v^-\phi): & -\frac{1}{M_{\text{P}}m}\frac{\langle 12\rangle^3\langle 13\rangle}{\langle 23\rangle} \\
     (h^+v^+\phi): & -\frac{1}{M_{\text{P}}m}\frac{[12]^3[13]}{[23]} \\
\end{array} \right.\\ \nn\\
\mathcal{B}_2 &\xrightarrow[]{\text{HE}}\left\{
\begin{array}{ll}
      (h^-h^-h^+): & \frac{1}{M_{\text{P}}}\frac{\langle 12\rangle^6}{\langle 13\rangle^2\langle 23\rangle^2} \\
     (h^-v^-v^+): & -\frac{2}{M_{\text{P}}}\frac{\langle 12\rangle^4}{\langle 23\rangle^2} \\(h^-\phi\phi): & \frac{3}{M_{\text{P}}}\frac{\langle 12\rangle^2\langle 13\rangle^2}{\langle 23\rangle^2}\\(h^+\phi\phi): & \frac{3}{M_{\text{P}}}\frac{[ 12]^2[ 13]^2}{[ 23]^2}\\
     (h^+v^+v^-): & -\frac{2}{M_{\text{P}}}\frac{[ 12]^4}{[ 23]^2} \\(h^+h^+h^-): & \frac{1}{M_{\text{P}}}\frac{[ 12]^6}{[ 13]^2[ 23]^2}
\end{array} \right.\\ \nn\\
\mathcal{B}_3 &\xrightarrow[]{\text{HE}}\left\{
\begin{array}{ll}
      (h^-h^-\phi): & \frac{2}{M_{\text{P}}m^2}\langle 12\rangle^4 \\
     (h^-v^-v^-): & -\frac{2}{M_{\text{P}}m^2}\langle 12\rangle^2\langle 13\rangle^2 \\(h^+v^+v^+): & -\frac{2}{M_{\text{P}}m^2}[ 12]^2[ 13]^2 \\(h^+h^+\phi): & \frac{2}{M_{\text{P}}m^2}[12]^4 
\end{array} \right.\\ \nn\\
\mathcal{B}_4 &\xrightarrow[]{\text{HE}}\left\{
\begin{array}{ll}
      (h^-h^-h^-): & \frac{1}{M_{\text{P}}m^4}\langle 12\rangle^2\langle 13\rangle^2\langle 23\rangle^2 \\
     (h^+h^+h^+): & \frac{1}{M_{\text{P}}m^4}[ 12]^2[ 13]^2[23]^2
\end{array} \right.\\ \nn\\
\mathcal{B}_5 &\xrightarrow[]{\text{HE}}\left\{
\begin{array}{ll}
      (h^-h^-\phi): & -\frac{1}{M_{\text{P}}m^2}\langle 12\rangle^4 \\
     (h^-v^-v^-): & \frac{1}{M_{\text{P}}m^2}\langle 12\rangle^2\langle 13\rangle^2 \\(h^+v^+v^+): & -\frac{1}{M_{\text{P}}m^2}[ 12]^2[ 13]^2 \\(h^+h^+\phi): & \frac{1}{M_{\text{P}}m^2}[12]^4 
\end{array} \right.\\ \nn\\
\mathcal{B}_6 &\xrightarrow[]{\text{HE}}\left\{
\begin{array}{ll}
      (h^-h^-h^-): & \frac{3}{M_{\text{P}}m^4}\langle 12\rangle^2\langle 13\rangle^2\langle 23\rangle^2 \\
     (h^+h^+h^+): & -\frac{3}{M_{\text{P}}m^4}[ 12]^2[ 13]^2[23]^2
\end{array} \right.
\end{align}
\endgroup
where $h^\pm$, $v^\pm$ and $\phi$ are the tensor, vector and scalar (Galileon) components respectively of the massive graviton. The high-energy limit of $\mathcal{B}_2$, in particular the non-universality of the coupling, agrees with the results of \cite{Arkani-Hamed:2017jhn}. For a detailed discussion of the physical interpretation of this result see \cite{deRham:2018qqo}.

Amplitudes in the massless limit must preserve at least the same number of supercharges as are preserved in the full massive amplitude. For example, for $\mathcal{B}_1$ to be $\mathcal{N}=1$ supersymmetrizable, $A_3\left(h^- v^- \phi\right)$ must be too. Conversely, if $A_3\left(h^- v^- \phi\right)$ isn't $\mathcal{N}=1$ supersymmetrizable, then $\mathcal{B}_1$ cannot be supersymmetrized. We can use this to rule out massive interactions via the supersymmetry incompatibility of their massless limits. To do this, we use the results of Table \ref{tab:decoupling} to identify which massless multiplets the scalar $\phi$, vector $v^\pm$ and tensor $h^\pm$ components of the massive graviton belong to. We find that which graviton interactions $\mathcal{B}_i$ are not $\mathcal{N}$-supersymemtrizable can be demonstrated already in the massless limit. Additionally, the ones that cannot be ruled out are exactly the supersymmetric interactions listed in Table \ref{tab:susycubicgraviton}.

Note that by studying interactions in the massless limit, we cannot determine their compatibility with massive supersymmetry, only their incompatibility. While the analysis of Section \ref{sec:cubicamps} is therefore strictly stronger, analyzing the high-energy limit is both a useful sanity check and reveals novel relations between supersymmetrizable interactions and the massless multiplets containing the vector and Galileon modes. Below, we go through the massless interactions one-by-one and understand their incompatibility with different amounts of supersymmetry.

\paragraph{Interaction $\mathbf{\mathcal{B}_1}$:}
In the following, we demonstrate the incompatibility of a $A_3\left(h^- v^- \phi\right)$ amplitude (which is generated in the massless limit of $\mathcal{B}_1$) with $\mathcal{N}=4$ supersymmetry. 

We assume that the tensor, vector and scalar modes lie in the $\mathcal{N}=4$ graviton, gravitino and vector multiplets respectively, as indicated in Table \ref{tab:decoupling}. This means that the states carry $R$-symmetry indices and there are actually two distinct classes of such interactions,
\begin{align}
    A_3\left(h^-, v_{abc}^-, \phi_{ab}\right) && A_3\left(h^-, v_{abc}^-, \phi_{ad}\right)\,.
\end{align}
These must vanish due to the following SWI,
\begin{align}
    Q_b \cdot A_3\left(h^-, v_{abc}^-, \psi^+_{a}\right) &= |3] A_3\left(h^-, v_{abc}^-, \phi_{ab}\right)= 0\,,\\
    Q_a \cdot A_3\left(h^-, v_{abc}^-, \psi^+_{d}\right) &= |3] A_3\left(h^-, v_{abc}^-, \phi_{ad}\right)= 0\,.
\end{align}

Note that this argument is evaded in the case of $\mathcal{N}=3$ only when the vector mode of the massive graviton has contributions from the gravitino multiplet in the massless limit. This is because in the gravitino multiplet, a similar Ward identity reads
\begin{align}
    Q_c \cdot A_3\left(h^-, v_{ab}^-, \psi^+\right) &= |2] A_3\left(h^-, \lambda_{abc}^-, \psi^+\right)+ |3] A_3\left(h^-, v_{ab}^-, \phi_c\right)= 0\,.
\end{align}
Since these amplitudes are supported on the branch of massless 3-particle kinematics with $|1]\propto |2] \propto |3]$, this equation can have a non-trivial solution. This places no constraints on the amplitude $A_3\left(h^-, v_{ab}^-, \phi_c\right)$. So the vector mode of a $\mathcal{N}=3$ massive graviton which interacts via vertex $\mathcal{B}_1$ cannot be solely in a $\mathcal{N}=3$ vector multiplet in the massless limit. Conversely, if the vector mode of an $\mathcal{N}=3$ massive graviton is a component of massless vector multiplets only, the interaction $\mathcal{B}_1$ must vanish.

\paragraph{Interaction $\mathbf{\mathcal{B}_2}$:}
Since the massless limit of $\mathcal{B}_2$ is compatible with $\mathcal{N}=1,2,3,4$, we cannot rule out the possibility of $\mathcal{B}_2$ being maximally supersymmetrizable. Indeed as we found in Section \ref{sec:cubicamps}, it is compatible with $\mathcal{N}=1,2,3,4$.

\paragraph{Interaction $\mathbf{\mathcal{B}_3}$:}
In the massless limit, $\mathcal{B}_3$ contributes to $A_3\left(h^-h^-\phi\right)$ which is incompatible with $\mathcal{N}=3,4$ as we show below.

Let us start with $\mathcal{N}=4$. Here the scalar and tensor modes lie in vector and graviton multiplets respectively. As a result they must satisfy the following Ward identity,
\begin{align}
    Q_b \cdot A_3\left(h^-, h^-, \psi^+_a\right)= |3]A_3\left(h^-, h^-, \phi_{ab}\right) = 0\,.
\end{align}
Since this is true for all $a$ and $b$, $\mathcal{B}_3$ is incompatible with $\mathcal{N}=4$.

In an $\mathcal{N}=3$ massive graviton theory, the scalar and tensor modes in the massless limit, once again belong to the vector and graviton multiplets respectively. There are two types of scalars in an $\mathcal{N}=3$ vector multiplet and so two types of amplitudes we need to consider,
\begin{align}
    A_3\left(h^-,h^-,\phi_a\right) && A_3\left(h^-,h^-,\phi_{ab}\right)\,.
\end{align}
Both of these must vanish due to the following Ward identities,
\begin{align}
Q_b \cdot A_3\left(h^-, h^-, \psi^+\right)= |3]A_3\left(h^-, h^-, \phi_{b}\right) = 0\,,\\
Q_b \cdot A_3\left(h^-, h^-, \psi_a^+\right)= |3]A_3\left(h^-, h^-, \phi_{ab}\right) = 0\,.
\end{align}

In the case of $\mathcal{N}=2$, the scalar mode either lies in a hyper multiplet or a vector multiplet. In the former case,
\begin{align}
    Q_b \cdot A_3\left(h^-, h^-, \psi^+\right)= |3]A_3\left(h^-, h^-, \phi_{b}\right) = 0\,,
\end{align}
and the argument goes through. So the Galileon of an $\mathcal{N}=2$ massive graviton which interacts via vertex $\mathcal{B}_3$ cannot be solely in an $\mathcal{N}=2$ hyper multiplet in the massless limit. Again conversely, if the Galileon mode of an $\mathcal{N}=2$ massive graviton is a component of massless hyper multiplets only, the interaction $\mathcal{B}_3$ must vanish.

\paragraph{Interaction $\mathbf{\mathcal{B}_4}$:}
The massless limit of the interaction $\mathcal{B}_4$ leads to a non-zero amplitude $A_3\left(h^\pm, h^\pm, h^\pm \right)$. Such amplitudes must vanish in any supersymmetric theory where $h^\pm\,$ lie in the graviton multiplet. This is due to the SWI,
\begin{align*}
    Q \cdot A_3\left(h^-, h^-, \psi^- \right)=|3] A_3\left(h^-, h^-, h^- \right) = 0\,.
\end{align*}
Therefore $\mathcal{B}_4$ is incompatible with $\mathcal{N}=1,2,3,4$.

\paragraph{Interaction $\mathbf{\mathcal{B}_5}$:}
Similar to $\mathcal{B}_3$, $\mathcal{B}_5$ also gives rise to $A_3\left(h^\pm, h^\pm, \phi\right)$ amplitudes that are incompatible with $\mathcal{N}=3$ and $\mathcal{N}=4$ supersymmetry. The argument is identical to the one discussed for $\mathcal{B}_3$ above. 

\paragraph{Interaction $\mathbf{\mathcal{B}_6}$:}
We see that $\mathcal{B}_6$ is non-supersymmetrizable for the same reason that $\mathcal{B}_4$ is: they both give rise to $A_3\left(h^\pm, h^\pm, h^\pm\right)$ amplitudes that are incompatible with any amount of supersymmetry.

\section{Double Copy}
\label{sec:doublecopy}

The double copy is a well-established map between products of amplitudes in (super) Yang-Mills theory and (super) gravity \cite{Kawai:1985xq,Bern:2019prr}. Recent efforts have extended the double copy to include amplitudes of massive particles \cite{Johansson:2019dnu,Momeni:2020hmc,Moynihan:2020ejh,Gonzalez:2021bes,Moynihan:2021rwh,Gonzalez:2021ztm,Gonzalez:2022mpa}. For the case of a particular dRGT massive gravity, cubic and quartic amplitudes were successfully recognized as the double copy of amplitudes in a theory of massive Yang-Mills \cite{Momeni:2020vvr,Johnson:2020pny}. Unfortunately, a KLT based massive double copy map generates spurious poles beginning at quintic order \cite{Johnson:2020pny}. It is currently unknown if some alternate version of the massive double copy is possible that avoids these pathologies. Nonetheless, cubic and quartic dRGT massive gravity amplitudes are double copies, and the possibility of simplifying extended supersymmetric dynamics by \textit{factoring} it into a product is very compelling. If an alternative, healthy, prescription for the double copy could be found, it would provide a powerful tool for constructing the models we are proposing at all multiplicity. In this section, we will discuss in detail how the double copy works for supersymmetric massive gravity at cubic order.

\subsection{Supersymmetric Massive Yang-Mills}
\label{sec:sYM}
In this section, we will discuss the particle content and possible interactions in a supersymmetric theory of massive Yang-Mills. Our goal is to parallel the discussion of massive gravity in previous sections. We define a \textit{massive gluon} as an (adjoint multiplet of) massive spin-1 particles transforming under a semi-simple Lie algebra $G$ which plays the role of color in the \textit{color-kinematics duality}. Apart from this color group the massive gluon is assumed to be a singlet under all symmetries. 

To follow massless color-kinematics duality as closely as possible, we will assume that the cubic amplitude is proportional to the (totally anti-symmetric) structure constant and define a \textit{color-stripped} amplitude\footnote{We could also consider a generalized double copy with operators constructed from symmetric color tensors $d^{abc}$; for massless models these interactions generically produce spurious singularities beginning at 5-point \cite{Chi:2021mio}. }
\begin{equation}
    A_3\left(g_a,g_b,g_c\right) \equiv f_{abc} A_3[g,g,g].
\end{equation}
For consistency with Bose symmetry the stripped amplitude defined in this way is totally anti-symmetric. At cubic order this definition is unambiguous and coincides with the usual definition of a color-ordered amplitude. The most general stripped cubic interaction between massive gluons is
\begin{align}
\label{eq:N0gluonamp}
    A_3\left[g,g,g\right]=&c_1\ \mathcal{C}_1+c_2\ \mathcal{C}_2+c_3\ \mathcal{C}_3\,,
\end{align}
where we use the following basis
\begin{align}
     \mathcal{C}_1 =& \frac{1}{m^2}\left(z_1\cdot p_2\right) \left(z_2\cdot p_3\right) \left(z_3\cdot p_1\right)\,,\nn\\
    \mathcal{C}_2 = &\left(z_2\cdot z_3\right)\left(z_1\cdot p_2\right)+ \left(z_1\cdot z_3\right) \left(z_2\cdot p_3\right) + \left(z_1\cdot z_2\right) \left(z_3\cdot p_1\right)\,,\nn\\
    \mathcal{C}_3 = &\frac{1}{m^2}\left[\left(z_1\cdot p_2\right)\epsilon\left(p_1 p_3 z_2 z_3\right) + \left(z_2\cdot p_3\right)\epsilon\left(p_2 p_1 z_3 z_1\right) + \left(z_3\cdot p_1\right)\epsilon\left(p_3 p_2 z_1 z_2\right)\right].
\end{align}
These amplitudes correspond to the following basis of local operators
\begin{align}
    \hat{\mathcal{L}}_1 &= \frac{f_{abc}}{m^2}{F^a}^{\mu\rho} {F^b_{\rho}}^\nu F^c_{\mu\nu}|_{(3)}&\rightarrow&\,\hspace{20mm}\mathcal{C}_1 \nonumber\\
    \hat{\mathcal{L}}_2 &= F^a_{\mu\nu} F_a^{\mu\nu} |_{(3)}&\rightarrow&\,\hspace{20mm}\mathcal{C}_2 \nonumber\\
    \hat{\mathcal{L}}_3 &= \frac{f_{abc}}{m^2}\epsilon_{\mu\nu\alpha\beta}{F^a}^{\mu\rho} F^{b\nu}_{\rho} F^{c\alpha\beta}|_{(3)}&\rightarrow&\,\hspace{20mm}\mathcal{C}_3.
\end{align}
where $F^a_{\mu\nu} = \partial_{[\mu} A^a_{\nu]}+i{f^a}_{bc}A^b_\mu A^c_\nu$, and $\vert_{(3)}$ denotes the $\mathcal{O}\left(A^3\right)$ part of the operator. In these formulae we are ignoring irrelevant numerical coefficients. The consistency of these interactions with supersymmetry can be analyzed using the same on-shell superspace methods introduced in Section \ref{sec:SWIsuperspace}. We want to use this to construct supersymmetric massive gravity as a double copy, so to avoid spin $>2$ states in the final result, we must restrict to spin $\leq 1$ in the massive gluon multiplets. Furthermore, to avoid a spin-2 degeneracy, the massive gluon must be in a long multiplet. This restricts the number of supersymmetries to be $\mathcal{N}\leq 2$; the corresponding multiplets are enumerated in Table \ref{tab:gluonmultiplets}. To construct cubic superamplitudes we impose a version of super-statistics to ensure total anti-symmetry of the color-stripped component amplitudes. Finally, as before, since we are only interested in the part of the superamplitude that contributes to the cubic massive gluon interaction it is sufficient to restrict to the R-singlet sector.
\begin{table}
    \begin{subtable}[h]{0.48\textwidth}
        \centering
\begin{tabular}{|c|c|c|c|}
		\hline
		Field & Spin & $U(1)_R$ & Dim.\\
		\hline
		\hline
		$\lambda^{I}$ & $\frac12$ & 1& 1\\
		\hline
		$g^{IJ}$ & 1 & 0& 1\\
		\hline
		$H$ & 0 & 0 & 1\\
		\hline
		$\tilde{\lambda}^{I}$ & $\frac12$ & -1 & 1\\
		\hline
	\end{tabular}
       \caption{$\mathcal{N}=1$ massive gluon multiplet}
       \label{tab:N1multipletgluon}
    \end{subtable}
    \hfill
    \begin{subtable}[h]{0.48\textwidth}
        \centering
	\begin{tabular}{|c|c|c|c|c|}
		\hline
		Field & Spin & $U(1)_R$ & $SU(2)_R$ & Dim.\\
		\hline
		\hline
		$\phi$ & 0 & 2 & $\bullet$ & $\mathbf{1}$\\
		\hline
		$\psi^I_a$ & $\frac12$ & 1 &\ytableausetup{boxsize=2mm,aligntableaux=top}\ydiagram{1}& $\mathbf{2}$\\
		\hline
		$g^{IJ}$ & 1 & 0 & $\bullet$& $\mathbf{1}$\\
		\hline
		$H_{ab}$ & 0 & 0 &\ydiagram{2}& $\mathbf{3}$\\
		\hline
		$\tilde{\psi}_a^I$ & $\frac{1}{2}$ & -1 &\ydiagram{1}& $\mathbf{2}$\\
		\hline
		$\tilde{\phi}$ & 0 & -2 &$\bullet$& $\mathbf{1}$\\
		\hline
	\end{tabular}
        \caption{$\mathcal{N}=2$ massive gluon multiplet\\
        \hspace{5mm}}
        \label{tab:N2multipletgluon}
     \end{subtable}
     \caption{
     \label{tab:gluonmultiplets}
     On-shell content of massive gluon multiplets with $\mathcal{N}\leq 2$ supersymmetry. States are labelled with capital Latin indices $I,J,...$ corresponding to $SU(2)_{\text{LG}}$ and lowercase Latin indices $a,b,...$ corresponding to $SU(\mathcal{N})_{R}$. The last and second-to-last columns gives the dimension and Young tableaux respectively of the $SU(\mathcal{N})_{R}$ representations in the conventions of \cite{Yamatsu:2015npn,Georgi:1982jb}.  }
\end{table}

\vspace{3mm}
\noindent\textbf{$\mathbf{\mathcal{N}=1}$ massive gluon:}
\begin{align}
    \Pi^{I} &=\lambda^{I}+\eta^{I} H + \eta_J g^{IJ}  + \frac{1}{2}\eta_J \eta^J \tilde{\lambda}^{I}.
\end{align}
The most general $\mathcal{N}=1$ massive Yang-Mills superamplitude
\begin{align}
\label{eq:N1gluonamp}
    \mathcal{A}_3[\Pi,\Pi,\Pi] =& \delta^{(2)}\left(Q^\dagger\right) \eta_{12,L_1} \beta_1\,\left[-2 \{\mathbf{1}1^{L_1}\} \br{23} + \br{21} \{\mathbf{3}1^{L_1}\} + \br{13} \{1^{L_1}\mathbf{2}\}\right.\nn\\
    &\hspace{32mm}\left.+ \br{12} \{\mathbf{3}1^{L_1}\} - \br{12}\{1^{L_1}\mathbf{3}\}\right]\,.
\end{align}
Projecting onto the cubic gluon amplitude and matching to (\ref{eq:N0gluonamp}) gives
\begin{align}
    c_2&=-16\sqrt{2}m^6 \beta_1\,,\nn\\
    c_1&=c_3=0\,.
\end{align}
Only the operator $\hat{\mathcal{L}}_2$ is consistent with $\mathcal{N}=1$ supersymmetry.

\vspace{3mm}
\noindent\textbf{$\mathbf{\mathcal{N}=2}$ massive gluon:}
\begin{align}
    \Theta &= \phi + \eta_{Ia}\psi^{Ia} +\frac{1}{2}\eta_{Ia} \eta^I_b H^{ab}+\frac{1}{2}\epsilon^{ab}\eta_{Ia} \eta_{Jb} g^{IJ} + \frac{1}{3}\eta_{Ia} \eta_J^a \eta^{I}_b \tilde{\psi}^{Jb} + \epsilon^{ab}\epsilon^{cd}\eta_{a}^I\eta_b^{J}\eta_{cI}\eta_{dJ} \tilde{\phi}.
\end{align}
The most general (R-singlet) massive Yang-Mills superamplitude
\begin{align}
\label{eq:N2gluonamp}
    \mathcal{A}_3[\Theta,\Theta,\Theta] =& \delta^{(4)}\left(Q^\dagger\right) \epsilon^{ab}\eta_{12,a,K_1} \eta_{12,b,L_1} \beta_1 \left[\{1^{K_1}1^{L_1}\}+\{1^{L_1}1^{K_1}\}\right].
\end{align}
Projecting onto the cubic gluon amplitude and matching to (\ref{eq:N0gluonamp}) gives
\begin{align}
    c_2&=16\sqrt{2}m^4 \beta_1\,,\nn\\
    c_1&=c_3=0\,.
\end{align}
Again, only the operator $\hat{\mathcal{L}}_2$ is consistent with $\mathcal{N}=2$ supersymmetry. These amplitudes and their massless limits were previously noted in \cite{Abhishek:2022nqv}.

Following the logic of Section \ref{sec:masslesscubic}, the fact that interactions $\mathcal{C}_1$ and $\mathcal{C}_3$ are not supersymmetrizable can be established in the high-energy limit:
\begingroup
\allowdisplaybreaks
\begin{align}
   \mathcal{C}_1 &\xrightarrow[]{\text{HE}}\left\{
\begin{array}{ll}
      (g^-g^-g^-): & \frac{1}{m^2}\langle 12\rangle\langle 13\rangle \langle 23\rangle \\
     (g^+g^+g^+): & \frac{1}{m^2}[ 12][ 13][  23] \\
\end{array} \right.\\ \nn\\
\mathcal{C}_2 &\xrightarrow[]{\text{HE}}\left\{
\begin{array}{ll}
      (g^-g^-g^+): &\frac{\langle 12\rangle^3}{\langle 13\rangle\langle 23\rangle}  \\(g^-\pi\pi): & -\frac{\langle 12\rangle\langle 13\rangle}{\langle 23\rangle}\\(g^+\pi\pi): & -\frac{[ 12]^[ 13]}{[ 23]} \\(g^+g^+g^-): & \frac{[ 12]^3}{[ 13][ 23]}
\end{array} \right.\\ \nn\\
\mathcal{C}_3 &\xrightarrow[]{\text{HE}}\left\{
\begin{array}{ll}
      (g^-g^-g^-): & -\frac{3}{m^2}\langle 12\rangle\langle 13\rangle \langle 23\rangle \\
     (g^+g^+g^+): & \frac{3}{m^2}[ 12][ 13][  23].
\end{array} \right.
\end{align}
\endgroup
Here we use $g^\pm$ and $\pi$ to denote the $h=\pm1$ and $h=0$ (pion) massless components respectively. Clearly the equal helicity amplitudes $A_3(g^\pm, g^\pm, g^\pm)$ are incompatible with any amount of supersymmetry. Thus $\mathcal{C}_1$ and $\mathcal{C}_2$ must also be incompatible.

\subsection{Double Copy in Massive Superspace}
\label{sec:DC}
In this section, we discuss the double copy construction of graviton supermultiplets and cubic superamplitudes. We begin with the double copy of free fields to understand the on-shell particle content of the double copy theories. The double copy on wave functions or one-particle states is given by simple multiplication followed by decomposition into irreducible representations of the little group $SU(2)_{\text{LG}}$. This wavefunction double copy also naturally generalizes to on-shell superfields
\begin{align}
    \mathcal{P}^{\{K\}}_{\{I,J\}}\alpha_{\mathcal{N}_A}^{\{I\}}\times \beta_{\mathcal{N}_B}^{\{J\}}=\gamma_{\mathcal{N}_A+\mathcal{N}_B}^{\{K\}}\,,
\end{align}
where $\alpha$ and $\beta$ are superfields of theories $A$ and $B$ with $\mathcal{N}_A$ and $\mathcal{N}_B$ amounts of supersymmetry respectively, while $\gamma$ is the superfield of a theory $A\otimes B$ with $\mathcal{N}=\mathcal{N}_A+\mathcal{N}_B$ and $\mathcal{P}$ is a projector onto the appropriate irreducible little group representation.

At the level of cubic amplitudes, the double copy takes a particularly simple form, the \textit{color factor} $f^{abc}$ is replaced with a \textit{kinematic factor}, the color-stripped cubic amplitude of a possibly distinct model
\begin{equation}
    A_3^{A\otimes B}\left(1,2,3\right) = \frac{1}{M_{\text{P}}}A_3^{A}\left[1,2,3\right] \times A_3^{B}\left[1,2,3\right].
\end{equation}
The dimensionful prefactor is included to ensure that the double copy amplitude has the correct units. Note that at cubic order the KLT and BCJ versions of the double copy are identical; it is difficult to imagine a generalization of the double copy that deviates from this prescription at this order. For supersymmetric theories, the double copy is defined analogously for superamplitudes. Including little group projectors
\begin{align}
    &\mathcal{A}^{A\otimes B}_3\left(\gamma_{\mathcal{N}_A+\mathcal{N}_B}^{\{K_1\}},\gamma_{\mathcal{N}_A+\mathcal{N}_B}^{\{K_2\}},\gamma_{\mathcal{N}_A+\mathcal{N}_B}^{\{K_3\}}\right)\nonumber\\
    &=\frac{1}{M_\text{P}}\mathcal{P}^{\{K_1\}}_{\{I_1,J_1\}}\mathcal{P}^{\{K_2\}}_{\{I_2,J_2\}}\mathcal{P}^{\{K_3\}}_{\{I_3,J_3\}}\mathcal{A}^A_3\left[\alpha_{\mathcal{N}_A}^{\{I_1\}},\alpha_{\mathcal{N}_A}^{\{I_2\}},\alpha_{\mathcal{N}_A}^{\{I_3\}}\right]\times \mathcal{A}^B_3\left[\beta_{\mathcal{N}_B}^{\{J_1\}},\beta_{\mathcal{N}_B}^{\{J_2\}},\beta_{\mathcal{N}_B}^{\{J_3\}}\right]\,.
\end{align}
All the possible double copy constructions of graviton supermultiplets are given in Table \ref{tab:DC}. Below, we study these cases in detail, first by constructing the superfields in the double copy and then the superamplitudes. We discuss which cubic massive graviton interactions are generated in each case and comment on physical implications.

\begin{table}
\begin{center}
\begin{tabular}{|c|c|c|c|}
\hline
     & Double Copy& Fields Generated   & R-Symmetry\\
\hline
\hline
    $\mathcal{N}=0+0$ & \begin{tabular}{c}
    $g^{(IJ}\otimes g^{KL)}$\Tstrut\Bstrut\\
    $\epsilon^{KL}g_{K}^{(I}\otimes g_{L}^{J)}$\Tstrut\Bstrut\\
    $g^{IJ}\otimes g_{IJ}$\Tstrut\Bstrut
    \end{tabular}& \begin{tabular}{cc}
        Graviton & $h^{IJKL}$ \\
        Massive 2-form & $B^{IJ}$\\
        Dilaton & $D$
    \end{tabular} 
    & $--$\\
\hline
     $\mathcal{N}=1+0$ & \begin{tabular}{c}
    $ \Pi^{(I}\otimes g^{JK)}$\Tstrut\Bstrut\\
   $\Pi_{J} \otimes g^{IJ}$\Tstrut\Bstrut
    \end{tabular}& \begin{tabular}{cc}
        Graviton superfield & $\Psi^{IJK}$\\
        Vector superfield & $V^I$ 
    \end{tabular} 
    &$U(1)$\\
\hline
     $\mathcal{N}=2+0$ & \begin{tabular}{c}
    $\Theta \otimes g^{IJ}$\Tstrut\Bstrut
    \end{tabular}& \begin{tabular}{cc}
        Graviton superfield & $\Gamma^{IJ}$
    \end{tabular} 
    &$U(2)$\\
\hline
     $\mathcal{N}=1+1$ & \begin{tabular}{c}
    $\Pi^{(I}\otimes \Pi^{J)}$\Tstrut\Bstrut\\
   $\epsilon_{IJ}\Pi^{I}\otimes \Pi^J$\Tstrut\Bstrut
    \end{tabular}& \begin{tabular}{cc}
        Graviton superfield & $\Gamma^{IJ}$\\
        Vector superfield & $W$ 
    \end{tabular} 
    &$U(1)\times U(1)$\\
\hline
     $\mathcal{N}=2+1$ & \begin{tabular}{c}
    $\Theta\otimes \Pi^{I}$\Tstrut\Bstrut
    \end{tabular}& \begin{tabular}{cc}
        Graviton superfield & $\Lambda^{I}$
    \end{tabular} 
    & $U(2)\times U(1)$\\
\hline
     $\mathcal{N}=2+2$  & \begin{tabular}{c}
    $\Theta\otimes \Theta$\Tstrut\Bstrut
    \end{tabular}& \begin{tabular}{cc}
        Graviton superfield & $\Phi$
    \end{tabular}
    & $U(2)\times U(2)$\\
\hline
\end{tabular}
\end{center}
\caption{To realize supermultiplets with different numbers of supersymmetries $\mathcal{N}$ as double copies, we use the double copy prescriptions listed in the table above. In both cases where a vector superfield is generated, the $\mathcal{N}=1$ and $\mathcal{N}=2$ vector multiplets $V^I$ and $W$, these contain both the dilaton and 2-form fields. Here $g^{IJ}$, $\Pi^I$ and $\Theta$ are the $\mathcal{N}=0$, $\mathcal{N}=1$ and $\mathcal{N}=2$ massive gluon multiplets given in Section \ref{sec:sYM}. Note that there are two ways to realize an $\mathcal{N}=2$ graviton supermultiplet as a double copy.}
\label{tab:DC}
\end{table}

\newpage
\noindent \textbf{$\mathbf{\mathcal{N}=0\otimes \mathcal{N}=0}$:}

\vspace{3mm}
\noindent In the non-supersymmetric case, there are three fields generated by the double copy: the graviton $h^{IJKL}$, the anti-symmetric 2-form (which is dual to a massive vector field) $B^{IJ}$ and the dilaton $D$. When discussing the supersymmetric double copy below, we will use the double copy listed in the first row of Table \ref{tab:DC} as the definition for $h^{IJKL}$, $B^{IJ}$ and $D$. This means that we identify these states as different projections of the massive gluon component field double copied with itself $g^{IJ}\otimes g^{KL}$.

Using a generic cubic massive gluon amplitude \eqref{eq:N0gluonamp} as the single copy, we can construct the most general double copy amplitudes. Projecting onto the cubic massive graviton amplitude gives \eqref{eq:generic3h} with coefficients
\begin{align}
    b_1&=-\frac34 M_\text{P} (c_2 d_1 + c_1 d_2 + 48\, c_3 d_3) \,,\nn\\
    b_2&=\frac14 M_\text{P}(c_1 d_2 + c_2 d_1 + 4\, c_2 d_2) \,,\nn\\
    b_3&=\frac14 M_\text{P}(c_2 d_1 + c_1 d_2 + 144\, c_3 d_3)\,,\nn\\
    b_4&=M_\text{P}(c_1 d_1 - 144\, c_3 d_3)\,,\nn\\
    b_5&=\frac32 M_\text{P}(c_3 d_2 + c_2 d_3)\,,\nn\\
    b_6&=-M_\text{P}(c_3 d_1 + c_1 d_3)\,.
\end{align}
Here $c_i$ and $d_i$ are the free coefficients in the non-supersymmetric massive gluon amplitudes of the left and right copies respectively.

The double copy of massive Yang-Mills i.e. $c_1=c_3=0$, has been studied in detail \cite{Johansson:2019dnu,Momeni:2020vvr}. In this case, the double copy produces an emergent $\mathds{Z}_2$ symmetry that acts on the 2-form as $B\rightarrow -B$. From the coefficient assignments above, we find that only the operator $3\mathcal{L}_2+\mathcal{L}_1$, corresponding to $\alpha_3=-1/2$, is generated in this case.

\vspace{3mm}
\noindent \textbf{$\mathbf{\mathcal{N}=1\otimes \mathcal{N}=0}$:}

\vspace{3mm}
\noindent The simplest non-trivial supersymmetric double copy is the construction of an $\mathcal{N}=1$ graviton superfield as given in Table \ref{tab:DC}. By using the full expression for the gluon and graviton superfields, we see how the product of superfields has non-zero projections onto two distinct irreducible representations: a graviton multiplet $\Psi^{IJK}$ and a vector multiplet $V^I$. The first double copy reduces to the following in terms of the component fields,
\begin{align}
    \Psi^{IJK} = & g^{(IJ} \otimes \Pi^{K)}\,,\nn\\
    \psi^{IJK}+\frac{1}{2\sqrt{3}}\eta^{(I}\gamma^{JK)}+\eta_Lh^{IJKL}+\frac{1}{2}\eta^2\tilde{\psi}^{IJK} = & g^{(IJ} {\otimes} \left[\lambda^{K)} + \eta_L g^{K)L} +\eta^{K)} H + \frac{1}{2}\eta^2 \tilde{\lambda}^{K)}\right]\,.
\label{eq:DC1}
\end{align}
This further allows us to identify the $B$-field and dilaton as belonging to the massive vector superfield generated by the double copy,
\begin{align}
    V^{I} = & g^{IJ} \otimes \Pi_{J}\,,\\
    \lambda^{I} + \eta_L B^{IL} +\eta^{I} D + \frac{1}{2}\eta^2 \tilde{\lambda}^{I}= & g^{IJ} \otimes \left(\lambda_{J} + \eta_L g_{J}^{L} +\eta_J H + \frac{1}{2}\eta^2 \tilde{\lambda}_{J}\right)\,.
\end{align}
Double copying the $\mathcal{N}=1$ cubic super amplitude with the $\mathcal{N}=0$ cubic gluon amplitudes give an $\mathcal{N}=1$ graviton amplitude,
\begin{align}
    &\mathcal{A}^{\mathcal{N}=1}_3\left(\Psi^{I_1J_1K_1},\Psi^{I_2J_2K_2},\Psi^{I_3J_3K_3}\right) \nonumber\\
    &=  \frac{1}{M_{\text{P}}}\mathcal{A}^{\mathcal{N}=0}_3\left[g^{(I_1J_1},g^{(I_2J_2},g^{(I_3J_3}\right]\times  \mathcal{A}_3^{\mathcal{N}=1}\left[\Pi^{K_1)},\Pi^{K_2)},\Pi^{K_3)}\right]\,.
\end{align}
Using the explicit expressions \eqref{eq:N0gluonamp} and \eqref{eq:N1gluonamp}, gives
\begin{align}
    &\mathcal{A}_3\left(\Psi,\Psi,\Psi\right)\nonumber\\
    &= \delta^{(2)}\left(Q^\dagger\right) \eta_{12,L_1} \beta_1\,\left[-2 \{\mathbf{1}1^{L_1}\} \br{23} + \br{21} \{\mathbf{3}1^{L_1}\} + \br{13} \{1^{L_1}\mathbf{2}\}\right.\nn\\
    &\hspace{43mm}\left.+ \br{12} (\{\mathbf{3}1^{L_1}\} - \{1^{L_1}\mathbf{3}\})\right]\times \left(c_1\, \mathcal{C}_1+c_2\, \mathcal{C}_2+c_3\, \mathcal{C}_3\right)\,.
\end{align}
Projecting onto the cubic graviton amplitude and matching with the general expression \eqref{eq:generic3h} gives
\begin{align}
    b_1 &= 12\, \sqrt{2}\, m^{6}\, M_{\text{P}}\, c_1 \beta_1\,,\nn\\
    b_2 &= -4\,\sqrt{2}\, m^{6}\, M_{\text{P}}\,\beta_1 (c_1+ 4\, c_2)\,,\nn\\
    b_3 &= -4\, \sqrt{2}\, m^{6}\, M_{\text{P}}\, c_1\, \beta_1\,,\nn\\
    b_5 &= -24\, \sqrt{2}\, m^{6}\, M_{\text{P}}\, c_3\, \beta_1\,,\nn\\
    b_4 &= b_6=0\,.
\end{align}
Matching this to the basis of local operators (\ref{cubicoperators}) we find that there are 3 independent $\mathcal{N}=1$ supersymmetrizable interactions $\mathcal{L}_1$, $\mathcal{L}_2$ and $\mathcal{L}_5$. Here we see the constraining power of the double copy, from the general 4 parameter family of $\mathcal{N}=1$ supersymmetrizable interactions we are reduced to a 3 parameter family. 

\vspace{3mm}
\noindent \textbf{$\mathbf{\mathcal{N}=2\otimes \mathcal{N}=0}$:}

\vspace{3mm}
\noindent From now on, we will implicitly use the superfield double copy listed in Table \ref{tab:DC} and begin with the cubic amplitude double copy directly. Note that in this first $\mathcal{N}=2$ construction, only one superfield is generated. Thus both the $B$-field and dilaton $D$ belong to the graviton superfield $\Gamma^{IJ}$. The corresponding $\mathcal{N}=2$ superamplitude constructed from the double copy is
\begin{align}
    &\mathcal{A}^{\mathcal{N}=2}_3(\Gamma,\Gamma,\Gamma) \nonumber\\
    &= \frac{1}{M_{\text{P}}}A_3^{\mathcal{N}=0}[g,g,g]\times  \mathcal{A}_3^{\mathcal{N}=2}[\Theta,\Theta,\Theta]\nn\\
    &= \delta^{(4)}\left(Q^\dagger\right) \epsilon^{ab}\eta_{12,a,K_1} \eta_{12,b,L_1} \beta_1 \left[\{1^{K_1}1^{L_1}\}+\{1^{L_1}1^{K_1}\}\right] \left(c_1 \mathcal{C}_1+c_2 \mathcal{C}_2+c_3 \mathcal{C}_3\right)\,.
\end{align}
Projecting onto the 3-graviton amplitude gives \eqref{eq:generic3h} with the coefficient assignments
\begin{align}
    b_1 &= -12\, \sqrt{2}\, c_1\, m^{4}\, M_{\text{P}}\, \beta_1\,,\nn\\
    b_2 &= 4\, \sqrt{2}\, m^{4}\, M_{\text{P}}\, \beta_1\, (4\, c_2 + c_1)\,,\nn\\
    b_3 &= 4\, \sqrt{2}\, c_1\, m^{4}\, M_{\text{P}}\, \beta_1\,,\nn\\
    b_5 &= 24\, \sqrt{2}\, c_3\, m^{4}\, M_{\text{P}}\, \beta_1\,,\nn\\
    b_4 &= b_6=0\,.
\end{align}
Matching this to the basis of local operators (\ref{cubicoperators}) we find the same supersymmetrizable local operators as $\mathcal{N}=1\otimes \mathcal{N}=0$. In this case, we start with an $R$-symmetry group of $SU(2)_R \times U(1)_R$ and so the resulting theory has maximal $R$-symmetry.

\vspace{3mm}
\noindent \textbf{$\mathbf{\mathcal{N}=1\otimes \mathcal{N}=1}$:}

\vspace{3mm}
\noindent The product of a $\mathcal{N}=1$ gluon superfield with itself can be projected onto two little group representations: a graviton superfield $\Gamma^{IJ}$ and a vector superfield $W^I$. The $B$-field belongs to the graviton multiplet, whereas the dilaton $D=g_{IJ}\otimes g^{IJ}$ belongs to a linear combination of these two multiplets.

The $\mathcal{N}=2$ superamplitude  from this double copy construction is
\begin{align}
    &\mathcal{A}^{\mathcal{N}=2}_3(\Gamma,\Gamma,\Gamma) \nonumber\\
    &= \frac{1}{M_{\text{P}}}\mathcal{A}_3^{\mathcal{N}=1}[\Pi,\Pi,\Pi]\times  \mathcal{A}_3^{\mathcal{N}=1}[\Pi,\Pi,\Pi]\nn\\
    &= \delta^{(4)}\left(Q^\dagger\right) \eta_{12,1,K_1} \eta_{12,2,L_1} \alpha_1\beta_1 \,\left[-2 \{\mathbf{1}1^{K_1}\} \br{23} + \br{21} \{\mathbf{3}1^{K_1}\}+ \br{13} \{1^{K_1}\mathbf{2}\} \right.\nn\\
    &\hspace{5mm}\left.+ \br{12} (\{\mathbf{3}1^{K_1}\} - \{1^{K_1}\mathbf{3}\})\right]\times \left[-2 \{\mathbf{1}1^{L_1}\} \br{23}+ \br{21} \{\mathbf{3}1^{L_1}\}+ \br{13} \{1^{L_1}\mathbf{2}\} \right.\nn\\
    &\hspace{5mm}\left. + \br{12} (\{\mathbf{3}1^{L_1}\} - \{1^{L_1}\mathbf{3}\})\right]\,.
\end{align}
Projecting onto the cubic graviton amplitude gives \eqref{eq:generic3h} with the coefficient assignments
\begin{align}
    b_2 &= 512\, m^{12}\, M_{\text{P}}\, \alpha_1\,\beta_1\,,\nn\\
    b_1 &= b_3=b_4=b_5=b_6=0\,.
\end{align}
Matching this to the basis of local operators (\ref{cubicoperators}) we find that only the operator $3\mathcal{L}_1+\mathcal{L}_2$ has been generated. 

Studying the double copy with extended supersymmetry, raises the natural question of whether $R$-symmetry is enhanced in any of the resulting theories. In the massless double copy, this is known to happen \textit{at two-derivative order}. For example, in the case of the double copy of $\mathcal{N}=4$ super Yang-Mills with itself, the manifest $SU(4)_R\times SU(4)_R$ symmetry is enhanced to $SU(8)_R$. For the massive double copy considered here, unlike in the case of $\mathcal{N}=2\otimes\mathcal{N}=0$, the $\mathcal{N}=1 \otimes \mathcal{N}=1$ double copy manifests a non-maximal $R$-symmetry group $U(1)_R\times U(1)_R$. Does this enhance to the maximal $\mathcal{N}=2$ $R$-symmetry group of $SU(2)_R\times U(1)_R$? Below we discuss why it cannot. 

The double copy of massive cubic interactions necessarily gives rise to a non-zero $hhD$ (graviton$^2$-dilaton) vertex \cite{Johnson:2020pny, Momeni:2020vvr}. In order for such a vertex to be present in a theory with $SU(2)_R\times U(1)_R$ symmetry, the dilaton must be an $R$-symmetry singlet. Therefore a necessary condition that must be satisfied in order for the double copy theory to be $R$-symmetric, is that the resulting supermultiplets must contain a scalar state that is uncharged under $R$-symmetry. Since the vector multiplet contributes to the dilaton and the $\mathcal{N}=2$ vector multiplet does not contain a $SU(2)_R$ singlet scalar state (see Table \ref{tab:gluonmultiplets}), we conclude that the theory produced by double copying a $\mathcal{N}=1$ gluon multiplet with itself cannot preserve $SU(2)_R\times U(1)_R$, but only $U(1)_R\times U(1)_R$.

\vspace{3mm}
\noindent \textbf{$\mathbf{\mathcal{N}=2\otimes \mathcal{N}=1}$:}

\vspace{3mm}
\noindent Since only one superfield is generated, the graviton superfield $\Lambda^I$ contains both the $B$-field and the dilaton. The resulting $\mathcal{N}=3$ superamplitude is
\begin{align}
    &\mathcal{A}^{\mathcal{N}=3}_3(\Lambda,\Lambda,\Lambda) \nonumber\\
    &= \frac{1}{M_{\text{P}}}\mathcal{A}_3^{\mathcal{N}=2}[\Theta,\Theta,\Theta]\times  \mathcal{A}_3^{\mathcal{N}=1}[\Pi\,\Pi\,\Pi]\nn\\
    &= \delta^{(6)}\left(Q^\dagger\right) \epsilon_{12}^{ab}\eta_{12,a,K_1} \eta_{12,b,L_1} \eta_{12,3,M_1} \alpha_1 \beta_1 \left[\{1^{K_1}1^{L_1}\}+\{1^{L_1}1^{K_1}\}\right]\nn\\
    & \hspace{5mm}\times\left[\br{21} \{\mathbf{3}1^{M_1}\}-2 \{\mathbf{1}1^{M_1}\} \br{23}+ \br{13} \{1^{M_1}\mathbf{2}\}+ \br{12} (\{\mathbf{3}1^{M_1}\} - \{1^{M_1}\mathbf{3}\})\right]\,,
\end{align}
where $\epsilon^{ab}_{12}$ corresponds to the Levi-Civita symbol for the $(1,2)$ R-index subspace. Projecting onto the cubic graviton amplitude gives \eqref{eq:generic3h} with the coefficient assignments
\begin{align}
    b_2 &= 512\, m^{10}\, M_{\text{P}}\, \alpha_1\,\beta_1\,,\nn\\
    b_1 &= b_3=b_4=b_5=b_6=0\,.
\end{align}
Again, matching to the basis of local operators (\ref{cubicoperators}), we find that only the operator $3\mathcal{L}_1+\mathcal{L}_2$ has been generated. Since the $\mathcal{N}=3$ graviton superfield does not have a candidate $R$-symmetric scalar state, by the same argument we used above, the double copy theory cannot be maximally $R$-symmetric. Nonetheless, it has a minimum of $SU(2)_R\times U(1)_R \times U(1)_R$ symmetry inherited from the single copies.

\vspace{3mm}
\noindent \textbf{$\mathbf{\mathcal{N}=2\otimes \mathcal{N}=2}$:}

\vspace{3mm}
\noindent Once again, only one superfield is generated, thus the graviton superfield $\Phi$ contains both the $B$-field and the dilaton. The resulting $\mathcal{N}=4$ superamplitude is
\begin{align}
    &\mathcal{A}^{\mathcal{N}=4}_3(\Phi,\Phi,\Phi) \nonumber\\
    &= \frac{1}{M_{\text{P}}}\mathcal{A}_3^{\mathcal{N}=2}[\Theta,\Theta,\Theta]\times  \mathcal{A}_3^{\mathcal{N}=2}[\Theta,\Theta,\Theta]\nn\\
    &= \delta^{(8)}\left(Q^\dagger\right) \epsilon_{12}^{ab}\eta_{12,a,K_1} \eta_{12,b,L_1} \epsilon_{34}^{cd}\eta_{12,c,M_1} \eta_{12,d,N_1}  \alpha_1\left[\{1^{K_1}1^{L_1}\}+\{1^{L_1}1^{K_1}\}\right]\nn\\
    &\hspace{6cm}\times  \beta_1  \left[\{1^{M_1}1^{N_1}\}+\{1^{N_1}1^{M_1}\}\right]\,.
\end{align}
Projecting onto the cubic graviton amplitude gives \eqref{eq:generic3h} with the coefficient assignments
\begin{align}
    b_2 &= 512\, m^{8}\, M_{\text{P}}\, \alpha_1\,\beta_1\,,\nn\\
    b_1 &= b_3=b_4=b_5=b_6=0\,.
\end{align}
Note that in Section \ref{sec:cubicamps}, we found a unique $\mathcal{N}=4$ supersymmetric cubic amplitude and so as expected this is the result obtained from the double copy. In some sense this construction could not have failed, and so the existence of an $\mathcal{N}=4$ supersymmetrizable cubic massive graviton interaction follows as a necessary consequence of the existence of an $\mathcal{N}=2$ supersymmetrizable cubic massive gluon interaction. 

$SU(4)_R$-invariance would require that the coefficient of the Grassmann polynomial above be fully symmetric in $I_1$, $J_1$, $K_1$ and $L_1$ which it is not. Indeed the $\mathcal{N}=4$ graviton superfield does not have a scalar $R$-singlet state, and so the double copy theory cannot be $SU(4)_R$-symmetric. Nonetheless, it has a minimum of $SU(2)_R \times U(1)_R \times SU(2)_R \times U(1)_R$ symmetry inherited from the single copies. As a consequence, unlike the maximally R-symmetric case, the pure massive graviton sector of the double copy is not a consistent truncation.

\section{Discussion}
\label{sec:discussion}
\noindent \textbf{Supersymmetry and non-renormalization of ghost-free interactions}
\vspace{3mm}

\noindent It is a special property of on-shell cubic amplitudes that, since there are no available Lorentz and $SU(2)_{\text{LG}}$ singlets, the number of possible kinematic structures is strictly finite. As a consequence, the classification of supersymmetric on-shell cubic amplitudes in Section \ref{sec:susy3point} is valid non-perturbatively. Importantly, this means that quantum corrections cannot generate operators forbidden by supersymmetry.  

One motivation for considering supersymmetrizations of massive gravity is to gain control over quantum corrections. One-loop corrections to massive gravity were considered in \cite{deRham:2013qqa}, and it was found that quantum corrections involving matter loops led to a renormalizing of the cosmological constant, as they do in general relativity, and gravitons running in the loops led to a detuning of the potential, although this detuning occurs in a way such that it never leads to a ghost with a mass smaller than the Planck scale. The results of Section \ref{sec:susy3point} may suggest that for $\mathcal{N}=3$ and $4$ no such detuning occurs at any scale.

We found that for $\mathcal{N}\geq 3$ supersymmetry, only the operators $\mathcal{L}_1$ and $\mathcal{L}_2$ (\ref{cubicoperators}) were consistent. These are precisely the interactions that appear in the ghost-free dRGT Lagrangian at cubic order (\ref{nonlinlag}). Therefore, not only is the restriction to the ghost-free interactions mandatory at this order, since these are the only interactions compatible with $\mathcal{N}=3$ supersymmetry, quantum effects cannot generate any new terms in the Lagrangian. Therefore at cubic order in the fields, the Boulware-Deser ghost must decouple non-perturbatively. In the case of $\mathcal{N}=4$, interactions are further restricted to the unique operator $3\mathcal{L}_1+\mathcal{L}_2$, and this specific tuning cannot be modified by quantum corrections.

This result is reminiscent of well-known non-renormalization theorems for supersymmetric models \cite{Seiberg:1994bp}. The usual argument for these theorems relies on holomorphy in off-shell superspace. Since we have constructed the interactions on-shell in this paper, it is not clear to what extent these arguments apply. It would be very interesting to construct an off-shell effective action for these models, perhaps using the $\mathcal{N}=1$ off-shell superfields described in \cite{Buchbinder:2002gh}, and determine if holomorphy implies similar non-renormalization theorems. Alternatively, it may be possible to re-derive these supersymmetric non-renormalization theorems directly in on-shell language.

In addition to dRGT massive gravity, pseudolinear massive gravity is also ghost free and enjoys a $\Lambda_3$ cutoff\footnote{Here we use the common notation for the cutoff scale $\Lambda_n \equiv \left(M_{\text{P}}m^{n-1}\right)^{1/n}$.} \cite{Hinterbichler:2013eza,Bonifacio:2018van}. It is also possible that there could be a parity-odd version of massive gravity, which may be ghost-free. The highest possible cutoff scale of such a theory would be $\Lambda_{7/2}$ \cite{Bonifacio:2018aon}. It would be an interesting question to investigate whether supersymmetry is picking out theories that are ghost-free. 

\vspace{3mm}
\noindent\textbf{Maximal supersymmetry, CTCs and asymptotic superluminality}

\vspace{3mm}
\noindent
Perhaps the most interesting result of the analysis of Section \ref{sec:susy3point} was the discovery that for $\mathcal{N}=4$ supersymmetry, there is a unique consistent cubic interaction corresponding to the value $\alpha_3=-\frac{1}{2}$ in the dRGT potential (\ref{nonlinlag}). 

Interestingly, this value has previously been noted to be special in a seemingly unrelated context. In \cite{Hinterbichler:2017qyt} it was shown to be the unique cubic interaction of a generic massive spin-2 particle free from an asymptotic Shapiro time advance, measured by the positivity of the eikonal phase. In addition, dRGT massive gravity apparently admits closed time-like curves unless $\alpha_3=-\frac{1}{2}$ \cite{Camanho:2016opx}. 
Thus we find that $\mathcal{N}=4$ supersymmetry constrains the dRGT parameter space to be exactly that which is compatible with avoiding these two distinct forms of pathological behavior. Coupled with our previous discussion about non-renormalization, we see that this tuning cannot be corrected by loop effects since this is the unique $\mathcal{N}=4$ interaction. 

This coincidence could be a hint that $\mathcal{N}=4$ massive gravity really does exist and can be embedded in a UV complete model with good causal behavior. 

\vspace{3mm}
\noindent\textbf{Maximal supersymmetry and partially massless symmetry}

\vspace{3mm}
\noindent
In addition to its appearance in connection to causality, the special $\alpha_3$ value also makes an appearance in the context of partially massless symmetry \cite{deRham:2012kf,deRham:2013wv,DeRham:2018axr}. Although there is no tuning of the parameters of dRGT massive gravity with partially massless symmetry, in the partial decoupling limit of massive gravity on a de Sitter background, the theory with $\alpha_2=-\frac{1}{2}$ and $\alpha_3=\frac{1}{8}$ has the special property that its strong coupling scale is raised from $\tilde{\Lambda}_4=(M_{\text{P}}\Delta^3)^{1/4}$ with $\alpha_3,\alpha_4$ free to $\tilde{\Lambda}_2=(M_{\text{P}}\Delta)^{1/2}$, where in de Sitter spacetime we can set, $m^2=2H^2+\Delta^2$. Here $H$ is the Hubble constant, $\Delta$ measures the distance from the Higuchi bound \cite{Higuchi:1986py}, and the partially massless limit is found by taking $\Delta\rightarrow 0$. This is analagous to the way the cutoff scale in flat space is raised from $\Lambda_5$ to $\Lambda_3$ when the parameters are tuned to those of dRGT. Although the flat space limit of this theory would not maintain an enhanced cutoff scale beyond $\Lambda_3$, there are interactions that vanish for these special tunings in the high energy limit. In flat space, the partially massless limit coincides with the massless limit. It is possible that $\mathcal{N}=3$ picks out the interactions that have enhanced behavior in the massless limit, while $\mathcal{N}=4$ picks out interactions that have enhanced behavior in the partially massless limit.  

\newpage
\noindent\textbf{Dimensional deconstruction of $11d$ supergravity}

\vspace{3mm}
\noindent If an $\mathcal{N}=4$ supersymmetric model of ghost-free massive gravity exists beyond cubic order, an obvious goal is to construct the complete off-shell effective action. This would be necessary to establish the decoupling of the Boulware-Deser ghost and study the properties of non-linear classical solutions. Important among these are the various known ``black hole" solutions of massive gravity \cite{Babichev:2015xha}. It would be particularly interesting to try and construct BPS black holes, and understand if they reduce to BPS solutions in massless supergravity in an appropriate limit. This may require coupling the massive graviton to additional, massless Maxwell fields (supermultiplets), along the lines of \cite{Babichev:2014fka}.

A possible path forward is to first construct a massive $\mathcal{N}=(1,0)$ supersymmetric model of massive gravity in $10d$ and then dimensionally reduce to $4d$. This is the strategy that was originally used to obtain the off-shell action of both $\mathcal{N}=4$ super Yang-Mills \cite{Brink:1976bc} and $\mathcal{N}=8$ supergravity \cite{Cremmer:1978km,Cremmer:1979up}. At the linearized level, the $4d$ $\mathcal{N}=4$ massive graviton multiplet has the same helicity content as the $\mathcal{N}=8$ massless graviton multiplet. This correspondence should lift to $10d$ where the massive $\mathcal{N}=(1,0)$ multiplet arises from a kind of Higgsing of (linearized) type-IIA supergravity where every state becomes massive. We can verify the self-consistency of this picture for the bosonic states
\begin{align}
    \text{Massless graviton } \oplus  \text{ R-R 1-form } \oplus \text{Dilaton} &\xrightarrow[]{\text{Higgs}} \text{Massive graviton} \nonumber\\
    \text{R-R 3-form} \oplus  \text{ NS-NS 2-form } &\xrightarrow[]{\text{Higgs}} \text{Massive 3-form}.
\end{align}
This massive spectrum suggests some sort of connection with $11d$ supergravity. In \cite{deRham:2013awa}, building on earlier work \cite{Arkani-Hamed:2001kyx,Arkani-Hamed:2003roe,Schwartz:2003vj}, it was shown that $d$-dimensional ghost-free massive gravity can be obtained by a discrete dimensional reduction (a \textit{dimensional deconstruction}) of Einstein gravity in $d+1$-dimensions. In \cite{Ondo:2016cdv}, this approach was extended to linearized supersymmetric massive gravity, obtaining the free $\mathcal{N}=1$ massive graviton in $4d$ from the $\mathcal{N}=2$ massive graviton in $5d$. Whether this approach remains viable and preserves the requisite (half) of the original supercharges in the presence of interactions remains unknown. A possible path to constructing the fully non-linear $10d$ $\mathcal{N}=(1,0)$ massive gravity model (and subsequently the $4d$ $\mathcal{N}=4$ model) is as a dimensional deconstruction of massless $11d$ supergravity. 

\vspace{3mm}
\noindent\textbf{Higher-multiplicity}

\vspace{3mm}
\noindent The analysis in this paper is restricted to cubic interactions. In Section \ref{sec:cubicstuff}, we found that $\mathcal{N}\geq 3$ supersymmetry automatically selects the cubic dRGT interactions. This could be a hint that sufficiently extended supersymmetry requires the decoupling of the Boulware-Deser ghost. In general, at quartic order there are two distinct (non-redundant on-shell) operators in the zero-derivative graviton potential, and only a specific linear combination is ghost-free. It would be interesting to see if sufficient supersymmetry uniquely selects this tuning.

The extension to higher-multiplicity is conceptually straightforward, although technically more challenging. At 4-point there are infinitely many non-redundant local operators that may appear in an effective action, and so the corresponding superspace analysis should be carried out order-by-order in a consistent derivative expansion. There is the additional complication that non-local contributions from tree-level Feynman diagrams can mix with local contact contributions, and so these should also be included in the construction of a 4-point superamplitude ansatz. In such a calculation the empirically observed maximally improved high-energy growth of ghost-free interactions may be an important organizing principle \cite{Schwartz:2003vj,Cheung:2016yqr,Bonifacio:2018vzv,Falkowski:2020mjq}. 

It should be noted that using the supersymmetric cubic amplitudes presented in Section \ref{sec:susy3point}, we can construct $n$-point amplitudes that are compatible with supersymmetry by gluing together 3-point amplitudes on poles. This procedure is ambiguous up to $n$-point supersymmetric contact terms. Such $n$-point amplitudes will correspond to a theory with the correct particle content and appropriately chosen supersymmetric ``contact'' contributions. Starting with the 3-point amplitude of a ghost-free theory $b_1 \mathcal{B}_1 + b_2\mathcal{B}_2$ however, does not guarantee that $n$-point amplitudes constructed in this way will also be ghost-free. For this, the pole terms may not be enough and contact contributions may have to be added. This entails a full higher-point analysis, which is more involved as discussed above.

Alternatively, as for massless supergravity, it might be possible to construct higher-multiplicity tree-level scattering amplitudes using a superspace \textit{on-shell recursion} \cite{Britto:2005fq,Drummond:2008cr}. BCFW-inspired recursion for massive states was described in \cite{Badger:2005zh,Schwinn:2007ee} and extended to massive on-shell superspace in \cite{Boels:2011zz,Herderschee:2019dmc}. There is the usual difficulty that establishing recursion relations requires exceptional high-energy growth for large complex momenta \cite{Arkani-Hamed:2008bsc}. Since, as discussed above, the dRGT model is uniquely characterized by its high-energy growth for physical momenta, it would be interesting to try and construct a complex momentum shift that exploits this. It is possible that an on-shell recursion relation \textit{only} exists for the supersymmetric model, or a consistent truncation thereof. For example, for massless Yang-Mills coupled to a massless complex (adjoint) scalar, BCFW recursion is possible only if a specifically tuned quartic scalar potential term is added to the minimal Lagrangian \cite{Elvang:2013cua}. The value of this coupling is precisely the value required for this model to be obtained as a consistent truncation of $\mathcal{N}=2$ super Yang-Mills. It would be interesting to see if an analogous connection between on-shell constructability and supersymmetry exists for massive gravity.

\vspace{3mm}
\noindent\textbf{Evading the $\mathbf{\mathcal{N}=4}$ Galileon no-go theorem}

\vspace{3mm}

\noindent Following the logic of Section \ref{sec:masslesscubic}, an easier approach to studying higher-multiplicity interactions may be to first construct a massless supersymmetric model describing the decoupling limit. The problem of constraining models of a supersymmetric Galileon has been considered \cite{Khoury:2011da,Koehn:2013hk,Farakos:2013fne,Elvang:2017mdq,Elvang:2018dco,Elvang:2021qhq}, and we might hope to make use of these results to constrain supersymmetric massive gravity. Of particular interest, \cite{Elvang:2021qhq} recently proved an interesting no-go theorem for Galileons in $\mathcal{N}=4$ supersymmetric models. Below we will re-derive this result in detail and discuss why it does not lead to any constraints on an $\mathcal{N}=4$ supersymmetric model of massive gravity.

From the results of Section \ref{sec:masslesslimitmultiplets} we know that the 4-particle scattering of the $\mathcal{N}=4$ Galileon must be a particular projection of the superamplitude describing the scattering of 4 massless vector multiplets. This has the general form \cite{Elvang:2013cua}
\begin{equation}
    \label{4vectorsuperamplitude}
    \mathcal{A}_4\left(\Gamma_{i_1 j_1},\Gamma_{i_2 j_2},\Gamma_{i_3 j_3},\Gamma_{i_4 j_4}\right) = \delta^{(8)}\left(Q^\dagger\right) \frac{A_4\left(\gamma_{i_1 j_1}^+,\gamma_{i_2 j_2}^+,\gamma_{i_3 j}^-,\gamma_{i_4 j_4}^-\right)}{\langle 34\rangle^4}.
\end{equation}
Here $A_4\left(\gamma_{i_1 j_1}^+,\gamma_{i_2 j_2}^+,\gamma_{i_3 j_3}^-,\gamma_{i_4 j_4}^-\right)$ is the component amplitude of 4 massless vectors, the (anti-symmetric) subscripts correspond to the $\mathbf{6}$ representation of $SU(4)_{\text{global}}$. The on-shell superfield for the CPT self-conjugate $\mathcal{N}=4$ massless vector multiplet is\footnote{See \cite{Elvang:2013cua} for a review of massless on-shell superspace.}
\begin{equation}
    \Gamma_{ij} = \gamma_{ij}^+ +\eta_a \lambda_{ij}^{+a}-\frac{1}{2} \eta_a \eta_b \phi_{ij}^{ab} - \frac{1}{6} \eta_a \eta_b \eta_c \tilde{\lambda}_{ij}^{-abc}+ \eta_1 \eta_2 \eta_3 \eta_4 \gamma^-_{ij}, 
\end{equation}
and the massless supersymmetric delta function is defined as
\begin{equation}
    \delta^{(8)}\left(Q^\dagger\right) = \frac{1}{16}\prod_{a=1}^4 \sum_{i,j=1}^4 \langle ij\rangle \eta_{ia}\eta_{ja}.
\end{equation}
In the non-supersymmetric dRGT model, the leading high-energy 2-to-2 (tree-level) scattering of Galileon modes takes the form\footnote{In this paper we use the Mandelstam convention $s=(p_1+p_2)^2$, $t=(p_1+p_3)^2$, and $u=(p_1+p_4)^2$.}
\begin{equation}
    \label{4gal}
    A_4\left(1_\phi,2_\phi,3_\phi,4_\phi\right) = -\frac{1+2\alpha_3+9\alpha_3^2-16\alpha_4}{6m^4 M_{\text{P}}^2}stu +\mathcal{O}\left(E^4\right).
\end{equation}
If we impose invariance under the maximal R-symmetry group $SU(4)_R \times U(1)_R$, restricting to the pure graviton sector of $\mathcal{N}=4$ massive gravity becomes a consistent truncation. Therefore the above quartic Galileon amplitude must be valid in that case also with the modification that we must set $\alpha_3=-1/2$ to satisfy the cubic constraints derived in Section \ref{sec:susy3point}. The no-go theorem of \cite{Elvang:2021qhq} is the claim that the amplitude (\ref{4gal}) cannot be obtained from an $\mathcal{N}=4$ supersymmetric model if the Galileon $\phi$ is a combination of scalars in a massless vector multiplet. This would seem to be consistent with a maximally R-symmetric model of $\mathcal{N}=4$ massive gravity only if the prefactor in the above expression vanishes, giving $\alpha_4=9/64$. This is a very intriguing possibility since the values $(\alpha_3,\alpha_4)=(-1/2,9/64)$ are almost uniquely singled out by combining causality constraints \cite{Camanho:2016opx,Bonifacio:2017nnt} and (a proposed refinement of) S-matrix positivity constraints \cite{Bellazzini:2017fep}. Unfortunately, there is actually a subtle difference between the assumptions made in \cite{Elvang:2021qhq} and the present analysis, and the no-go theorem does not apply in this case. To see this we will explicitly construct the most general form of the superamplitude (\ref{4vectorsuperamplitude}).

The strategy will be to first enumerate a list of properties satisfied by the component amplitude $A_4\left(\gamma_{i_1 j_1}^+,\gamma_{i_2 j_2 }^+,\gamma_{i_3 j_3}^-,\gamma_{i_4 j_4}^-\right)$, and use this to write down the general form of this object up to a set of undetermined coefficients. We then impose the additional super-statistics constraints on this expression and find a solution only if (\ref{4vectorsuperamplitude}) has non-trivial dependence on additional flavor structure. We begin with some basic dimensional analysis. Since the quartic Galileon (\ref{4gal}) corresponds to a 6-derivative operator we need only consider contributions to the 4-vector component amplitude at this order. These can be either local contact contributions from operators of the schematic form $\partial^2 F^4$ or from tree-level exchange diagrams. For the latter we can immediately rule out the possibility of massless exchange in the $t$- or $u$-channels. If such exchanges were present then on the corresponding singularity the amplitude would factor into a product of 3-point amplitudes of the form
\begin{equation}
    A_3\left(\gamma_{ij}^+, \gamma_{kl}^-, X^{+h}\right) \propto \frac{1}{M^{h-1}}\frac{[13]^2}{[12]^2}\left(\frac{[13][23]}{[12]}\right)^h,
\end{equation}
where $M$ is some mass scale controlling the derivative expansion. Requiring this interaction to mix with 6-derivative contact terms uniquely fixes the the helicity of the exchanged state to be $h=2$ and therefore this state must be identified with the massless limit of the tensor mode of the massive graviton. Such an interaction describes a long-range gravitational force between matter particles. A model containing this interaction is mathematically self-consistent only if it also contains a cubic graviton self-interaction of the same strength, the on-shell statement of the Einstein equivalence principle \cite{Weinberg:1964ew}. Such self-interactions for the tensor mode are set to zero in the decoupling limit by taking $M_{\text{P}}\rightarrow \infty$ and so we conclude that $t$- and $u$-channel singularities must be absent. Massless exchange in the $s$-channel cannot be excluded by such an argument since, due to the helicity structure, the corresponding 3-point amplitude is non-minimal and \textit{a priori} could survive the decoupling limit. Next, since the Galileon is an $SU(4)_R\times U(1)_R$ singlet we can restrict to a superamplitude in this sector. The massless vectors $\gamma_{ij}$ are singlets of $SU(4)_R^{m=0}\times U(1)_R^{m=0} \times U(1)_{\text{global}}$ and so this condition is the same as the 4-vector component amplitude being $SU(4)_{\text{global}}$ invariant. Finally, we require Bose symmetry for the exchange of particles $1\leftrightarrow 2$ and $3\leftrightarrow 4$. The general form of the component amplitude satisfying these constraints is 
\begin{align}
    \label{masslessvector4pt}
    &A_4\left(\gamma_{i_1 j_1}^+,\gamma_{i_2 j_2 }^+,\gamma_{i_3 j_3}^-,\gamma_{i_4 j_4}^-\right) \nonumber\\
    &= [12]^2 \langle 34\rangle^2 \left[b_1\epsilon_{i_1 j_1 i_2 j_2} \epsilon_{i_3 j_3 i_4 j_4} s + b_2\left(\epsilon_{i_1 j_1 i_3 j_3}\epsilon_{i_2 j_2 i_4 j_4}+\epsilon_{i_1 j_1 i_4 j_4}\epsilon_{i_2 j_2 i_3 j_3}\right)s\right.\nonumber\\
    &\hspace{22mm}\left.+b_3\left(\epsilon_{i_1 j_1 i_3 j_3}\epsilon_{i_2 j_2 i_4 j_4}t+\epsilon_{i_1 j_1 i_4 j_4}\epsilon_{i_2 j_2 i_3 j_3}u\right)\right. \nonumber\\
    &\hspace{22mm} \left. +b_4\epsilon_{i_1 j_1 i_2 j_2}\epsilon_{i_3 j_3 i_4 j_4}\frac{tu}{s}+b_5\left(\epsilon_{i_1 j_1 i_3 j_3}\epsilon_{i_2 j_2 i_4 j_4}\frac{t^2}{s}+\epsilon_{i_1 j_1 i_4 j_4}\epsilon_{i_2 j_2 i_3 j_3}\frac{u^2}{s}\right)\right].
\end{align}
Not every choice of coefficients $b_i$ defines a consistent superamplitude however, there are additional super-statistics constraints that we must impose on the component amplitude to ensure that the superamplitude (\ref{4vectorsuperamplitude}) is completely symmetric under relabelling. These take the form
\begin{align}
    \label{N=4simpleidentities}
    A_4\left(\gamma_{i_1 j_1}^+,\gamma_{i_2 j_2}^+,\gamma_{i_3 j_3}^-,\gamma_{i_4 j_4}^-\right) &= \frac{\langle 34\rangle^4}{\langle 14\rangle^4 }  \left[A_4\left(\gamma_{i_1 j_1}^+,\gamma_{i_2 j_2}^+,\gamma_{i_3 j_3}^-,\gamma_{i_4 j_4}^-\right)\right]_{1\leftrightarrow 3} \nonumber\\
    A_4\left(\gamma_{i_1 j_1}^+,\gamma_{i_2 j_2}^+,\gamma_{i_3 j_3}^-,\gamma_{i_4 j_4}^-\right) &= \frac{\langle 34\rangle^4}{\langle 13\rangle^4 }  \left[A_4\left(\gamma_{i_1 j_1}^+,\gamma_{i_2 j_2}^+,\gamma_{i_3 j_3}^-,\gamma_{i_4 j_4}^-\right)\right]_{1\leftrightarrow 4}.
\end{align}
Imposing these constraints on (\ref{masslessvector4pt}), we find a unique non-trivial solution
\begin{equation}
    b_1 = b_3, \hspace{5mm} b_2 = b_4 = b_5 =0,
\end{equation}
or equivalently the complete superamplitude
\begin{align}
    &\mathcal{A}_4\left(\Gamma_{i_1 j_1},\Gamma_{i_2 j_2 },\Gamma_{i_3 j_3},\Gamma_{i_4 j_4}\right) \nonumber\\
    &\propto \delta^{(8)}\left(Q^\dagger\right)\frac{[12][34]}{\langle 12 \rangle \langle 34\rangle }\left[\epsilon_{i_1 j_1 i_2 j_2} \epsilon_{i_3 j_3 i_4 j_4} s+\epsilon_{i_1 j_1 i_3 j_3}\epsilon_{i_2 j_2 i_4 j_4}t+\epsilon_{i_1 j_1 i_4 j_4}\epsilon_{i_2 j_2 i_3 j_3}u\right].
\end{align}
Here we see how to evade the no-go theorem of \cite{Elvang:2021qhq}. In the absence of the $SU(4)_{\text{global}}$ tensor structures the above analysis gives the same result but without the Levi-Civita symbols, and so the unique $\mathcal{N}=4$ compatible amplitude at six-derivative order is proportional to $s+t+u=0$. 

This argument does \textit{not} imply that $\mathcal{N}=4$ massive gravity is supersymmetric at quartic order, or that it is consistent with any choice of $\alpha_4$. Rather we have shown that the quartic Galileon interactions in the decoupling limit do not by themselves give any additional constraints. To determine the true quartic order constraints requires a detailed massive superspace analysis that is left to future work.

\vspace{3mm}
\noindent \textbf{Acknowledgements}

\vspace{3mm}
We would like to thank James Bonifacio, Claudia de Rham, Henriette Elvang, Lavinia Heisenberg, Aidan Herderschee, and Kurt Hinterbichler for useful comments and discussions. The work of SP was supported by the U.S. Department of Energy grant DE-SC0009999, funds from the University of California and a Barbour Scholarship from Rackham at the University of Michigan. CRTJ would like to thank QMAP at UC Davis for their hospitality during the completion of this work.

\appendix

\section{Conventions}
\label{appendix:conventions}
\noindent \textbf{Lorentz and supersymmetry conventions}
\vspace{3mm}

\noindent We will assume the mostly-plus metric convention 
\begin{equation}
    \eta^{\mu\nu} = 
    \begin{pmatrix}
        -1 & 0 & 0 & 0 \\
        0 & +1 & 0 & 0 \\
        0 & 0 & +1 & 0 \\
        0 & 0 & 0 & +1
    \end{pmatrix}.
\end{equation}
The on-shell condition for a massive momentum $p^\mu$ is therefore $p^2 = -m^2$. Our spinor conventions follow \cite{Elvang:2009wd,Srednicki:2007qs}, and also the mostly-plus version of \cite{Dreiner:2008tw}. In particular our Pauli matrices take the form
\begin{align}
    \sigma^0 &= \overline{\sigma}^0 = \begin{pmatrix}
        1 & 0 \\
        0 & 1
    \end{pmatrix}\,, \hspace{15mm} \sigma^1 = -\overline{\sigma}^1 = \begin{pmatrix}
        0 & 1 \\
        1 & 0
    \end{pmatrix}\,, \nonumber\\
    \sigma^2 &= -\overline{\sigma}^2 = \begin{pmatrix}
        0 & -i \\
        i & 0
    \end{pmatrix}\,, \hspace{10mm} \sigma^3 = -\overline{\sigma}^3 = \begin{pmatrix}
        1 & 0 \\
        0 & -1
    \end{pmatrix}.
\end{align}
For both Lorentz spinors and $SU(2)_{\text{LG}}$ the Levi-Civita symbol is defined as
\begin{align}
\label{levicivita}
   \epsilon^{\alpha \beta}= \epsilon^{\dot{\alpha} \dot{\beta}}= \epsilon^{IJ}=  \begin{pmatrix}
    0 & 1 \\
    -1 & 0 \\
    \end{pmatrix},\hspace{10mm}   \epsilon_{\alpha \beta}= \epsilon_{\dot{\alpha} \dot{\beta}}= \epsilon_{IJ}=  \begin{pmatrix}
    0 & -1 \\
    1 & 0 \\
    \end{pmatrix}.
\end{align}
Our supersymmetry conventions follow \cite{Elvang:2009wd,Srednicki:2007qs}. In particular we assume the following form of the super Poincar\'{e} algebra
\begin{align}
    \label{susyalgebra}
    [M^{\mu\nu},P^\rho] &= i\left(P^\mu \eta^{\nu\rho} - P^\nu \eta^{\mu\rho}\right)\,, \nonumber\\
    [M^{\mu\nu},M^{\rho\sigma}] &= i\left(M^{\mu\sigma}\eta^{\nu\rho}-M^{\nu\sigma}\eta^{\mu\rho}+M^{\nu\rho}\eta^{\mu\sigma}-M^{\mu\rho}\eta^{\nu\sigma}\right)\,, \nonumber\\
    \{Q^a_\alpha,Q_{b\dot{\alpha}}^\dagger\} &=- 2\delta^a_b \sigma^\mu_{\alpha \dot{\alpha}}P_\mu\,,\nonumber\\
    \{Q^a_\alpha,Q^b_{\beta}\} &=0\,,\nonumber\\
    \{Q^\dagger_{a\dot{\alpha}},Q^\dagger_{b\dot{\beta}}\} &=0\,,\nonumber\\
    [Q^a_\alpha,M^{\mu\nu}] &= {(\sigma^{\mu\nu})_{\alpha}}^\beta Q^a_\beta\,, \nonumber\\
    [Q^\dagger_{a\dot{\alpha}},M^{\mu\nu}] &= {(\overline{\sigma}^{\mu\nu})_{\dot{\alpha}}}^{\dot{\beta}}Q_{a\dot{\beta}}^\dagger\,.
\end{align}
As discussed in Section (\ref{sec:multiplets}), we are assuming the absence of central charges.

\vspace{3mm}
\noindent \textbf{Massive Spinors}

\vspace{3mm}
\noindent For a massive particle, we define the massive spinors from the bispinor representation of the momentum
\begin{equation}
\label{massspinor}
    p_{i\mu}\sigma^\mu_{\alpha\dot{\alpha}} = p_{i\alpha\dot{\alpha}} = |i_I]_\alpha \langle i^I |_{\dot{\alpha}}, \hspace{10mm} p_{i\mu}\overline{\sigma}^{\mu \dot{\alpha}\alpha} = p_i^{\dot{\alpha}\alpha} = -|i_I\rangle^{\dot{\alpha}}[i^I|^\alpha.
\end{equation}
The little-group indices can be raised and lowered with the Levi-Civita symbol 
\begin{equation}
    |i^I] = \epsilon^{IJ}|i_J], \hspace{5mm} [i^I| = \epsilon^{IJ}[i_J|, \hspace{5mm} |i^I\rangle = \epsilon^{IJ}|i_J\rangle, \hspace{5mm} \langle i^I| = \epsilon^{IJ}\langle i_J|.
\end{equation}
In addition to the redundancy corresponding to the action of the massive little group $SU(2)_{\text{LG}}$, the definition (\ref{massspinor}) has an additional redundancy of the form 
\begin{equation}
    |i^I] \rightarrow t |i^I], \hspace{10mm} |i^I\rangle \rightarrow t^{-1}|i^I\rangle,
\end{equation}
for arbitrary $t\neq 0$. We fix this by imposing the normalization conditions
\begin{equation}
\label{normalization}
    \langle i^I i^J\rangle = m\epsilon^{IJ}, \hspace{10mm} [i^I i^J] = -m\epsilon^{IJ}.
\end{equation}
By making this choice we can derive the Weyl equations 
\begin{align}
\label{Weyl}
    &p_i|i^I] = -m|i^I\rangle\,, &&p_i|i^I\rangle = -m|i^I]\,, \nonumber\\
     &[i^I|p_i = m\langle i^I|\,, &&\langle i^I|p_i = m[i^I|,
\end{align}
and the completeness relations
\begin{equation}
    |i_I]_\alpha [i^I|^\beta = m\delta^\beta_\alpha\,, \hspace{10mm} |i_I\rangle^{\dot{\alpha}} \langle i^I|_{\dot{\beta}} = - m\delta^{\dot{\alpha}}_{\dot{\beta}}\,.
\end{equation}
For real momenta the massive spinors satisfy the following Hermiticity conditions
\begin{equation}
    \left(|p_I]_\alpha\right)^{\dagger} = -\langle p^I|_{\dot{\alpha}}\,, \hspace{10mm} \left(\langle p_I|_{\dot{\alpha}}\right)^\dagger = |p^I]_\alpha\,.
\end{equation}
In this formalism, particles of spin $s$ are represented as symmetric tensors with $2s+1$ indices. Thus we will often need to explicitly symmetrize over little group indices. We will use the conventions
\begin{align}
    X^{(I_1,\cdots,I_n)}=\sum_{\sigma\in S_n} X^{I_{\sigma_1},\cdots,I_{\sigma_n}}
\end{align}
where $S_n$ is the symmetric group, i.e. the set of all $n!$ permutations of $n$ labels.

\vspace{3mm}
\noindent \textbf{High-Energy Limit}

\vspace{3mm}
\noindent Following the results of \cite{Arkani-Hamed:2017jhn}, we can take the massless limit of massive amplitudes by defining the massive spinors as
\begin{align}
    &|i^I\rangle^{\dot{\alpha}} =\left(|i\rangle^{\dot{\alpha}}\,\,|\eta_i\rangle^{\dot{\alpha}}\right),\quad \langle i^I|_{\dot{\alpha}} =\left(\langle i|_{\dot{\alpha}}\,\,\langle\eta_i|_{\dot{\alpha}}\right),\nonumber\\
    &|i_I]_{\alpha} =\left(|i]_{\alpha}\,\,|\eta_i]_{\alpha}\right),\quad [i_I|^{\alpha} =\left([i|^{\alpha}\,\,[\eta_i|^{\alpha}\right),
\end{align}
where the little group indices can be raised and lowered with the Levi-Civita tensors (\ref{levicivita}). For convenience, the spinors with lowered little group indices are given here:
\begin{align}
    &|i_I\rangle^{\dot{\alpha}} =\left(-|\eta_i\rangle^{\dot{\alpha}}\,\,|i\rangle^{\dot{\alpha}}\right),\quad \langle i_I|_{\dot{\alpha}} =\left(-\langle\eta_i|_{\dot{\alpha}}\,\,\langle i|_{\dot{\alpha}}\right),\nonumber\\
    &|i^I]_{\alpha} =\left(|\eta_i]_{\alpha}\,\,-|i]_{\alpha}\right),\quad [i^I|^{\alpha} =\left([\eta_i|^{\alpha}\,\,-[i|^{\alpha}\right).
\end{align}

Although for most of the results given here, we won't need explicit expressions in terms of momentum and their angles, the spinors can be given explicitly for momentum $p^\mu=(E,p\textrm{ sin}\,\theta\, \textrm{cos}\,\phi,p\textrm{ sin}\,\theta\, \textrm{sin}\,\phi,p \textrm{ cos}\,\theta)$ as 
\begin{align}
|i\rangle^{\dot{\alpha}}&=\sqrt{E+p}\begin{pmatrix}c \\s\end{pmatrix}, && |\eta_i\rangle^{\dot{\alpha}}=\sqrt{E-p}\begin{pmatrix}-s^* \\c\end{pmatrix},\nn\\
\langle i|_{\dot{\alpha}}&=\sqrt{E+p}\begin{pmatrix}-s \\c\end{pmatrix}, && \langle\eta_i|_{\dot{\alpha}}=\sqrt{E-p}\begin{pmatrix}-c \\-s^*\end{pmatrix}\nn\\
|i]_{\alpha}&=\sqrt{E+p}\begin{pmatrix}-s^* \\c\end{pmatrix}, && |\eta_i]_{\alpha}=\sqrt{E-p}\begin{pmatrix}-c \\-s\end{pmatrix},\nn\\
[ i|^{\alpha}&=\sqrt{E+p}\begin{pmatrix}c \\s^*\end{pmatrix}, && [\eta_i|^{\alpha}=\sqrt{E-p}\begin{pmatrix}-s \\c\end{pmatrix},
\end{align}
where $c=\textrm{cos}\left(\frac{\theta}{2}\right)$ and $s=\textrm{sin}\left(\frac{\theta}{2}\right)e^{i\phi}$.
From our definitions, we see that 
\begin{align}
    \langle i^Ii^J\rangle=\begin{pmatrix}
    \langle i\,i\rangle & \langle i\,\eta_i\rangle \\
    \langle \eta_i\,i\rangle & \langle \eta_i\,\eta_i\rangle \\
    \end{pmatrix}=m\,\epsilon^{IJ}
\end{align}
and 
\begin{align}
    [i^Ii^J]=\begin{pmatrix}
    [ i\,i] & [ i\,\eta_i] \\
    [ \eta_i\,i] & [ \eta_i\,\eta_i] \\
    \end{pmatrix}=-m\,\epsilon^{IJ},
\end{align}
consistent with our normalization conditions (\ref{normalization}). 
For massive spin 2 amplitudes, each external state is a totally symmetric rank-4 $SU(2)_{\text{LG}}$ tensor. There are five physical polarizations for a massive spin-2 and in this formalism they correspond to the possible little group index combinations:
\begin{equation}
    \{(1,1,1,1)\,,(1,1,1,2)\,,(1,1,2,2)\,,(1,2,2,2)\,,(2,2,2,2)\}\,.
\end{equation}
In the massless limit, these correspond to the helicity states $(h^-,v^-,\phi,v^+,h^+)$. After symmetrizing over all the particles' little group indices and considering all possible helicities, we will have an array of size $5\times5\times5$, each element consisting of functions of $\langle ij\rangle,[ij],\langle i\eta_j\rangle,[ i\eta_j],\langle \eta_i j\rangle,[ \eta_i j],\langle \eta_i\eta_j\rangle$ and $[ \eta_i\eta_j]$. When doing the high energy expansion, 
\begin{align}
    \sqrt{E+p}\approx \sqrt{2E}\left(1-\frac{m^2}{8E^2}+...\right),\quad \sqrt{E-p}\approx \sqrt{2E}\left(\frac{m}{2E}+\frac{m^3}{16E^3}+...\right).
\end{align}
Due to this, the leading part of $|i\rangle,|i]$ will be their massless counterparts, while $|\eta_i\rangle,|\eta_i]$ will be higher order in m. The caveat is that due to special 3-particle kinematics, for the high energy limit, we have to pick whether to work with either angles or squares as one or the other will vanish. For example, if we pick the angles to be non-vanishing, something like $[ij]$ can be smaller than $[i\eta_j]$ just due to special kinematics. We can remove square brackets using identities from momentum conservation such as
\begin{align}
    [ij]\langle jk\rangle&=-[i\eta_j]\langle \eta_j k\rangle-[i\eta_i]\langle \eta_i k\rangle-[i\eta_k]\langle \eta_k k\rangle\nn\\
   \rightarrow [ij] &=(m \langle\eta_i k\rangle +m[i \eta_k]-[i \eta_j]\langle \eta_j k\rangle)/\langle jk\rangle
\end{align}
as well as
\begin{equation}
    [i,\eta_j]=-m\frac{\langle jk\rangle}{\langle ki\rangle},\quad \textrm{and} \quad [\eta_i\eta_j]=-\frac{m^2 \langle ij \rangle}{4 E_i E_j}, \quad \textrm{for }i\neq j .
\end{equation}
Similar identities can be used when the square brackets are non-vanishing.

\section{Spinor Braces}
\noindent\textbf{Proof that brace polynomials form a spanning set}
\label{bracesequivalent}

\vspace{3mm}
\noindent
In this Appendix we will give a short argument justifying the claim that polynomials in spinor braces (\ref{braces}) span the space of all massive 3-particle amplitudes and $F$-functions.  

The objects we consider are functions of 3-particle, massive, on-shell kinematics, with each external state labelled by an $SU(2)_{\text{LG}}$ tensor which we will not assume corresponds to an irreducible representation. The only objects available with $SU(2)_{\text{LG}}$ indices are the Levi-Civita symbols $\epsilon^{IJ}$ and massive spinors $[i^I|$ and $\langle i^I|$. Using the normalization condition (\ref{normalization}) and the Weyl equations (\ref{Weyl}), the expression can always be rewritten so that the $SU(2)_{\text{LG}}$ indices appear only in \textit{square} massive spinors. The general form of such an expression for rank-$n_i$ little-group tensors for particle $i$ is 
\begin{equation}
    \label{Hexp}
    \sim [1|^{n_1} [2|^{n_2} [3|^{n_3} H(p_1,p_2,p_3), 
\end{equation}
where $H$ is an object with $n_1+n_2+n_3$ lowered, undotted Lorentz spinor indices. The only available objects with these spinor indices are
\begin{equation}
    \epsilon_{\alpha \beta},\hspace{5mm} \sigma^\mu_{\alpha\dot{\alpha}} \overline{\sigma}^{\nu \dot{\alpha}\beta}\epsilon_{\beta\gamma},\hspace{5mm} \sigma^\mu_{\alpha\dot{\alpha}} \overline{\sigma}^{\nu \dot{\alpha}\beta}\sigma^\rho_{\beta\dot{\beta}} \overline{\sigma}^{\kappa \dot{\beta}\gamma}\epsilon_{\gamma\delta}, \hspace{5mm}...
\end{equation}
and so on with longer strings of $\sigma \overline{\sigma}\sigma \overline{\sigma}\sigma...$. The free Lorentz indices must be contracted with either a momentum $p_{i\mu}$ or an invariant tensor $\eta_{\mu\nu}$ or $\epsilon_{\mu\nu\rho\sigma}$. In the latter two cases well-known identities always allow us to rewrite the expression as products of lower-order monomials of this kind with Lorentz indices contracted with momenta \cite{Dreiner:2008tw}. The general expression (\ref{Hexp}) can therefore always be written as a polynomial of Lorentz-invariant (and little group-covariant) monomials of the form
\begin{equation}
    [i^{I}|\slashed{p}_{a_1} ...\slashed{p}_{a_{2n}}|j^J],
\end{equation}
where $a_k \in \{1,2,3\}$ for cubic amplitudes. Further, we can always remove $p_3$ using momentum conservation and so it is sufficient to consider binary strings of $\slashed{p}_1$ and $\slashed{p}_2$. Next we make use of the Clifford algebra identity 
\begin{equation}
\label{Clifford}
    \slashed{p}_a \slashed{p}_b + \slashed{p}_b \slashed{p}_a = -2 (p_a \cdot p_b),
\end{equation}
suppressing spinor indices. If $n=0$ then this can be used to reintroduce momenta and express the monomial in terms of braces\footnote{In the exceptional case $m_3^2=m_1^2+m_2^2$ a different pair of internal momenta should be chosen to define the braces, \textit{e.g.} $\{ij\}\equiv [i|\slashed{p}_2 \slashed{p}_3|j]$. This is irrelevant for this paper since every state is mass degenerate. }
\begin{equation}
    [i^I j^J] = -\frac{1}{m_1^2+m_2^2-m_3^3}\left(\{i^I j^J\}-\{j^J i^I\}\right).
\end{equation}
For $n>0$, (\ref{Clifford}) can be used to either delete pairs of adjacent identical momenta or re-order non-identical pairs. By some finite sequence of moves of this type, each monomial can always be reduced to spinor braces multiplied by a scalar function of masses. We therefore conclude that polynomials in spinor braces give a spanning set for 3-particle massive scattering amplitudes and $F$-functions.

\vspace{3mm}
\noindent\textbf{Notation translations}
\label{zptospinor}

\vspace{3mm}
\noindent In Section \ref{sec:susy3point}, we wrote a basis of massive spin-2 3-point amplitudes (\ref{cubicamps}) in a simple form by introducing a set of auxiliary vectors $z_i^\mu$. To compare these amplitudes to the explicit projections of the 3-particle superamplitudes constructed in Section \ref{sec:susy3point} we need the following replacement rules derived using (\ref{ztoe}). In index suppressed notation, we find for the $z_i\cdot z_j$
\begin{align}
    z_1\cdot z_2 \rightarrow &\frac{1}{4m^6}\left(\{\mathbf{12}\}^2-\{\mathbf{12}\}\{\mathbf{21}\}\right)\,,\nonumber\\
    z_1\cdot z_3 \rightarrow &\frac{1}{4m^6}\left(\{\mathbf{31}\}^2-\{\mathbf{13}\}\{\mathbf{31}\}\right)\,,\nonumber\\
    z_2\cdot z_3 \rightarrow &\frac{1}{4m^6}\left(\{\mathbf{23}\}^2-\{\mathbf{23}\}\{\mathbf{32}\}\right)\,.
\end{align}
Similarly for the $p_i\cdot z_j$ we have
\begin{align}
    p_2\cdot z_1 \rightarrow &\frac{1}{2\sqrt{2}m^2}\{\mathbf{11}\}\,,\nonumber\\
    p_3\cdot z_2\rightarrow &\frac{1}{2\sqrt{2}m^2}\{\mathbf{22}\}\,,\nonumber\\
    p_1\cdot z_3 \rightarrow &\frac{1}{2\sqrt{2}m^2}\{\mathbf{33}\}\,.
\end{align}
Finally, for the odd-parity structures, we have 
\begin{align}
       \epsilon\left(p_1,p_2,\epsilon_1,\epsilon_2\right) &\rightarrow \frac{i}{2m^4}\left[-\{\mathbf{12}\}^2+3 \{\mathbf{21}\} \{\mathbf{12}\}-2 \{\mathbf{21}\}^2+\{\mathbf{11}\} \{\mathbf{22}\}\right]\nonumber\\
       \epsilon\left(p_1,p_2,\epsilon_1,\epsilon_3\right) &\rightarrow \frac{-i}{2m^4}\left[-\{\mathbf{31}\}^2+3 \{\mathbf{31}\} \{\mathbf{13}\}-2 \{\mathbf{13}\}^2+\{\mathbf{11}\} \{\mathbf{33}\}\right]\nonumber\\
       \epsilon\left(p_1,p_2,\epsilon_2,\epsilon_3\right) &\rightarrow\frac{i}{2m^4}\left[-\{\mathbf{23}\}^2+3 \{\mathbf{32}\} \{\mathbf{23}\}-2 \{\mathbf{32}\}^2+\{\mathbf{22}\} \{\mathbf{33}\}\right].
\end{align}

\vspace{3mm}
\noindent\textbf{Bose/Fermi symmetry}
\label{BFrelation}

\vspace{3mm}
\noindent
To impose the super-statistics constraints (\ref{superstatistics12}) on a general ansatz for the $F$-functions constructed from spinor braces we need the following relations.

\noindent For $1\leftrightarrow 2$: 
\begin{align}
    &\{1^I 1^J\} \xleftrightarrow{\;12\;\;} -\{2^J 2^I\}, \hspace{5mm} \{3^I 3^J\} \xleftrightarrow[]{\;12\;\;} -\{3^J 3^I\}, \hspace{5mm}\{1^I 2^J\} \xleftrightarrow[]{\;12\;\;} -\{1^J 2^I\}, \nonumber\\
    &\{1^I 3^J\} \xleftrightarrow[]{\;12\;\;} -\{3^J 2^I\}, \hspace{5mm} \{2^I 1^J\} \xleftrightarrow[]{\;12\;\;} -\{2^J 1^I\}, \hspace{5mm} \{2^I 3^J\} \xleftrightarrow[]{\;12\;\;} -\{3^J 1^I\}.
\end{align}
For $1\leftrightarrow 3$:
\begin{align}
    &\{1^I 1^J\} \xleftrightarrow[]{\;13\;\;} -\{3^J 3^I\}, \hspace{5mm} \{2^I 2^J\} \xleftrightarrow[]{\;13\;\;} -\{2^J 2^I\}, \hspace{5mm} \{1^I 2^J\} \xleftrightarrow[]{\;13\;\;} -\{2^J 3^I\}\nonumber\\
    &\{1^I 3^J\} \xleftrightarrow[]{\;13\;\;} -\{1^J 3^I\},\hspace{5mm} \{3^I 1^J\} \xleftrightarrow[]{\;13\;\;} -\{3^J 1^I\}, \hspace{5mm} \{3^I 2^J\} \xleftrightarrow[]{\;13\;\;} -\{2^J 1^I\}.
\end{align}
For $2\leftrightarrow 3$:
\begin{align}
    &\{1^I 1^J\} \xleftrightarrow[]{\;23\;\;} -\{1^J 1^I\}, \hspace{5mm} \{2^I 2^J\} \xleftrightarrow[]{\;23\;\;} -\{3^J 3^I\}, \hspace{5mm} \{1^I 2^J\} \xleftrightarrow[]{\;23\;\;} -\{3^J 1^I\}\nonumber\\
    &\{2^I 1^J\} \xleftrightarrow[]{\;23\;\;} -\{1^J 3^I\},\hspace{5mm} \{2^I 3^J\} \xleftrightarrow[]{\;23\;\;} -\{2^J 3^I\},\hspace{5mm} \{3^I 2^J\} \xleftrightarrow[]{\;23\;\;} -\{3^J 2^I\}.
\end{align}

\section{Constructing Superfields}
\label{appendix:superfields}
\noindent\textbf{Normalization and completeness}

\vspace{3mm}
\noindent
Given the charges of a component field under $U(1)_R$ and $SU(\mathcal{N})_R$, there is a unique choice for the Grassmann polynomial with an appropriate little group weight that multiplies the component field in the superfield. On the other hand, the normalization of this field is not fixed by any symmetries. Here we describe how to fix the normalization of the component fields to get the expressions \eqref{eq:N1sfield}, \eqref{eq:N2sfield}, \eqref{eq:N3sfield} and \eqref{eq:N4sfield}.

Consider for example an $\mathcal{N}=1$ massive scalar superfield $\Phi$, which has been fixed up to component field normalization,
\begin{align}
    \Phi = \alpha_1\phi +\alpha_2\eta_I \lambda^I+\alpha_3 \eta_I \eta^I \bar{\phi}\,.
\end{align}
The normalization condition we choose for the propagator of $\Phi$ is
\begin{align}
    \int d^2 \eta \frac{1}{p^2}\Phi_p \widetilde{\left(\Phi^\dagger_{-p}\right)}\,.
\end{align}
where the $\widetilde{\Box}$ denotes a Grassmann Fourier transform. This must equal the sum of the individual component field propagators,
\begin{align}
    \int d^2 \eta \Phi_p \widetilde{\left(\Phi^\dagger_{-p}\right)}= \phi_p\bar{\phi}_{-p}+\bar{\phi}_p\phi_{-p}+\lambda_{I,p}\lambda^I_{-p}\,.
\end{align}
Thus if a proposed normalization gives us the correct state sum above, the superfield is correct, up to phases that cannot be determined by this method. 

Let us see how this works. First we take the Grassmann Fourier transform of the superfield,
\begin{align}
    \widetilde{\left(\Phi^\dagger_{-p}\right)} &= \mathcal{FT}\left(\alpha_1 \bar{\phi}_{-p} +\alpha_2\eta^\dagger_I \lambda_{-p}^I+ \alpha_3 \eta^\dagger_I \eta^{\dagger I} \phi_{-p}\right)\nonumber\\
    &=-\frac12 \alpha_1 \eta_I \eta^I\bar{\phi}_{-p} -\alpha_2\eta_I \lambda_{-p}^I +2\alpha_3\phi_{-p}\,.
\end{align}
Plugging this into the Grassmann integral, only terms quadratic in Grassmann variables will contribute,
\begin{align}
    \int d^2 \eta \left[-\frac12 \alpha_1^2 \eta_I \eta^I\phi_p\bar{\phi}_{-p}+ 2\alpha_3^2 \eta_I \eta^I\bar{\phi}_p\phi_{-p}- \alpha_2^2 \eta_I \eta_J \lambda_p^I \lambda_{-p}^J\right]\,.
\end{align}
Using the following integrals,
\begin{align*}
    \int d^2 \eta \left[\eta_I \eta^I\right] =&2\,,\\
    \int d^2 \eta \left[\eta_K \eta_L\right] =&- \epsilon_{KL}\,,
\end{align*}
we determine that the correctly normalized $\mathcal{N}=1$ massive scalar superfield is
\begin{align}
    \Phi = \phi + \eta_I \lambda^I+ \frac12 \eta_I \eta^I \bar{\phi}\,.
\end{align}

With the introduction of non-trivial $SU(\mathcal{N})_R$ representations, the calculation is more involved. Nonetheless we follow the same logic to normalize the superfields \eqref{eq:N1sfield}, \eqref{eq:N2sfield}, \eqref{eq:N3sfield} and \eqref{eq:N4sfield}.

\vspace{3mm}
\noindent\textbf{Phases and CPT}

\vspace{3mm}
\noindent
Finally there is also an ambiguity in the relative phases between the components of the superfield that is not fixed by the completeness relation. These phases are in fact completely non-physical since they can always be removed by a suitable unitary transformation on the Hilbert space. This is equivalent to the statement that the representation of the supersymmetry algebra on the space of one-particle states (\ref{N=1oneparticlesusyrep}) is unique up to a unitary isomorphism. 

The phase choice we make here is however correlated with the choice of phases that appear in the action of certain discrete symmetries. For example, any physical model must have an anti-unitary CPT symmetry which, in a supersymmetric model acts on the supercharges as
\begin{equation}
    \label{CPTsusy}
    (CPT)^{-1} Q_\alpha (CPT) = -i Q^\dagger_{\dot{\alpha}}, \hspace{10mm} (CPT)^{-1} Q^\dagger_{\dot{\alpha}} (CPT) = iQ_{\alpha}.
\end{equation}
The action of CPT on a given species of one-un state is defined up to an overall phase \cite{Weinberg:1995mt}. For example for the $\mathcal{N}=1$ massive gravity multiplet, CPT acts on the states as
\begin{align}
    \label{CPTrep}
    CPT|\psi^{IJK}(\vec{p})\rangle &=  \zeta_\psi |\tilde{\psi}^{IJK}(-\vec{p})\rangle\,, \nonumber\\
    CPT|\tilde{\psi}^{IJK}(\vec{p})\rangle &= \zeta_{\tilde{\psi}} |\psi^{IJK}(-\vec{p})\rangle\,,  \nonumber\\
    CPT|\gamma^{IJ}(\vec{p})\rangle &= \zeta_\gamma|\gamma^{IJ}(-\vec{p})\rangle\,, \nonumber\\
    CPT|h^{IJKL}(\vec{p})\rangle &= \zeta_h |h^{IJKL}(-\vec{p})\rangle\,,
\end{align}
where the phases are further constrained by the requirement that $(CPT)^2 = (-1)^F$ \cite{Weinberg:1995mt}. By explicit calculation, (\ref{N=1oneparticlesusyrep}) is not compatible with (\ref{CPTsusy}) and (\ref{CPTrep}) unless some of the CPT phases are non-trivial. Contrarily, if the CPT phases are chosen to be trivial, then compatibility with supersymmetry requires introducing non-trivial phases in the superfield. These two options are related by a unitary isomorphism and therefore lead to the same physics. In this paper, since it is simplest for our purposes, we choose to set the superfield phases to unity.

\bibliographystyle{JHEP}
\bibliography{Draft.bib}
\end{document}